\documentclass[a4paper,11pt]{article}

\usepackage{amsmath, amsfonts, amssymb}
\usepackage{mathtools}
\usepackage{graphicx}
\usepackage{float}
\usepackage{natbib}
\usepackage{geometry}
\usepackage{subcaption}
\usepackage{enumerate}
\usepackage{orcidlink}

\usepackage{dcolumn,booktabs}
\newcolumntype{d}[1]{D{.}{.}{#1}}

\usepackage[symbol]{footmisc}

\usepackage{authblk}

\usepackage{hyperref}

\newcommand{\vect}[1]{\boldsymbol{#1}}

\newcommand{\keywords}[1]{
    \noindent
    \textbf{Keywords:} #1
}

\DeclareRobustCommand{\lcroof}[1]{
  \hbox{\vtop{\vbox{%
      \hrule\kern 1pt\hbox{%
        $\scriptstyle #1$%
        \kern 1pt}}\kern1pt}%
    \vrule\kern1pt}}
\DeclareRobustCommand{\angle}[1]{
  {\lcroof{#1}}}

\title{Structured dispersion modelling for mortality data using a Conway--Maxwell--Poisson specification}

\author[a]{Jackie Siaw Tze Wong\, \orcidlink{0000-0002-0314-6684}}
\author[b]{Emiliano A. Valdez \orcidlink{0000-0003-4037-8385}}

\affil[a]{\small
School of Mathematics, Statistics, and Actuarial Science, University of Essex,  
Wivenhoe Park, Colchester, CO4 3SQ, UK. 
Email: \href{mailto:jw19203@essex.ac.uk}{\texttt{jw19203@essex.ac.uk}}.  
\emph{Corresponding author.}
}

\affil[b]{\small
Department of Mathematics, University of Connecticut,  
Storrs, CT 06269-1009, USA.
Email: \href{mailto:emiliano.valdez@uconn.edu}{\texttt{emiliano.valdez@uconn.edu}}.
}

\begin{document}

\maketitle

\begin{abstract} 

Mortality models that attempt to capture dispersion typically assume a fixed dispersion structure,
an assumption that is rarely satisfied in practice and that can lead to miscalibrated uncertainty and
poor predictive performance. In this paper, we introduce a flexible framework for explicitly modelling dispersion in mortality data using the Conway--Maxwell--Poisson (CMP) distribution, which accommodates underdispersion, equidispersion, and overdispersion within a unified specification. Rather than imposing a global dispersion parameter, the framework allows both the type and degree of dispersion to vary by age and over time, thus capturing structural heterogeneity that simpler models may overlook. A Bayesian formulation treats dispersion as unknown, with prior structures that coherently propagate parameter, process, and distributional uncertainty. Estimation is carried out via Markov chain Monte Carlo (MCMC) methods. Using empirical death data for males in England and Wales, we show that variability in mortality counts differs substantially across ages and across time periods. This has meaningful implications for the calibration of longevity risk and the pricing of annuity products.

\end{abstract}

\vspace{0.2cm}



\section{Introduction}
\label{sec:intro} 
 

Mortality forecasting plays a central role in demography, actuarial science, and public policy, as it informs projections of population dynamics, the sustainability of pension and social security systems, and the financial stability of insurance institutions. A major breakthrough in this field was the introduction of the Lee--Carter (LC) model (\citealp{lca}), which provides a parsimonious yet powerful stochastic framework for decomposing and projecting age-specific mortality rates. Other improved models have since emerged, including the Carins--Blake--Dowd (CBD) family of models (\citealp{cairns2006two,cairns2009quantitative}), model with cohort extensions (\citealp{RH_cohort}), and numerous variants designed to capture other features of mortality dynamics. Comprehensive reviews of subsequent methodological developments can be found in \cite{basellini2023} and \cite{booth2008}. While foundational models such as LC typically assume simple distributions for death counts, most notably the Poisson distribution, a growing body of research has highlighted the importance of appropriately modelling dispersion, namely, the additional variability in death data not captured by the underlying model structure. Failure to account for overdispersion can lead to overly narrow prediction intervals, resulting in an underestimation of longevity risk and potentially significant financial implications, as pointed out by \cite{barrieu2012understanding}.


A substantial body of research has focused on accommodating overdispersion in mortality and longevity modelling. For example, \cite{RH_cohort} employed a quasi-Poisson likelihood framework, introducing a dispersion parameter to capture extra-Poisson variation and thereby increasing the flexibility of the model through several alternative error specifications. \cite{delwarde2007negative} considered a Poisson-Gamma mixture formulation, which gives rise to a negative-binomial version of the Lee--Carter model. Along similar lines, \cite{li2009uncertainty} incorporated an age-specific latent variable to represent mortality heterogeneity; marginalizing over this latent component likewise yields a negative-binomial Lee--Carter model. \cite{wong_2017} presented two overdispersed mortality models based on negative-binomial and Poisson log-normal specifications, respectively, both of which employ a single global dispersion parameter to capture overall mortality heterogeneity. More recently, \cite{rakhmawan2024unifying} and \cite{wong2026bayesianCMP} presented the use of the Conway--Maxwell--Poisson (CMP) distribution for modelling death counts as a more flexible way of accounting for not only overdispersion, but also the possibility of considering equidispersion and underdispersion, depending on the nature of the data.

While overdispersion in death count modelling has been widely acknowledged, it is typically addressed through one global dispersion parameter or latent frailty component. As will be detailed in Section \ref{sec:dispersion_modeling}, there is substantial evidence that dispersion due to heterogeneity should be viewed as a structured quantity that varies across age and time in order to enable more accurate and localised  error calibration in mortality data. This paper contributes to the literature by explicitly introducing age-specific and period-specific dispersion parameters within a flexible CMP framework, thereby capturing residual heterogeneity not explained by standard overdispersion mortality models.

There are similar attempts at capturing such age/period-specific heterogeneity in the actuarial/demographic literature, most notably through frailty-based models, where unobserved individual differences are incorporated as latent variables affecting mortality risks (e.g. \citealp{caswell2014matrix, colchero2020beyond}). In the context of Lee-Carter modelling framework, \cite{carannante2024frailty} proposed several extensions to further allow such heterogeneity in the form of age/time-dependent latent factors. In contrast, relatively little attention has been given to modelling heterogeneity directly through dispersion, particularly in a manner that varies systematically across age or time. Our proposed approach addresses this particular aspect, allowing the structure of heterogeneity to be inferred directly from the mortality data, without necessitating additional external information or the use of latent factors. To the best of our knowledge, this has not been investigated before.

In this paper, we present three main contributions. First, we extend the approach of \cite{wong2026bayesianCMP} by further modifying the CMP specification to allow for a more flexible dispersion structure rather than a single global dispersion parameter. This allows users to infer the underlying dispersion structure, both the type (underdispersion, equidispersion, or overdispersion) and extent, in more detail without pre-specifying it or making any assumptions.

Second, we present a Bayesian implementation of the models specified by adapting the MCMC schemes developed in \cite{wong2026bayesianCMP}. The rationale for considering a Bayesian framework in our context is mainly for coherence in the integration of various sources of uncertainty, and to avoid the two-stage estimation procedure which is prevalent in a frequentist mortality modelling framework (see for example, \citealp{bayespoisson,wong_2017}). Bayesian methods in mortality and longevity modelling have recently received considerable attention; e.g. \cite{Diana_Wong_Pittea_2025}, \cite{goes2025}, \cite{shi2024multi}. A further advantage of Bayesian methods in our context is the role of prior distributions in facilitating the automated estimation of the highly dynamic structure of the dispersion parameters, the structure of which is treated as an unknown entity to be estimated. We also present a prior distribution that preserves the temporal pattern of the estimated dispersion, which can be projected forward in time, allowing the incorporation of such temporal trend and correlation in heterogeneity into the corresponding mortality projection. 

Finally, we demonstrate the proposed methodology using male death data from England and Wales. The empirical results show that the underlying structure of the dispersion derived from the mortality data displays highly irregular behaviour, highlighting the importance of the flexibility offered by the proposed approach relative to conventional overdispersion models with one fixed global dispersion parameter. We also assess the model performance in terms of actuarial quantities to illustrate its practical relevance and significance in actuarial applications.

The rest of this paper is organised as follows. The remainder of Section \ref{sec:intro} introduces the data, notation, and some typical distributions used to model death counts. Section \ref{sec:dispersion_modeling} presents empirical evidence of structural dispersion and introduces a flexible CMP modelling framework to accommodate such features. Section \ref{sec:bayesian} outlines the Bayesian implementation, including the prior specification and the corresponding estimation algorithm. Empirical results based on England and Wales male mortality data, including an actuarial analysis using annuity-due values, are reported in Section \ref{sec:results}. Finally, the paper concludes with the forecasting of estimated time-varying dispersion parameters in Section \ref{sec:forecast_nu_vary_year}, followed by some closing remarks in Section \ref{sec:conclusion}.

\subsection{Data and notation}
\label{sec:data}

Let $d_{xt}$ denote the observed number of deaths for age group $x$ in year $t$, where $x=\{x_1,\ldots,x_A\}$ and $t=\{t_1,\ldots,t_T\}$ represent a set of $A$ different age groups and $T$ different years, respectively. Suppose further that $e_{xt}$ and $\mu_{xt}$ respectively denote the corresponding central exposure to risk and central mortality rate.

The data chosen for illustrative purposes are male death data and the corresponding exposures of England and Wales, extracted from the Human Mortality Database (HMD)\footnote[2]{ See http://www.mortality.org.}. They are classified by single year of age from $0$ to $99$, and years ranging from $1961$ to $2002$. Hence, here we have $\{x_1,\ldots,x_A\}=\{0,\ldots,99\}$ and $\{t_1,\ldots,t_T\}=\{1961,\ldots,2002\}$ with $A=100$ and $T=42$. The data for years 2003-2021 were reserved as a validation set (see Section \ref{sec:results}).

\subsection{Poisson distribution for modelling death counts}
\label{sec:poisson_death}

The Poisson distribution is commonly employed to model death counts (see, for example, \citealp{mccullagh2019generalized} or \citealp{selvin2004statistical}). More recent applications include \cite{hayat2014understanding}, \cite{zero_inflate_poisson}, \cite{covid_poisson_arima} etc. Let $D_{xt}$ be the random variable denoting the number of deaths of age $x$ year $t$, then
\begin{eqnarray}
D_{xt} \vert \mu_{xt} \sim {\rm Poisson}(e_{xt}\mu_{xt}), 
\label{eqn:poisson_Dxt}
\end{eqnarray}
with the corresponding probability mass function (PMF) 
\begin{eqnarray}
\mathbb{P}[D_{xt}=d_{xt} \vert \mu_{xt}] = \frac{(e_{xt}\mu_{xt})^{d_{xt}}\exp(-e_{xt}\mu_{xt})}{d_{xt}!} \propto \mu_{xt}^{d_{xt}}\exp(-e_{xt}\mu_{xt}).
\label{eqn:poisson_pmf_Dxt}
\end{eqnarray}
The full data likelihood is thus
\begin{equation}
f(\boldsymbol{d}|\boldsymbol{\alpha},\boldsymbol{\beta},\boldsymbol{\kappa})=\prod_{x,t}\mathbb{P}[D_{xt}=d_{xt}|\mu_{xt}] \propto \left(\prod_{x,t} \mu_{xt}^{d_{xt}}\right)\exp\left(-\sum_{x,t}e_{xt}\mu_{xt}\right).
\label{eqn:poisson_likelihood}
\end{equation}
This formulation imposes the mean-variance equality, i.e. $\mathbb{E}[D_{xt}]=\text{Var}[D_{xt}]=e_{xt}\mu_{xt}$. This is generally not true for mortality data, which are known to contain substantial excess variations due to mortality heterogeneity.

\subsection{CMP distribution for modelling death counts}
\label{sec:CMP_death}

The issue of overdispersion has been discussed extensively by \cite{wong_2017} in the context of mortality modelling/forecasting. \cite{wong2026bayesianCMP} introduced the use of CMP for modelling death counts in the presence of such overdispersion issue. In particular, they proposed
\begin{equation}
    D_{xt} \vert \mu_{xt},\nu \sim \text{CMP }\left( (e_{xt}\mu_{xt}+\frac{1}{2}-\frac{1}{2\nu})^\nu , \nu\right)\ ,
    \label{eqn:cmp_dxt}
\end{equation}
where $\nu$ is the dispersion parameter to be estimated. Suppose we write $\lambda_{xt}=(e_{xt}\mu_{xt}+\frac{1}{2}-\frac{1}{2\nu})^\nu$ for simplicity, the corresponding PMF is
\[
\mathbb{P}[D_{xt}=d_{xt} | \mu_{xt},\nu] = \frac{1}{Z(\lambda_{xt},\nu)}\frac{\lambda_{xt}^{d_{xt}}}{(d_{xt}!)^\nu}\ , 
\]
where  
\[
Z(\lambda_{xt},\nu)=\sum_{j=0}^{\infty}\frac{\lambda_{xt}^j}{(j!)^\nu}\ .
\]
The full data likelihood is thus
\begin{equation}
f(\boldsymbol{d}|\boldsymbol{\alpha},\boldsymbol{\beta},\boldsymbol{\kappa},\nu)=\prod_{x,t}\mathbb{P}[D_{xt}=d_{xt}|\mu_{xt},\nu] =\prod_{x,t} \frac{1}{Z(\lambda_{xt},\nu)}\frac{(\lambda_{xt})^{d_{xt}}}{(d_{xt}!)^\nu}.
\label{eqn:cmp_likelihood}
\end{equation}


As explained by \cite{wong2026bayesianCMP}, this formulation is designed to give the expected number of deaths as 
\begin{eqnarray}
\mathbb{E}[D_{xt} ] &\approx& e_{xt}\mu_{xt},
\label{eqn:cmp_dxt_mean_var}
\end{eqnarray}
and the implied variance as
\begin{eqnarray}
\text{Var}[D_{xt} ] &\approx& \frac{e_{xt}\mu_{xt}+\frac{1}{2}-\frac{1}{2\nu}}{\nu} = \frac{\mathbb{E}[D_{xt}]+\frac{1}{2}-\frac{1}{2\nu}}{\nu}.
 \label{eqn:var_dxt_cmp}
\end{eqnarray}
So $\nu$ governs the level of overdispersion, where the term $1/\nu$ can also be (approximately) interpreted as the multiplicative increase in the variance of $D_{xt}$ relative to its mean. 
For instance, if $\nu < 1$, we have $\text{Var}[D_{xt} \vert \mu_{xt},\nu] > \mathbb{E}[D_{xt}\vert \mu_{xt},\nu]$ which is the case for overdispersion, and vice versa. 
And when $\nu = 1$, $\text{Var}[D_{xt}\vert \mu_{xt},\nu] = \mathbb{E}[D_{xt}\vert \mu_{xt},\nu]$ indicating the case of equidispersion. 

\subsection{The rate models}
\label{sec:rate_models}

Following \cite{wong2026bayesianCMP}, we consider two commonly-used models for $\mu_{xt}$,
\begin{equation}
\log \mu_{xt} =
\begin{cases}
\alpha_x + \beta_x \kappa_t, \\
\alpha_x + \beta_x \kappa_t + \gamma_c,
\end{cases}
\label{eqn:rate_model}
\end{equation}
coupled with the use of (\ref{eqn:cmp_dxt}), where $\gamma_c$ are cohort parameters with $c=t-x\in\{1,\ldots,C\}$ denoting the cohort index, which represent cohorts born in $\{1861,\ldots,2001\}$, and $C=A+T-1$. For model identifiability, the following constraints are adopted 
\begin{equation}
\begin{cases}
\sum_{x}\beta_x=1 \mbox{ \ \  and  \ \ } \sum_{t}\kappa_t=0, \\
\sum_{x}\beta_x=1 \mbox{ \ \  and  \ \ } \sum_{t}\kappa_t=\sum_c \gamma_c=\sum_c c\gamma_c=\sum_c c^2\gamma_c=0.
\end{cases}
\label{eqn:LC_constraints}
\end{equation}
We term the former the LC-based models and the latter the LCC-based models. Combined with the specification in (\ref{eqn:poisson_Dxt}), these yield the Poisson--LC and Poisson--LCC models, respectively. Likewise, combining the specification in (\ref{eqn:cmp_dxt}) with the LC and LCC rate structures gives rise to the CMP--LC and CMP--LCC models, respectively. The parameters can be interpreted as follows: 
\begin{eqnarray*}
\alpha_x &:& \mbox{ denotes the average log mortality rate at age $x$ over time (i.e., $\alpha_x=\frac{\sum_t \log \mu_{xt}}{T}$).} \\
\beta_x &:& \mbox{ captures the age-specific pattern of mortality improvement, measuring the responsiveness} \\
&& \mbox{ of mortality at each age to overall changes in mortality levels.} \\
\kappa_t &:& \mbox{ represents the overall time trend in mortality, with its impact varying across ages through $\beta_x$.}
\end{eqnarray*}

\section{Structural modelling of dispersion}
\label{sec:dispersion_modeling}

A single global dispersion parameter provides only an overall measure of mortality heterogeneity and may fail to capture important structural features underlying the data. As we demonstrate below, residual variability exhibits systematic patterns across age and calendar year, indicating that the degree of dispersion is not constant throughout the mortality surface. These observations motivate the development of more flexible dispersion specifications that allow heterogeneity to be modelled and calibrated in a localised manner. 

\begin{figure}[htbp]
\includegraphics[scale=0.60]{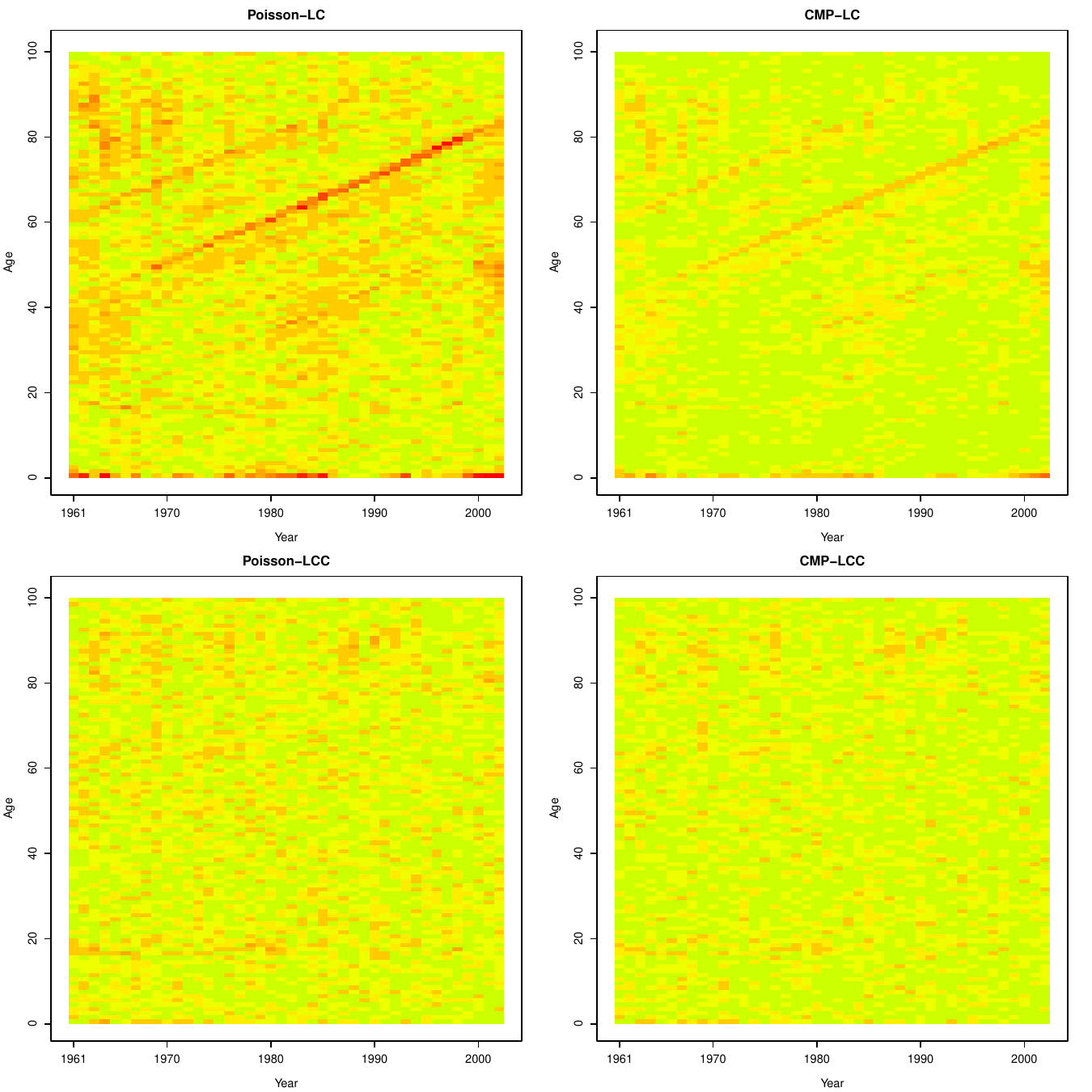}
\includegraphics[width=13.6cm,height=2.0cm]{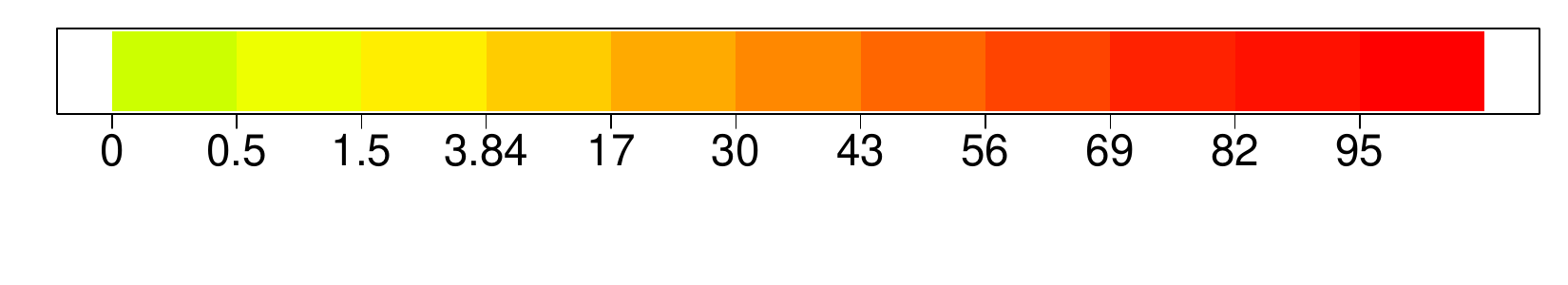}
\caption{Heatmap of $r_{xt}^2(d)$ for the Poisson--LC, CMP--LC, Poisson--LCC, and CMP--LCC models.}
\label{fig:heatmap_compare_LC_LCC_male}
\end{figure}

A visualization of age-specific and period-specific dispersion can be constructed using a heatmap of squared Pearson residuals, defined as 
\begin{equation}
\label{eqn:pearson}
r_{xt}^2(d)
=
\left.
\frac{\bigl(d_{xt}-\mathbb{E}[D_{xt}]\bigr)^2}
     {\operatorname{Var}(D_{xt})}
\right|_{\mu_{xt}=\hat{\mu}_{xt},\,\nu=\hat{\nu}}
=
\begin{cases}
\displaystyle
\frac{(d_{xt}-e_{xt}\hat{\mu}_{xt})^2}
     {e_{xt}\hat{\mu}_{xt}},
& \text{for Poisson}, \\[2ex]
\displaystyle
\frac{(d_{xt}-e_{xt}\hat{\mu}_{xt})^2}
     {\frac{1}{\hat{\nu}}
      \left(
        e_{xt}\hat{\mu}_{xt}
        +\frac{1}{2}
        -\frac{1}{2\hat{\nu}}
      \right)},
& \text{for CMP}.
\end{cases}
\end{equation}
where
\begin{equation*}
\hat{\mu}_{xt} =
\begin{cases}
\exp\!\left(\hat{\alpha}_x + \hat{\beta}_x \hat{\kappa}_t\right),
& \text{for LC--based models}, \\[4pt]
\exp\!\left(\hat{\alpha}_x + \hat{\beta}_x \hat{\kappa}_t + \hat{\gamma}_c\right),
& \text{for LCC--based models}.
\end{cases}
\end{equation*}
Here, $\hat{\alpha}_x$, $\hat{\beta}_x$, $\hat{\kappa}_t$, $\hat{\gamma}_c$, $\hat{\nu}$ denote the corresponding parameter estimates\footnote{In the Bayesian setting, point estimates are taken as posterior means computed from the MCMC samples.}. Figure \ref{fig:heatmap_compare_LC_LCC_male} presents colour-coded heatmaps of $r_{xt}^2(d)$ under the different model specifications, computed according to (\ref{eqn:pearson}), thereby illustrating the distribution of residual dispersion across ages and calendar years.

According to Figure \ref{fig:heatmap_compare_LC_LCC_male}, the residual discrepancies display erratic yet structured patterns across both age and calendar year in all model specifications, with particularly pronounced effects under the LC-based model. Distinct clusters of red and orange patches are visible along both the horizontal (age-specific) and vertical (period-specific) dimensions. Diagonal patterns, attributed to cohort effects (previously discussed in \citealp{wong_2023}), are also evident.

More specifically, pronounced discrepancies at infant ages indicate the presence of additional variability not captured. Horizontal bands around ages $18-23$, $45-50$, $60-65$, $80-95$, and near age 99 provide evidence of age-specific dispersion. Similarly, vertical clusters observed in years such as 1963, 1969, 1985, 1994, and 2000-2002 suggest the presence of period-specific dispersion. Although the inclusion of cohort effects in the LCC-based specification attenuates several of these structured discrepancies, some residual patterns persist, most notably around ages $18-23$, infancy, $80-95$, and in years such as 1969, 1994, and 2002, indicating that additional sources of heterogeneity remain unaccounted for.

The CMP-based models achieve improved overall calibration of data variability. Nevertheless, certain clusters of excess variation persist, particularly at infant ages and around ages $18-23$ and $80-95$. As will be detailed later, these residual patterns can be attributed to the use of a single global dispersion parameter, which effectively averages dispersion across all ages. While accounting for overdispersion at an aggregate level through the CMP specification (as proposed by \citealp{wong2026bayesianCMP}) partially mitigates these discrepancies, it does not fully eliminate the structured residual patterns. In contrast, allowing a more flexible specification of dispersion structures enables a more accurate and localised calibration of the heterogeneous variation present in the data.

We therefore introduce two structural specifications for the dispersion component. These models relax the homogeneity assumption by incorporating additional parameters, thereby permitting a richer representation of the underlying dispersion structure, such as the age-specific and period-specific patterns observed in Figure \ref{fig:heatmap_compare_LC_LCC_male}.

\subsection{Age-specific dispersion levels}
\label{sec:vary_age}

Systematic differences in mortality risk among individuals within the same age group can arise from both observed and unobserved characteristics, a phenomenon we refer to as age-specific heterogeneity. Even conditional on age, individuals are not homogeneous in their survival prospects. Crucially, human mortality can exhibit vastly different levels of heterogeneity across ages, reflecting the evolving influence of biological, behavioural, environmental, and socioeconomic factors over the life course, as described by \cite{barrieu2012understanding}. 

Differences in underlying health status, such as the presence of chronic conditions (e.g., cardiovascular disease or diabetes), generate substantial variation in mortality risk and remaining life expectancy at a given age. Socioeconomic factors such as income, education, and occupational background further contribute to longevity differentials through their influence on access to healthcare, living conditions, and long-term health trajectories. Lifestyle behaviours, such as smoking history, alcohol consumption, diet, and physical activity, also induce persistent mortality disparities within age groups. In addition, genetic predisposition creates variation in susceptibility to age-related diseases, affecting survival outcomes. Beyond these observable characteristics, individuals may differ in latent frailty (unobserved propensity toward higher mortality risk), which leads to excess variability in death counts relative to standard homogeneous mortality models. Geographic location and ethnicity can likewise shape mortality experiences through environmental, structural, and social determinants of health. 

Accounting for age-specific heterogeneity is therefore crucial in mortality and longevity modelling, as failure to do so may underestimate uncertainty, distort projected life expectancy, and obscure the true dynamics of survival improvements across the age distribution.

The model we propose for this purpose extends the CMP specification in (\ref{eqn:cmp_dxt}) by allowing the dispersion parameter to vary across ages. That is,
\begin{equation}
    D_{xt} \vert \mu_{xt},\nu_x \sim \text{CMP }\left( (e_{xt}\mu_{xt}+\frac{1}{2}-\frac{1}{2\nu_x})^{\nu_x} , \nu_x\right)\ ,
    \label{eqn:cmp_dxt_vary_age}
\end{equation}
where $\nu_x$ ($x=1,\ldots,A$) is the set of age-specific dispersion parameters to be estimated. The full data likelihood is
\begin{equation}
f(\boldsymbol{d}|\boldsymbol{\alpha},\boldsymbol{\beta},\boldsymbol{\kappa},\boldsymbol{\nu}_{\text{age}})=\prod_{x,t}\mathbb{P}[D_{xt}=d_{xt}|\mu_{xt},\nu_x] =\prod_{x,t} \frac{1}{Z(\lambda_{xt},\nu_x)}\frac{(\lambda_{xt})^{d_{xt}}}{(d_{xt}!)^{\nu_x}},
\label{eqn:cmp_likelihood_vary_age}
\end{equation} 
where $\boldsymbol{\nu}_{\text{age}}=(\nu_1,\ldots,\nu_A)^\top$ is the vector of age-specific dispersion parameters. These models will be referred to as the CMP--LC--$x$ and CMP--LCC--$x$ models.

\subsection{Period-specific dispersion levels}
\label{sec:vary_year}

Calendar year/period-specific heterogeneity is conceptually similar to age-specific heterogeneity but operates along the temporal dimension rather than across individuals within an age group. It refers to residual variation in mortality risk across calendar years that remains after accounting for systematic period effects. Although stochastic mortality models such as LC incorporate a time-varying index $k_t$ to capture broad mortality trends, this component may not fully explain all year-to-year fluctuations in death counts. Short-term shocks, unobserved aggregate influences, or structural changes\footnote{For example, caused by public health interventions, macroeconomic conditions, environmental shocks, or in an extreme case, epidemic outbreaks.} can induce additional variability that is not otherwise absorbed by the period index. Such residual temporal heterogeneity manifests as excess dispersion in mortality data and motivates the use of flexible distributional frameworks capable of accommodating period-specific variability.

Similarly to (\ref{eqn:cmp_dxt_vary_age}), the proposed model is 
\begin{equation}
    D_{xt} \vert \mu_{xt},\nu_t \sim \text{CMP }\left( (e_{xt}\mu_{xt}+\frac{1}{2}-\frac{1}{2\nu_t})^{\nu_t} , \nu_t\right)\ ,
    \label{eqn:cmp_dxt_vary_year}
\end{equation}
where $\nu_t$ ($t=1,\ldots,T$) is the set of period-specific dispersion parameters to be estimated. The full data likelihood is
\begin{equation}
f(\boldsymbol{d}|\boldsymbol{\alpha},\boldsymbol{\beta},\boldsymbol{\kappa},\boldsymbol{\nu}_{\text{year}})=\prod_{x,t}\mathbb{P}[D_{xt}=d_{xt}|\mu_{xt},\nu_t] =\prod_{x,t} \frac{1}{Z(\lambda_{xt},\nu_t)}\frac{(\lambda_{xt})^{d_{xt}}}{(d_{xt}!)^{\nu_t}},
\label{eqn:cmp_likelihood_vary_year}
\end{equation} 
where $\boldsymbol{\nu}_{\text{year}}=(\nu_1,\ldots,\nu_T)^\top$ is the vector of period-specific dispersion parameters. These models will be referred to as the CMP--LC--$t$ and CMP--LCC--$t$ models.

\section{Bayesian estimation}
\label{sec:bayesian}

The motivation for considering a Bayesian framework has been extensively discussed in the literature (e.g., \citealp{wong_2017}). In our setting, this approach allows the dispersion parameters $\nu_x$ ($x=1,\ldots,A$) and $\nu_t$ ($t=1,\ldots,T$) to be treated as unknown random variables to be estimated naturally and directly from the data. This data-driven specification enables inference on not only the type of dispersion, but also the extent of dispersion, without the need to pre-specify fixed values. In other words, we can infer the underlying age/period structure of dispersion from the data, as is intended by the design of the framework. This also forms a coherent framework for uncertainty calibration, including that due to dispersion and other sources of uncertainty, such as parameter uncertainty through prior distributions. In the following subsections, we describe the prior specification, derive the joint posterior distributions, and outline our proposed MCMC sampling procedure. 

\subsection{Prior specification}
\label{sec:prior}

In this paper, we follow the prior specification of \cite{wong2026bayesianCMP} for most model parameters, while assigning vague prior distributions to the dispersion parameters. This reflects the absence of strong prior knowledge regarding the underlying dispersion structure and allows the corresponding inference to be informed predominantly by the data. Consequently, the Bayesian framework facilitates the estimation of all unknown parameters, while providing a coherent mechanism for learning about the dispersion structure from the observed mortality experience. 

A summary of the priors specified is as follows.
\begin{align*}
\vect{\alpha}&\sim \hspace{0.3cm} N(-5\cdot\vect{1}_{A},4\cdot{I}_{A}), \\
\vect{\beta}&\sim \hspace{0.3cm} N\left(\frac{1}{A}\cdot\vect{1}_{A-1},0.005\cdot\left({I}_{A-1}-\frac{1}{A}\cdot{J}_{A-1}\right)\right), \\
\beta_1 &= 1-\sum_{x=2}^{A}\beta_x, 
\end{align*}
where $\vect{\alpha}=(\alpha_1,\ldots,\alpha_A)^\top$, $\vect{\beta}=(\beta_2,\ldots,\beta_A)^\top$, $\vect{1}_{A}$ is a length$-A$ vector of ones, ${I}_{A}$ and ${J}_{A}$ are respectively the identity matrix and a matrix of ones with dimension $A\times A$.

The prior distribution for $\kappa_t$ is
\begin{eqnarray}
\left.\begin{array}{l c}
\kappa_{t}-\eta_{t}=\rho(\kappa_{t-1}-\eta_{t-1})+\epsilon_{t} , & \mbox{ for } t=2,3,\ldots,T, \\
\kappa_1=\eta_1+\epsilon_1, & \end{array}\right\}
\label{eqn:ar1}
\end{eqnarray}
where $\eta_{t}=\psi_1+\psi_2 t$ denotes the linear drift and $\epsilon_t\stackrel{{\rm ind}}{\sim} N(0,\sigma_{\kappa}^2)$. Note that the constraint $\sum_t \kappa_t=0$ can be applied on (\ref{eqn:ar1}) and expressed jointly as a multivariate distribution. See Appendix A for more details. For other hyperparameters,
\begin{align*}
\frac{\rho+1}{2} &\sim \hspace{0.3cm} {\rm Beta}(3,2), \mbox{ where $\rho \in (-1,1)$}, \\
\sigma_\kappa^{-2} &\sim \hspace{0.3cm} \mbox{Gamma}(1,0.0001), \\
\vect{\psi} &\sim \hspace{0.3cm} N\left(\left(\begin{array}{c}
0 \\
0
\end{array}\right),\left(\begin{array}{c c}
2000 & 0 \\
0 & 2
\end{array}\right)\right), 
\end{align*}

For the cohort parameters $\gamma_c$, an ARIMA(1,1,0) prior is specified, i.e.
\begin{eqnarray}
\left.\begin{array}{l l}
(\gamma_c-\gamma_{c-1})=\rho_\gamma (\gamma_{c-1}-\gamma_{c-2})+\epsilon_c^\gamma , & \mbox{ for } c=3,\ldots,C, \\
\gamma_2-\gamma_1=\frac{1}{\sqrt{1-\rho_\gamma^2}}\epsilon_2^\gamma, & \\
\gamma_1= 100\epsilon_1^\gamma,
\end{array}\right\}
\label{eqn:gamma_projection_model}
\end{eqnarray}
where $\epsilon_c^{\gamma}\stackrel{{\rm ind}}{\sim} N(0,\sigma_\gamma^2)$ for $c=1,\ldots,C$, and $\rho_\gamma$ and $\sigma_\gamma^2$ are hyperparameters. Similarly, the constraints in (\ref{eqn:LC_constraints}) can be applied on (\ref{eqn:gamma_projection_model}) to avoid deterministic adjustment within the MCMC scheme (see Appendix A). We use the following priors for the hyperparameters,
$$\rho_\gamma\sim N(0,1) \mbox{\hspace{1cm} and \hspace{1cm}} \sigma_\gamma\sim \mbox{Uniform}(0.1).$$

As recommended by \cite{wong2026bayesianCMP}, we specify vague prior distributions for the dispersion parameters as
\begin{eqnarray}
    \nu_x \sim \text{Gamma}(1,0.01) \mbox{\ \ \ and \ \ \ } \nu_t \sim \text{Gamma}(1,0.01),
    \label{eqn:prior_nu}
\end{eqnarray}
for $x=1,\ldots,A$ and $t=1,\ldots,T$. This is in the spirit of data-driven inferences, where we allow the data to ``speak for itself''. If there is sufficient information in the data to estimate the dispersion parameters, we can derive the structural distribution of the dispersion of the mortality data.

If the dispersions are assumed to be independent \textit{apriori}, then 
\begin{align*}
\pi(\boldsymbol{\nu}_{\text{age}})&=\prod_{x=1}^{A}\pi(\nu_x)=0.01^A \times \exp\left(-0.01\sum_{x=1}^{A}\nu_x\right), \\
\pi(\boldsymbol{\nu}_{\text{year}})&=\prod_{x=1}^{T}\pi(\nu_t)=0.01^T \times \exp\left(-0.01\sum_{t=1}^{T}\nu_t\right).
\end{align*}

We remark that due to the temporal nature of $\nu_t$, it is sensible to consider (auto-correlated) time-series priors. That way, the dispersion modelled and the corresponding temporal correlations learned from the data can be projected into the future, resembling the act of forecasting of heterogeneity in the mortality data. This will be revisited in Section \ref{sec:forecast_nu_vary_year}.

\subsection{Posterior distributions}

First, denote $\boldsymbol{\theta}_{\text{CMP--LC--}x}=\{\boldsymbol{\alpha},\boldsymbol{\beta},\boldsymbol{\kappa},\vect{\psi},\sigma_\kappa^2,\rho,\boldsymbol{\nu}_{\text{age}}\}$ as the full set of parameter for the CMP--LC--$x$ model. The joint posterior distribution is then
\begin{eqnarray*}
\pi(\boldsymbol{\theta}_{\text{CMP--LC--}x}|\boldsymbol{d}) &\propto & f(\boldsymbol{d}|\boldsymbol{\alpha},\boldsymbol{\beta},\boldsymbol{\kappa},\boldsymbol{\nu}_{\text{age}}) \times \pi(\boldsymbol{\theta}) \\ 
&=& f(\boldsymbol{d}|\boldsymbol{\alpha},\boldsymbol{\beta},\boldsymbol{\kappa},\boldsymbol{\nu}_{\text{age}}) \times \pi(\boldsymbol{\alpha})  \pi(\boldsymbol{\beta})  \pi(\boldsymbol{\kappa}|\vect{\psi},\sigma_\kappa^2,\rho)  \pi(\vect{\psi})  \pi(\sigma_\kappa^2) \pi(\rho)  \pi(\boldsymbol{\nu}_{\text{age}}),
\end{eqnarray*}
where $f(\boldsymbol{d}|\boldsymbol{\alpha},\boldsymbol{\beta},\boldsymbol{\kappa},\boldsymbol{\nu}_{\text{age}})$ is the data likelihood as given in (\ref{eqn:cmp_likelihood_vary_age}), and the rest are the prior distributions as presented in Section \ref{sec:prior}. 
Similarly, for the CMP--LCC-$x$ model, $\boldsymbol{\theta}_{\text{CMP--LCC--}x}=\{\boldsymbol{\alpha},\boldsymbol{\beta},\boldsymbol{\kappa},\vect{\psi},\sigma_\kappa^2,\rho,\boldsymbol{\gamma},\sigma_\gamma^2,\rho_\gamma,\boldsymbol{\nu}_{\text{age}}\}$, giving the joint posterior
\begin{eqnarray*}
\pi(\boldsymbol{\theta}_{\text{CMP--LCC--}x}|\boldsymbol{d}) &\propto& f(\boldsymbol{d}|\boldsymbol{\alpha},\boldsymbol{\beta},\boldsymbol{\kappa},\boldsymbol{\gamma},\boldsymbol{\nu}_{\text{age}})  \pi(\boldsymbol{\alpha})  \pi(\boldsymbol{\beta})  \pi(\boldsymbol{\kappa}|\vect{\psi},\sigma_\kappa^2,\rho)  \pi(\vect{\psi})  \pi(\sigma_\kappa^2) \\
&& \ \ \times \pi(\rho) \pi(\boldsymbol{\gamma}|\sigma_\gamma^2,\rho_\gamma) \pi(\sigma_\gamma^2)  \pi(\rho_\gamma)  \pi(\boldsymbol{\nu}_{\text{age}}).
\end{eqnarray*}
The full parameter variables $\boldsymbol{\theta}_{\text{CMP--LC--}t}$ and $\boldsymbol{\theta}_{\text{CMP--LCC--}t}$ are defined analogously for the CMP--LC--$t$ and CMP--LCC--$t$ models. In what follows, the subscripts corresponding to the model type are suppressed appropriately for brevity.

\subsection{The MCMC updating scheme}

The MCMC updating schemes used in this paper are broadly similar to those proposed in \cite{wong2026bayesianCMP}. For a brief description of the MCMC updating schemes of all other parameters please refer to Appendix B in the Supplementary Materials. Here we describe in detail the MCMC updating step for the dispersion parameters, $\nu_x$ and $\nu_t$. 

When dispersion parameters are independent of each other, it is convenient to update them sequentially within the MCMC scheme. For example for $\nu_x$ under the CMP--LC--$x$, we first note that the conditional posterior distribution of $\nu_x$ is proportional to the likelihood function multiplied by their prior distributions, i.e. 
\begin{equation}
\pi(\nu_x \vert \boldsymbol{d},\boldsymbol{\theta}_{-\nu_x}) \propto f(\boldsymbol{d}|\boldsymbol{\alpha},\boldsymbol{\beta},\boldsymbol{\kappa},\nu_x)  \times \pi(\nu_x), 
\label{eqn:cond_post_nu}
\end{equation}
where $\boldsymbol{\theta}_{-\nu_x}=\boldsymbol{\theta}\backslash\{\nu_x\}=\{\boldsymbol{\alpha},\boldsymbol{\beta},\boldsymbol{\kappa},\vect{\psi},\sigma_\kappa^2,\rho\}$ is the full set of parameters excluding $\nu_x$. MCMC updating scheme is then applied iteratively to sample from (\ref{eqn:cond_post_nu}), which can be combined with other posterior samples to form samples of the joint posterior distribution, $\pi(\boldsymbol{\theta} \vert \boldsymbol{d})$.

To avoid proposing negative values, we perform the updating scheme on a log scale, i. e. 
$$\log (\nu_x^*) \sim N(\log (\nu_x^{(i)}), \sigma^2_{\nu_x}),$$
where $\nu_x^{(i)}$ is the current iterate, $\sigma_{\nu_x}^2$ is the proposal variance calibrated so that the resulting acceptance rate is between 0.15-0.45 (\citealp{acceptrate}). From our pilot runs, a set of appropriate values of $\sigma_{\nu_x}^2$ and $\sigma_{\nu_t}^2$ can be numerically determined, and is presented in Appendix B.

The acceptance ratio for $\nu_x^*$ is
\begin{equation}
\alpha(\nu_x^* \vert \boldsymbol{\theta}_{-\nu_x}^{(i)},\nu_x^{(i)})=\min\left\{1,\frac{f(\boldsymbol{d}|\boldsymbol{\alpha}^{(i)},\boldsymbol{\beta}^{(i)},\boldsymbol{\kappa}^{(i)},\nu_x^*)  \times \pi(\nu_x^*)}{f(\boldsymbol{d}|\boldsymbol{\alpha}^{(i)},\boldsymbol{\beta}^{(i)},\boldsymbol{\kappa}^{(i)},\nu_x^{(i)})  \times \pi(\nu_x^{(i)})}\right\}. 
\label{eqn:accept_nu}
\end{equation}
The MCMC updating scheme proceeds similarly for $\nu_t$ when the independent prior in (\ref{eqn:prior_nu}) is used.

\subsection{Initialisation and Convergence Diagnostics}
\label{sec:initialisation}
    For more details about the initialisation strategy of the parameters for the MCMC procedure, please refer to \cite{wong2026bayesianCMP}. Additionally, the dispersion parameters are initialised as $\nu_x^{(0)}=0.5$ $(x=1,\ldots,A)$ and $\nu_t^{(0)}=0.5$ $(t=1,\ldots,T)$. Note that the initialisation is proposed on the basis of values close to the MLE to speed up convergence, but it should have little effect on the estimation procedure once convergence is attained. 

We apply a burn-in phase of $1{,}000$ iterations to mitigate the effect of initialisation. In addition, we apply 50$^{\rm th}$ posterior sample thinning (collecting one realization every 50 iterations) to each of the parameters to reduce the autocorrelation of these series. After discarding burn-in iterations and applying thinning, we obtain a sample of size $10{,}000$ for each of the parameters under all the models, i.e. the CMP--LC--$x$, CMP--LC--$t$, CMP--LCC--$x$, and CMP--LCC--$t$ models.  	

Before making any inferential comparisons, trace plots and auto-correlation plots (see for example, \citealp[Chapter~3]{bugs}) can be used as diagnostic tools for detecting anomalies in the MCMC generated posterior samples. As an illustration, the trace plots of some of the dispersion parameters, $\nu_x$ and $\nu_t$, and other parameters are presented in Appendix B. These plots indicate strong evidence of MCMC convergence, with proper mixing and no apparent anomaly. The sample auto-correlations also appear to decay fairly quickly after applying thinning, except perhaps $\gamma_c$, which are relatively more correlated. A few commonly used convergence diagnostics such as those proposed by \cite{rubin} and \cite{geweke1991evaluating}, all of which are presented in Appendix C. Specifically, the Gelman's statistic, potential scale reduction factor (PSRF) computed from the posterior samples under all the models considered are all very close to 1, indicating that there is no convergence issue. The Z statistics computed using the approach by \cite{geweke1991evaluating} also satisfy the convergence criterion.

In summary, the MCMC-generated posterior samples seem to be well-behaved and, thus, are ready to be used to perform subsequent computations for accurate inferences to be drawn.

\section{Results}
\label{sec:results}

When presenting our results, we also include, where appropriate, those obtained under the Poisson and CMP specifications (see \citealp{bayespoisson} and \citealp{wong2026bayesianCMP}, respectively) for $D_{xt}$ using Bayesian methods, for comparison purposes. The Poisson specification is as described in Section \ref{sec:poisson_death} and the CMP specification is as described in Section \ref{sec:CMP_death}. The comparison is intended to highlight the significance of accounting for, firstly, overdispersion, and secondly, for incorporating more complex structures of dispersion levels.

\subsection{Estimated parameters}

\begin{figure}[!htbp]
\centering\includegraphics[scale=0.615]{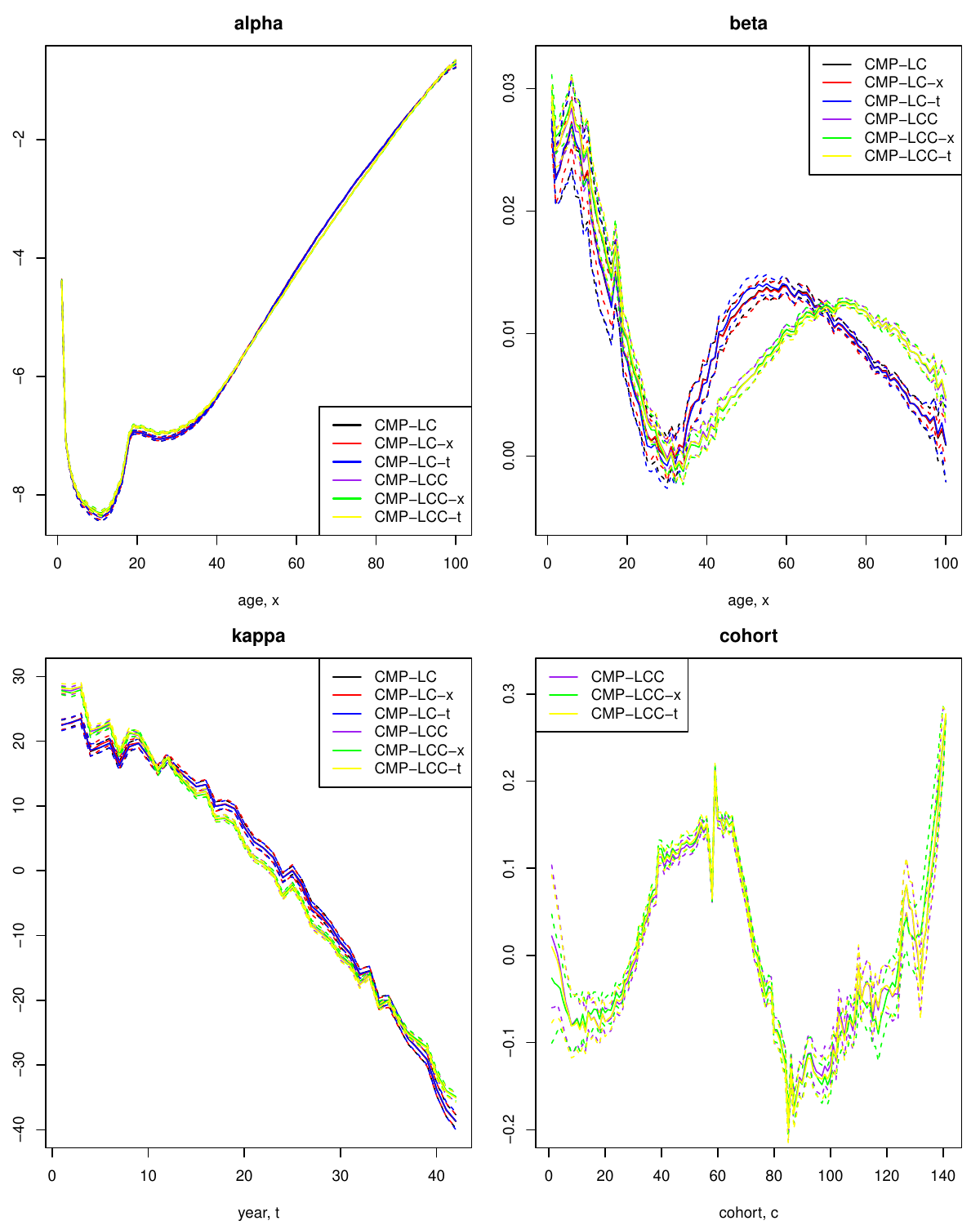}
\caption{\label{fig:estimated_abkc_LC_LCC_male_vary_dispersion}Plot of the medians (solid lines) and $95\%$ credible intervals (dashed lines) of the estimated $\boldsymbol{\alpha}$, $\boldsymbol{\beta}$, $\boldsymbol{\kappa}$, and $\boldsymbol{\gamma}$ under the CMP--LC, CMP--LC--$x$, CMP--LC--$t$, CMP--LCC, CMP--LCC--$x$, and CMP--LCC--$t$ models.}
\end{figure}

Figure \ref{fig:estimated_abkc_LC_LCC_male_vary_dispersion} presents the fitted values (posterior medians) of $\boldsymbol{\alpha}$, $\boldsymbol{\beta}$, $\boldsymbol{\kappa}$, and $\boldsymbol{\gamma}$, together with their corresponding $95\%$ credible intervals, for the CMP--LC, CMP--LC--$x$, CMP--LC--$t$, CMP--LCC, CMP--LCC-- $x$ and CMP--LCC--$t$ models. Overall, the fitted values of are broadly similar within the LC-based and LCC-based model classes (particularly for $\boldsymbol{\alpha}$), which is expected as the primary differences between these specifications lie in the modelling of dispersion rather than the mean structure. Some deviations are observed for $\vect{\beta}$, where the CMP--LC--$x$ yields noticeably different estimates compared to the CMP--LC and CMP--LC--$t$ models. Similarly, the cohort parameters $\vect{\gamma}$ under the CMP--LCC--$x$ specification differ visibly from those obtained under the CMP--LCC and CMP--LCC--$t$ models. While these differences are relatively modest, they highlight the impact of introducing flexible, structure-specific dispersion into the model. Such flexibility can influence parameter estimation and may induce differences in other aspects of the model calibration, as discussed later.

\begin{figure}[!htbp]
\centering
\includegraphics[scale=0.7]{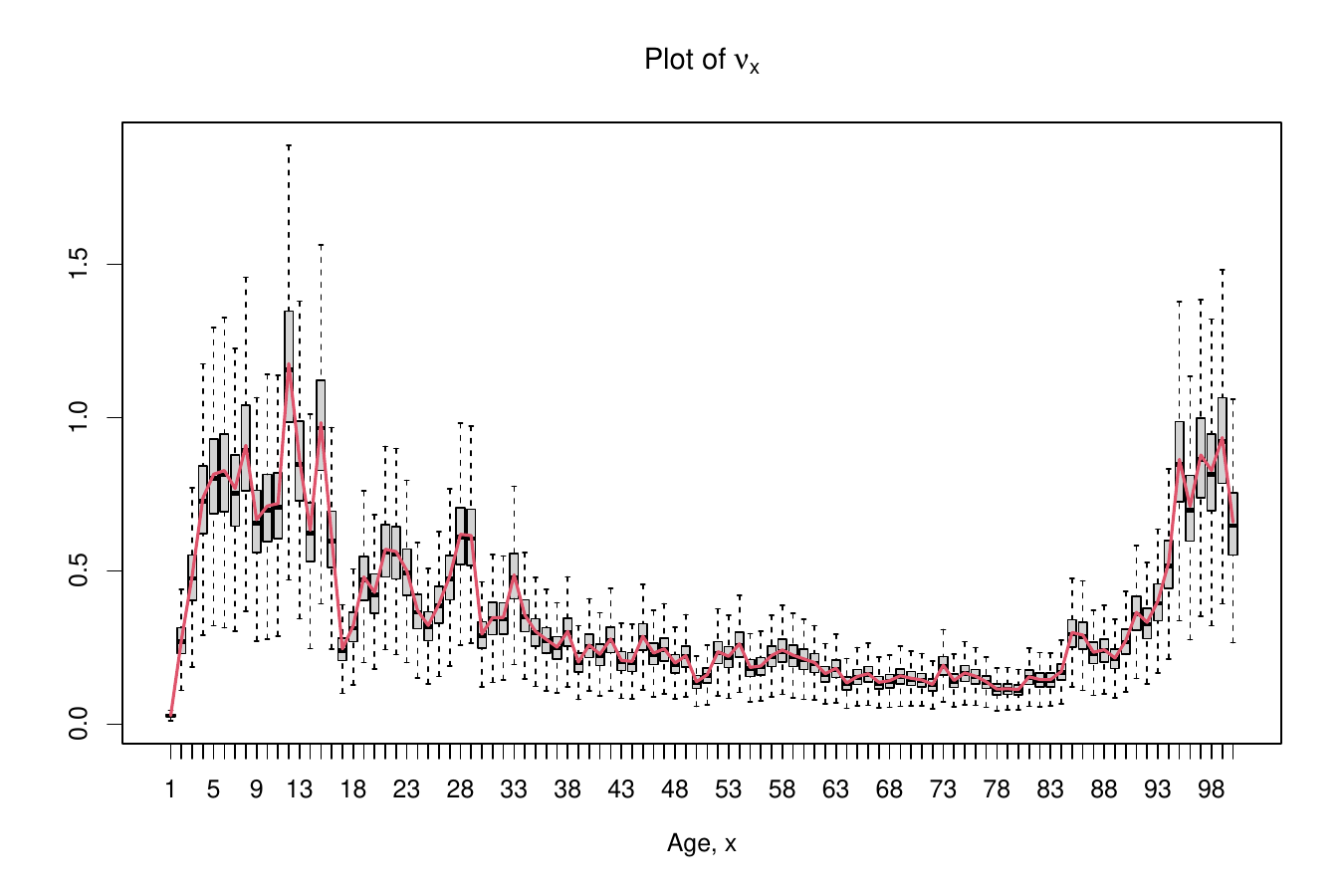}
\includegraphics[scale=0.7]{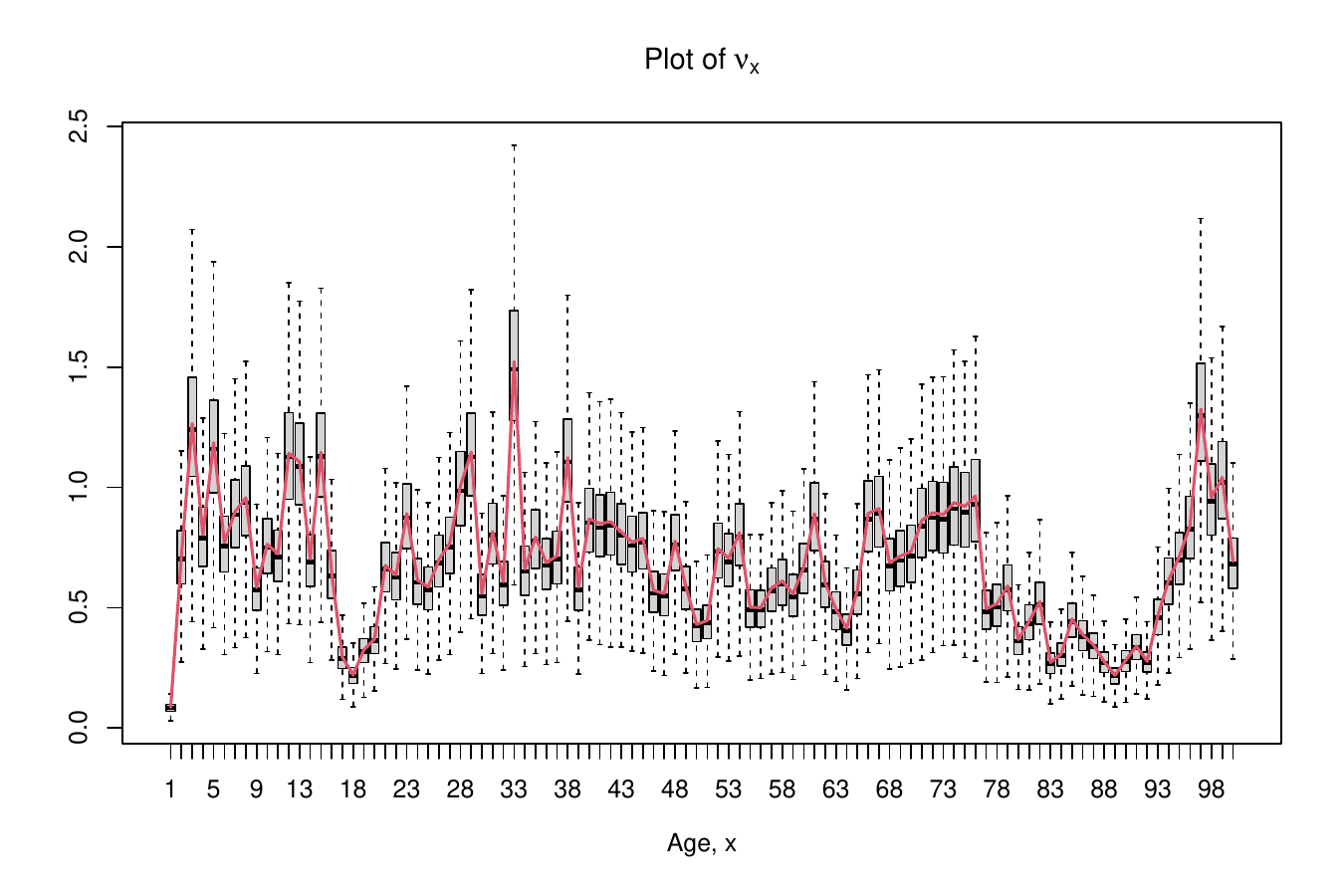}
\caption{\label{fig:CMP_LC_LCC_nu_vary_age_boxplot}Boxplots of the estimated age-specific dispersion parameters $\nu_x$ ($x=1,\ldots,A$) under the CMP--LC--$x$ (top) and CMP--LCC--$x$ (bottom) models. Posterior means are included in red.}
\end{figure}

\begin{figure}[!htbp]
\centering\includegraphics[scale=0.5]{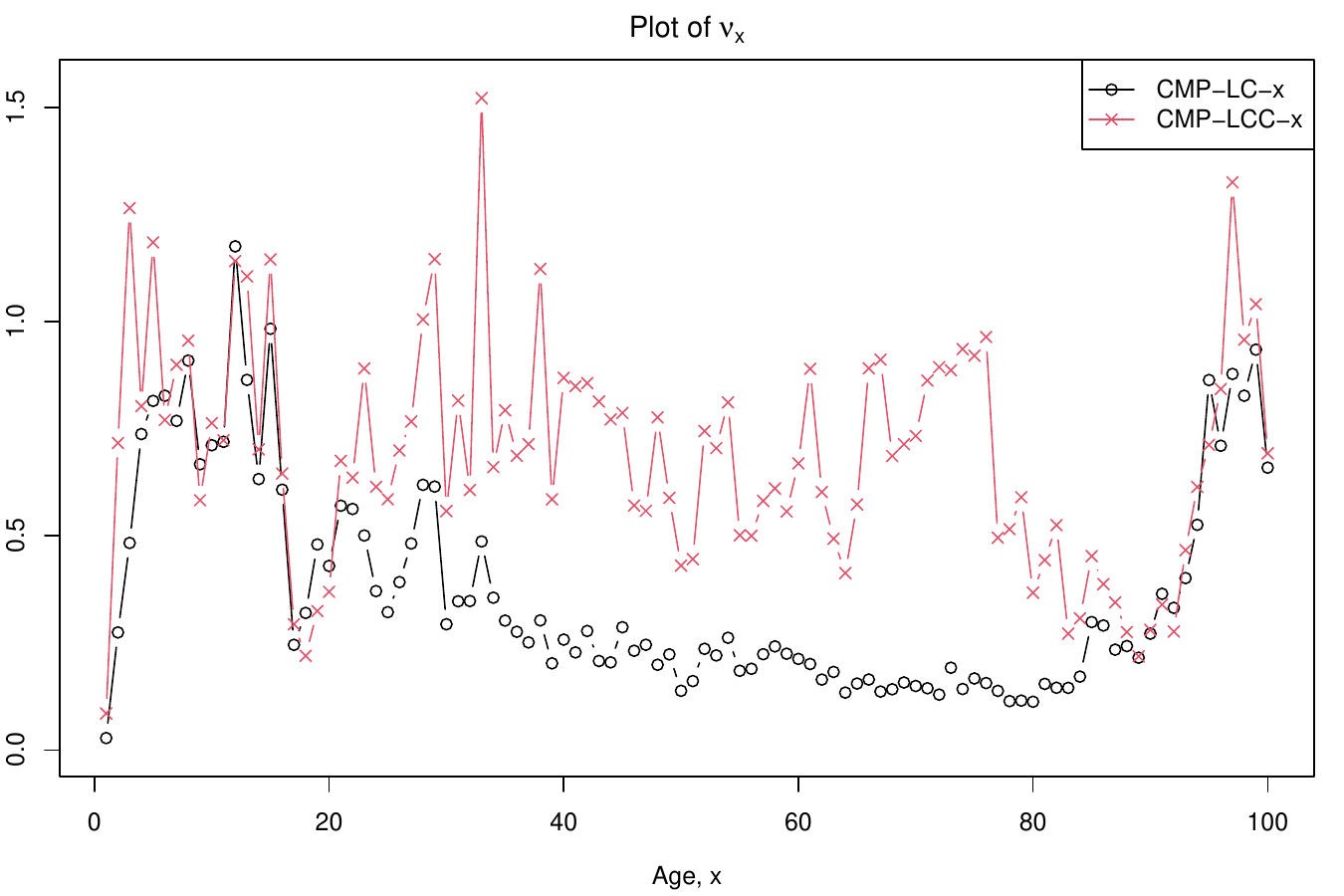}
\caption{\label{fig:posterior_CMP_LC_LCC_nu_vary_age}Posterior means of the age-specific dispersion parameters $\nu_x$ ($x=1,\ldots,A$) under the CMP--LC--$x$ and CMP--LCC--$x$ models.}
\end{figure}

\begin{figure}[!htbp]
\centering
\includegraphics[scale=0.7]{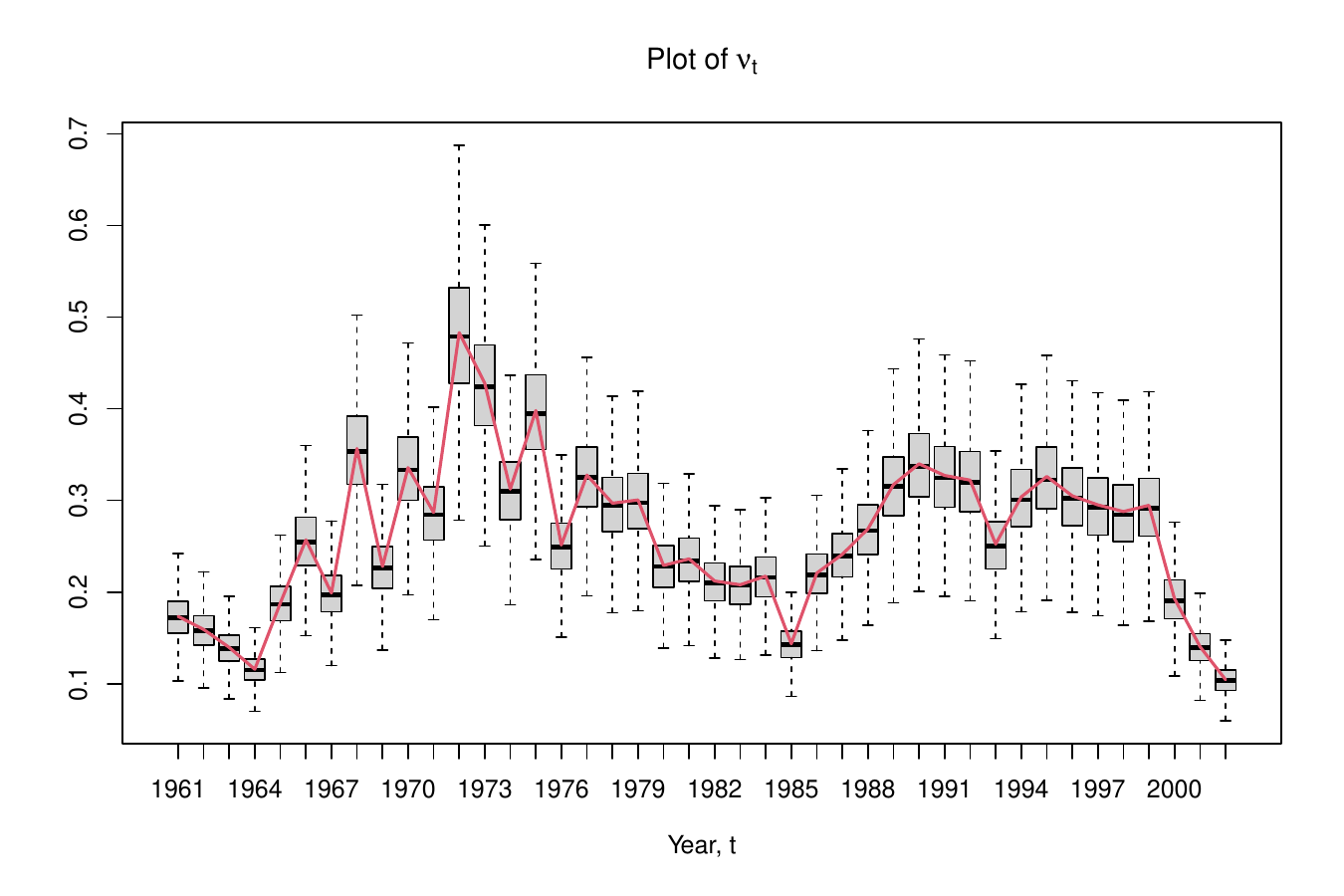}
\includegraphics[scale=0.7]{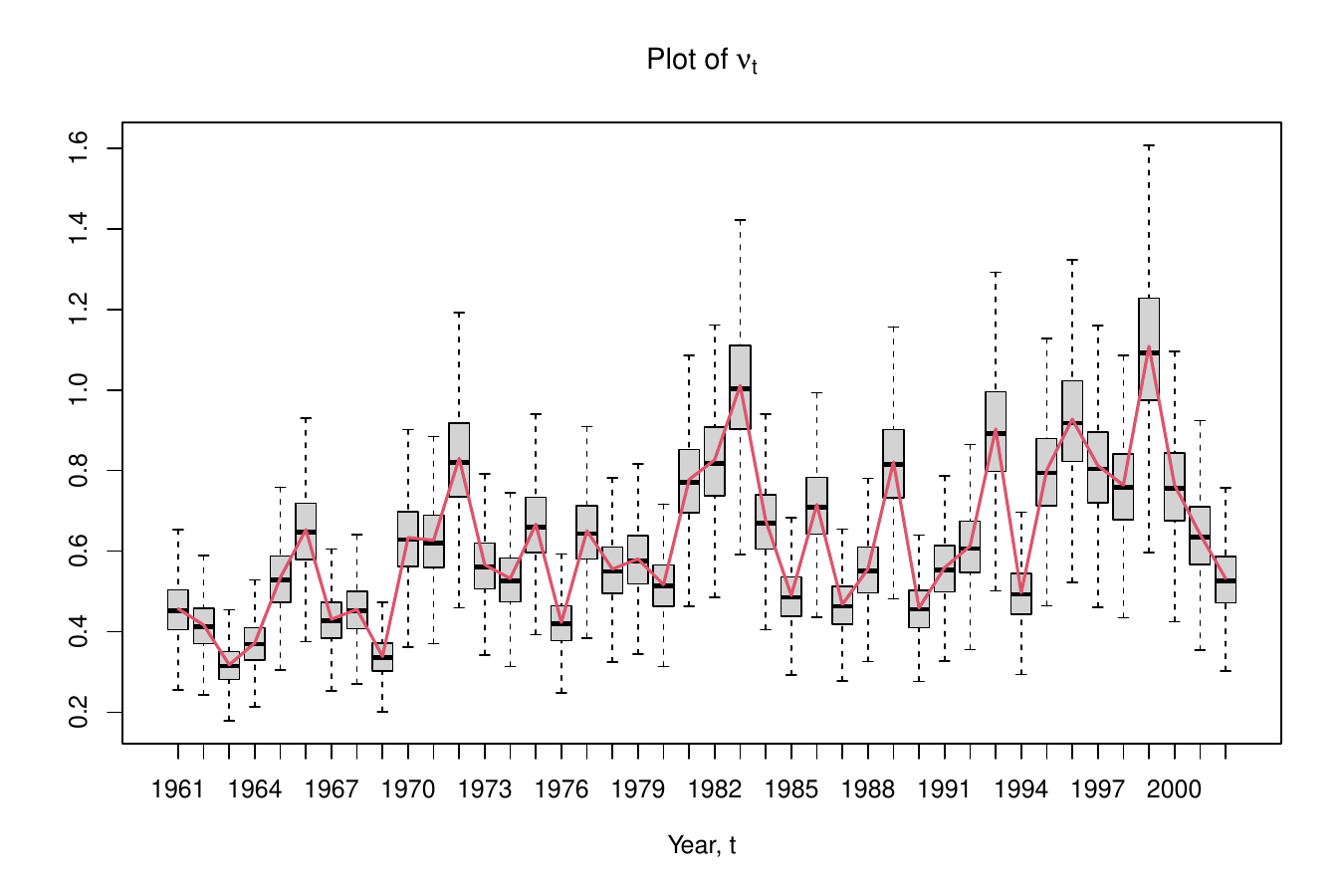}
\caption{\label{fig:CMP_LC_LCC_nu_vary_year_boxplot}Boxplots of the estimated period-specific dispersion parameters $\nu_t$ ($t=1,\ldots,T$) under the CMP--LC--$t$ (top) and CMP--LCC--$t$ (bottom) models. Posterior means are included in red.}
\end{figure}

\begin{figure}[!htbp]
\centering\includegraphics[scale=0.65]{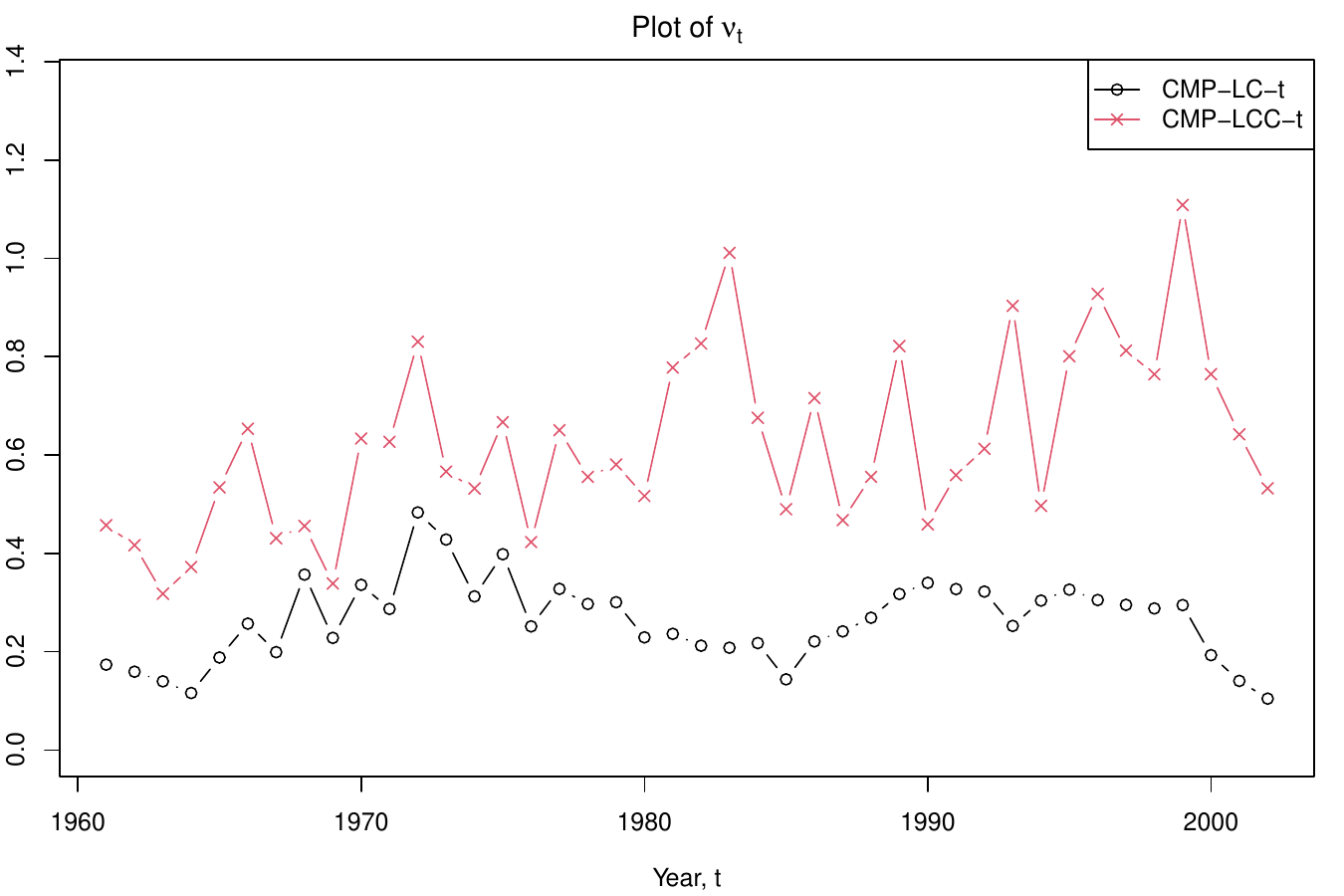}
\caption{\label{fig:posterior_CMP_LC_LCC_nu_vary_year}Posterior means of the period-specific dispersion parameters $\nu_t$ ($t=1,\ldots,T$) under the CMP--LC--$t$ and CMP--LCC--$t$ models.}
\end{figure}

Figure \ref{fig:CMP_LC_LCC_nu_vary_age_boxplot} illustrates that the estimated age-specific dispersion parameters $\nu_x$ vary substantially across ages, providing clear evidence that dispersion levels differ by age rather than being adequately represented by a single shared parameter. First, the estimated $\nu_x$ are generally larger in magnitude under the CMP--LC--$x$ model than under the CMP--LCC--$x$ model. This is expected, as unaccounted cohort effects in the LC specification can inflate the estimated dispersion by attributing systematic cohort patterns to random variability. The inclusion of cohort components in the LCC framework therefore enables a more accurate calibration of uncertainty by reducing the misclassification of structural data signals as stochastic noise (\citealp{wong_2023}).

In addition, the dispersion patterns observed in Figure \ref{fig:CMP_LC_LCC_nu_vary_age_boxplot} are broadly consistent with the evidence presented in Section \ref{sec:dispersion_modeling}. In particular, most values of $\nu_x$ are below one, indicating the presence of overdispersion across many ages. Infant mortality exhibits especially high levels of overdispersion, with other clusters appearing around ages $18-23$, $45-50$, $60-65$, $80-95$, and near age $99$. These patterns closely mirror the residual structures previously identified in Section \ref{sec:dispersion_modeling}. A small number of ages, such as around ages 35 and 96, display mild underdispersion ($\nu_x > 1$). More broadly, the age-specific pattern of dispersion revealed here provides valuable insights into the underlying dynamics of variability in mortality data, highlighting how the degree of stochastic variation differs systematically across the age spectrum.

Similar features can be observed in Figure \ref{fig:CMP_LC_LCC_nu_vary_year_boxplot}, which displays the estimated period-specific dispersion parameters $\nu_t$ for the CMP--LC--$t$ and CMP--LCC--$t$ models. As with age-specific results, most values of $\nu_t$ are below one, indicating that overdispersion is prevalent in many calendar years. The inclusion of cohort effects again improves the calibration of the dispersion by separating systematic cohort-related patterns from stochastic variability.

Some years exhibit near equidispersion, such as $1983$ and $1999$. In contrast, several years, including $1963$, $1969$, $1985$, $1994$, and $2000-2002$, show strong evidence of overdispersion, with posterior mean estimates of $\nu_t$ falling below 0.5. The most recent two years also display a particularly high degree of dispersion, which may plausibly be attributed to the effects of the pandemic.

Interestingly, the sequence of estimated values of $\nu_t$ fluctuates around a gradually increasing trend over time. This temporal structure suggests that the dispersion process itself may exhibit time dependence. Consequently, the pattern observed in Figure \ref{fig:CMP_LC_LCC_nu_vary_year_boxplot} motivates the specification of a time-series prior for $\nu_t$, which will be detailed in Section \ref{sec:forecast_nu_vary_year}.

\subsection{Model assessments}

\begin{figure}[htbp]
\includegraphics[scale=0.60]{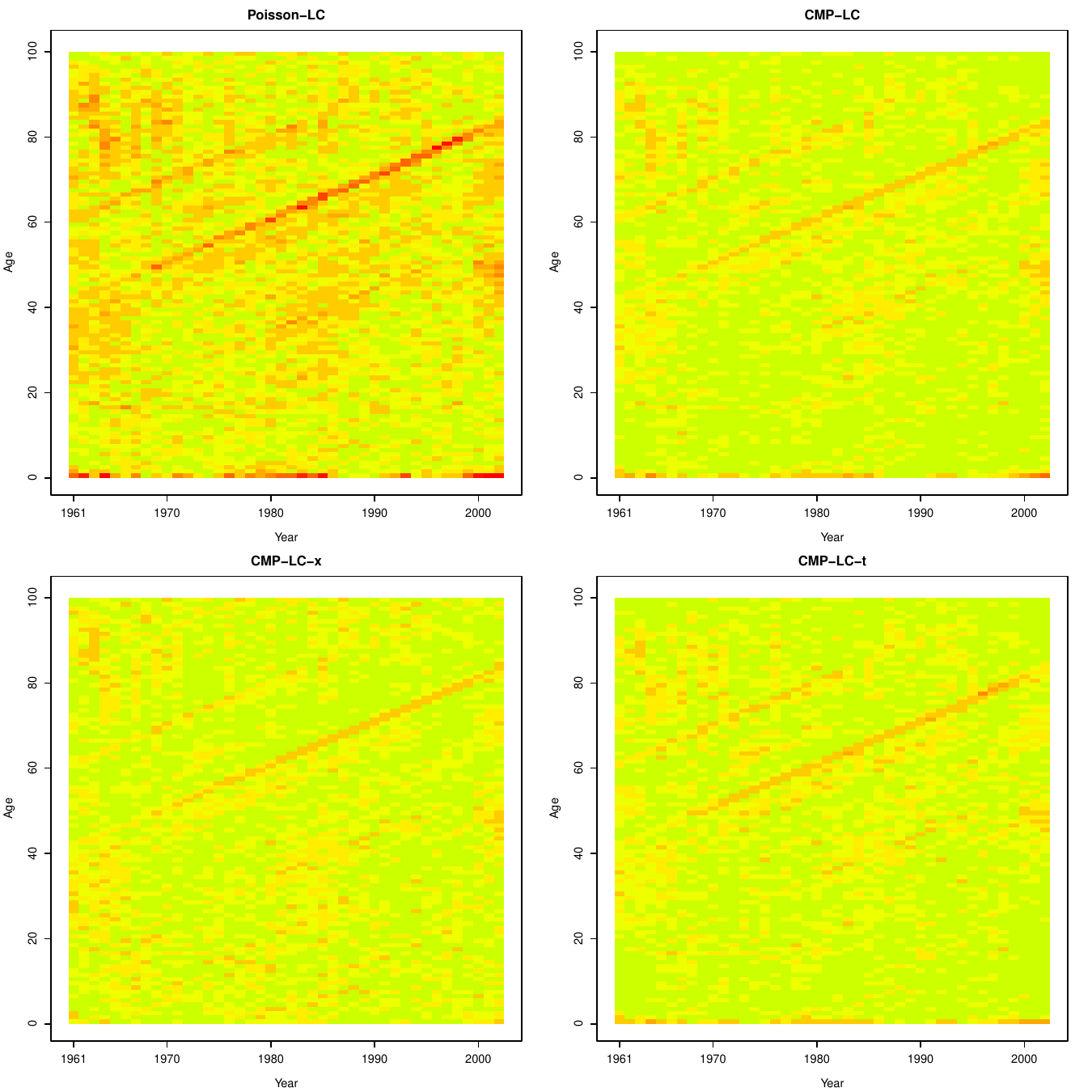}
\includegraphics[width=13.6cm,height=2.0cm]{heatmap_label_male.pdf}
\caption{Heatmap of $r_{xt}^2(d)$ for the LC models.}
\label{fig:heatmap_compare_LC_male}
\end{figure}

\begin{figure}[htbp]
\includegraphics[scale=0.60]{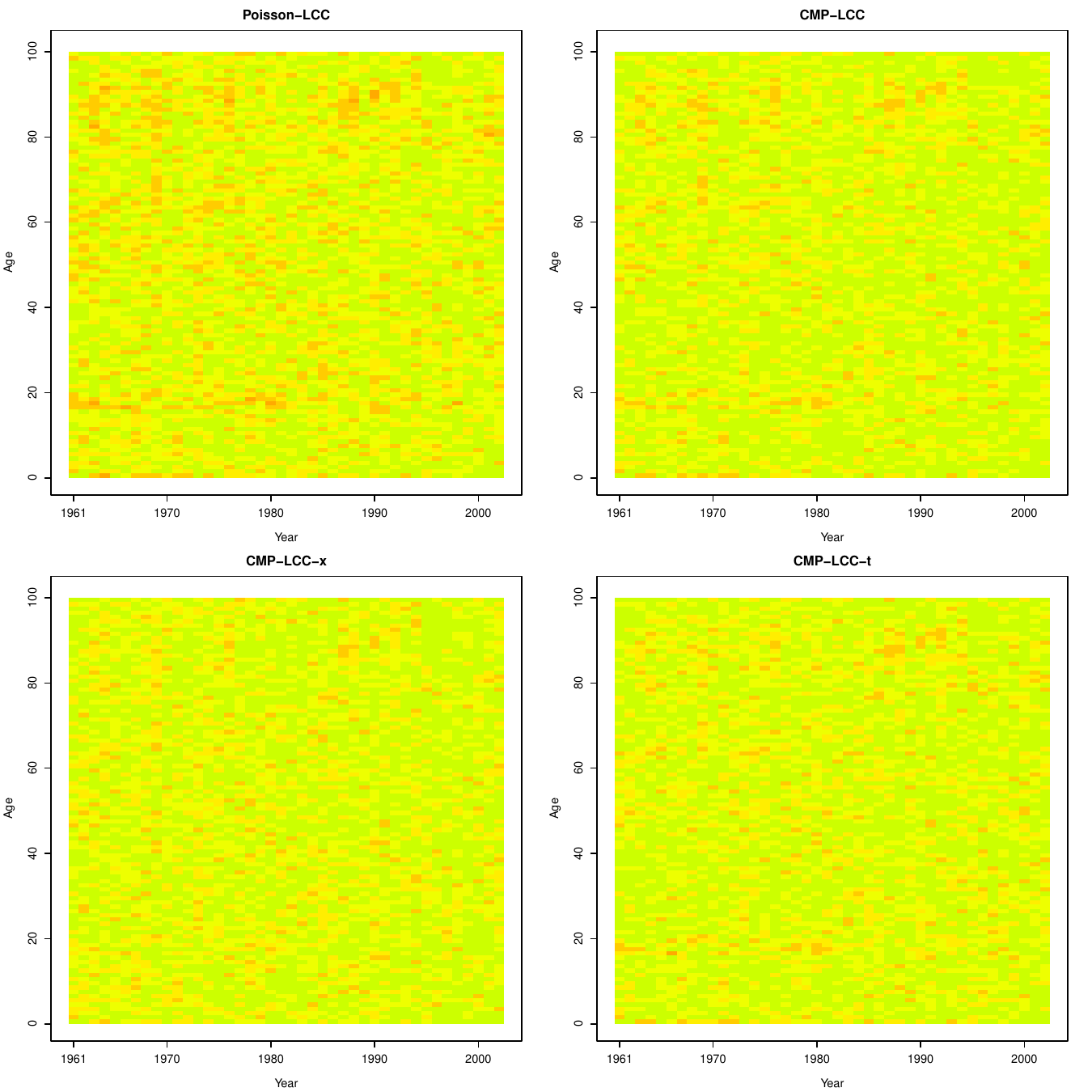}
\includegraphics[width=13.6cm,height=2.0cm]{heatmap_label_male.pdf}
\caption{Heatmap of $r_{xt}^2(d)$ for the LCC models.}
\label{fig:heatmap_compare_LCC_male}
\end{figure}

Heatmaps of residuals by age and time are presented in Figures \ref{fig:heatmap_compare_LC_male} and \ref{fig:heatmap_compare_LCC_male} to demonstrate in-sample goodness of fit. These are constructed similarly as in (\ref{eqn:pearson}). For example for the CMP--LC--$x$ and CMP--LCC--$x$ models, we have 
\begin{eqnarray}
r_{xt}^2(d)=\left.\frac{(d_{xt}-\mathbb{E}[D_{xt}])^2}{\mathrm{Var}[D_{xt}]}\right|_{\mu_{xt}=\hat{\mu}_{xt},\nu_x=\hat{\nu_x}}=\frac{(d_{xt}-e_{xt}\hat{\mu}_{xt})^2}{\frac{1}{\hat{\nu}_x}\left(e_{xt}\hat{\mu}_{xt}+\frac{1}{2}-\frac{1}{2\hat{\nu}_x}\right)},
\label{eqn:pearson_vary_age}
\end{eqnarray}
The estimated total model residuals $r^2(d)$ in terms of number of deaths can be computed as 
\begin{equation}
    r^2(d) = \sum_{x,t}^{}r^2_{xt}(d),
\end{equation}
to serve as an indication of overall goodness of fit, as reported in Table \ref{tab:pearson_residuals}.  Evidently, the CMP specifications with age-varying dispersion outperform their counterparts by producing smaller overall deviance. The best performing model in terms of total overall residuals $r^2(d)$ is the CMP--LCC--$x$.

In general, all the CMP specifications yield residuals that are better calibrated than the Poisson specifications, as expected due to the existence of overdispersion. Models with cohort effects included produce heatmaps that are more well-behaved so we shall focus our analysis on them. At a micro level, the CMP--LCC--$x$ model manages to alleviates clusters of discrepancies mentioned earlier in Section \ref{sec:dispersion_modeling}, particularly those around ages $0$, $18-23$ and $80-95$, leading to the best result. The CMP--LCC--$t$ produces better dispersion calibration in the temporal dimension, though it does not alleviate residuals arising due to age-specific heterogeneity. Overall, the residual patterns observed suggest that the CMP specifications with varying levels of dispersion, both age-specific and period-specific, provide more solid results by offering a more localised calibration.

\begin{table}[!htbp]
\caption{The estimated total model residuals $r^2(d)$ under each of the model specifications. Also included in parentheses are the percentages of cells with poor fit, i.e. $r^2_{xt}>3.84$. Models that generate less than $5\%$ poor fit cells are good.}
  \label{tab:pearson_residuals}
  \centering
\begin{tabular}{r rrrr}
\hline
Model & Poisson & CMP & CMP (age-specific) & CMP (period-specific) \\
\hline
LC & 16709.85 & 3970.60 & \textbf{3970.10} & 3971.62 \\
 & $(26.86\%)$ & $(4.40\%)$ & $\boldsymbol{(4.24\%)}$ & ${(4.33\%)}$ \\
 \hline
LCC & 6628.97 & 3850.75 & \textbf{3845.20} & 3857.22 \\
 & $(11.29\%)$ & $(4.33\%)$ & $\boldsymbol{(3.88\%)}$ & ${(4.14\%)}$ \\
\hline
\end{tabular}
\end{table}



\subsection{Fitted and projected crude mortality rates}
\label{sec:fitted_projected_rates}

We also assess the performance of the models using crude mortality rates (observable) rather than the underlying mortality rates, $\mu_{xt}$ (unobservable). To obtain the fitted crude mortality rates, the fitted number of deaths, $D_{xt}^F$, is first generated through Equation (\ref{eqn:cmp_dxt}), where joint posterior samples of $\alpha_x$, $\beta_x$, $\kappa_t$, $\gamma_c$, $\nu_x$ ($x=1,\ldots,A$), and $\nu_t$ ($t=1,\ldots,T$) are substituted as appropriate. The crude fitted mortality rate for each $x=1,\ldots,A$ and $t=1,\ldots,T$ is then computed as $\hat{\mu}_{xt}=\frac{D_{xt}^F}{e_{xt}}$.

For mortality projection, we note that the posterior predictive distribution of 1-year ahead number of deaths for each age (with age parameters held fixed), under the CMP--LCC--$x$ model for instance, is
\begin{eqnarray}
&& f(D_{x\,T+1}|\vect{d}) \nonumber \\ 
&=&\int f(D_{x\, T+1}|\alpha_x,\beta_x,\kappa_{T+1},\gamma_{T+1-x},\nu_x)\times f(\kappa_{T+1}|\kappa_T,\vect{\psi},\rho,\sigma_\kappa^2)\times f(\gamma_{T+1-x}|\gamma_{T-x},\gamma_{T-x-1},\rho_\gamma,\sigma_\gamma^2) \nonumber\\
&&\times \pi(\alpha_x,\beta_x,\kappa_T,\vect{\psi},\rho,\sigma_\kappa^2,\gamma_{T+1-x},\rho_\gamma,\sigma_\gamma^2,\nu_x|\vect{d}) {\rm d}\alpha_x \ldots {\rm d}\nu_x,
\end{eqnarray}
where $\pi(\alpha_x,\beta_x,\kappa_T,\vect{\psi},\rho,\sigma_\kappa^2,\gamma_{T+1-x},\rho_\gamma,\sigma_\gamma^2,\nu_x|\vect{d})$ is the joint posterior distribution. This suggests the following generation procedure for the projected crude mortality rates:

\begin{enumerate}[(i)]
\item Generate $\kappa_{T+1}$ from Equation (\ref{eqn:ar1}) where joint posterior samples of $(\vect{\psi},\rho,\kappa_T,\sigma_\kappa^2)$ are appropriately substituted.

\item Generate $\gamma_{T}$ from Equation (\ref{eqn:gamma_projection_model}) where joint posterior samples of $(\gamma_{T-1},\gamma_{T-2},\rho_\gamma,\sigma_\gamma^2)$ are substituted. Other ``observed'' cohort components $(\gamma_{T-99},\ldots,\gamma_{T-3})$ are also available from posterior samples.

\item Generate for each $x$ the forecasted number of deaths using (\ref{eqn:cmp_dxt}), i.e.,
$$D^F_{x\,T+1} \sim \text{CMP }\left((e_{x\, T+1}\exp(\alpha_x+\beta_x\kappa_{T+1}+\gamma_{T+1-x})+\frac{1}{2}-\frac{1}{2\nu_x})^{\nu_x} , \nu_x\right),$$
where $e_{x\, T+1}$ are holdout central exposed to risks, joint posterior samples of $(\alpha_x,\beta_x,\nu_x)$ are substituted, $\kappa_{T+1}$ and $\gamma_{T+1-x}$ are respectively from steps (i) and (ii).

\item Compute for each $x$ the projected crude mortality rates as $\hat{\mu}_{x\, T+1}=\frac{D^F_{x\,T+1}}{e_{x\,T+1}}$.
\end{enumerate}
By analogy, $h$-year ahead projections can be obtained by recursive implementation of the above algorithm. Having generated samples for $\hat{\mu}_{xt}$ for $x=1,\ldots,A$ and $t=1,\ldots,T+h$, sample percentiles can be constructed to form median and 95$\%$ intervals. 

Under the Poisson and CMP specifications, we generate from $D_{xt}^F\sim \mbox{ Poisson}(e_{xt}\mu_{xt})$ and $D_{xt}^F\sim \mbox{ CMP}((e_{xt}\mu_{xt}+\frac{1}{2}-\frac{1}{2\nu})^\nu),\nu)$, using posterior samples of the relevant parameters, and then compute the fitted crude mortality rates as $\hat{\mu}_{xt}=\frac{D_{xt}^F}{e_{xt}}$. Note that the above generation procedure of fitted and projected mortality rates builds in uncertainties due to the data generating mechanisms (Poisson or CMP), parameter, and forecast uncertainties, allowing us to better visualize the overall performances of each distributional specification for $D_{xt}$ in the presence of various sources of uncertainties.

\begin{figure}[htbp]
\centering
\includegraphics[scale=0.8]{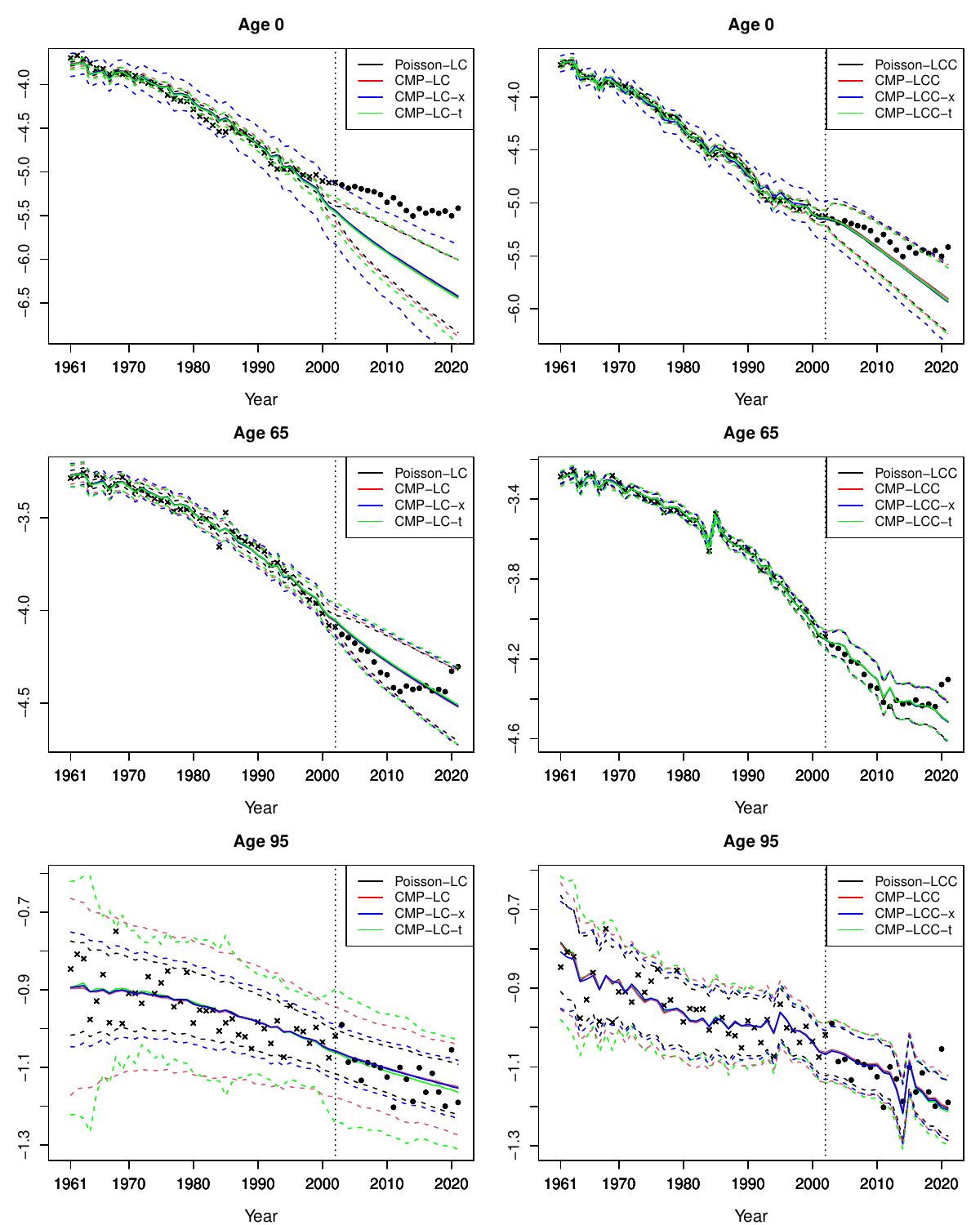}
\caption{\label{fig:projection_logrates_age5_age65_age90}Plots of observed and holdout log crude death rates, fitted log crude death rates and the associated $19$-year ahead projection of the crude log death rates for several chosen ages under the LC (left panels) and LCC (right panels) rate models, accompanied by $95\%$ credible intervals in dashed lines.}
\end{figure}

Figure \ref{fig:projection_logrates_age5_age65_age90} shows the fitted and projected crude death rates for several chosen ages when $h=19$. Also included are the observed and holdout crude mortality rates. Generally speaking, all the CMP specifications (with and without varying dispersion) yield mortality forecasts with good properties, with close adherence to the holdout crude rates and 95\% credible intervals that have high coverage rates. By contrast, the Poisson specification yields credible intervals that are overly narrow, highlighting the importance of accounting for overdispersion. As detailed in \cite{wong_2023}, the incorporation of cohort effects generally leads to a better description  of the underlying mortality trend for both fitted and projected rates. In particular, the $95\%$ prediction intervals of the cohort models (all specifications) offer realistic coverage rates for including the holdout rates, despite having narrower widths. 

Interestingly, the conventional CMP specification yields generally wider intervals than the CMP--LCC--$x$, which can sometimes be suboptimal because it ignores the local structure of dispersion. In particular, it may overestimate the level dispersion at some ages (e.g. age 90) because there is insufficient model flexibility to account for such localised structure, leading to unnecessarily wide intervals. On the contrary, the CMP--LCC--$x$ model offers more flexibility for the calibration of interval widths across various ages, the effect of which is strongly visible for some ages. In particular,  the CMP--LCC--$x$ produces intervals of largest width for infants; while yielding narrower intervals for age 95. The CMP--LCC--$x$ allows such ability to adapt to the width size according to the data, resulting in a pronounced improvement in fit for these ages. If the model has only one global dispersion, it will need to compromise by having one-for-all value, and hence, no additional flexibility for the dispersion to be calibrated for different ages. For age 65, all models with cohort parameters produce similar output. Note that the two most recent mortality rates were observed to be unusually high due to Covid-19 effects, which we do not address here. Overall, simultaneous inclusion of cohort parameters and structural dispersion further facilitates the calibration between signals and errors, reducing the bias in the mortality projection with sufficiently wide intervals to characterize uncertainty.


\subsection{Actuarial analysis using annuity-due}
\label{sec:fitted_projected_axn}

For a given year $t$ and interest rate $i$, the temporary annuity-due function is defined for all ages $x = 1,\ldots,A$ by
\begin{equation*}
\ddot{a}_{x t:\angle{n_x}} \coloneqq \sum_{k=0}^{n_x-1} v^k \, {}_{k}p_{x t},
\label{eqn:axn}
\end{equation*}
where
\[
p_{x t} = \exp(-\mu_{x t}), 
\qquad
{}_{k}p_{x t} = \prod_{l=0}^{k-1} p_{x+l,\, t},
\qquad
v = (1+i)^{-1},
\]
and $n_x = 100 - x$ (ceasing at age 100). The calculation is then repeated for all years in the dataset, $t = 1, \ldots, T$. The result of this is a matrix of annuity-due values by age and time as below,
\begin{equation*}
    \bordermatrix{  \text{Age/Year} &t_{1}&t_{2}&\ldots&t_{T}\cr
									  x_{1}& \ddot{a}_{1\,1:\angle{n_1}} & \ddot{a}_{1\,2:\angle{n_1}} &\ldots& \ddot{a}_{1\,T:\angle{n_1}} \cr
									  x_{2}& \ddot{a}_{2\,1:\angle{n_2}} & \ddot{a}_{2\,2:\angle{n_2}} &\ldots& \ddot{a}_{2\,T:\angle{n_2}} \cr
									  \vdots &\vdots &\vdots &\ddots &\vdots\cr
									  x_{A}& \ddot{a}_{A\,1:\angle{n_A}} & \ddot{a}_{A\,2:\angle{n_A}} &\ldots& \ddot{a}_{A\,T:\angle{n_A}}}
\end{equation*}

Recall that we have $m=10{,}000$ posterior samples of death rates from our MCMC iterations, i.e. we have $\mu_{xt}^{(j)}$ where $j=1,\ldots,m$. Thus, we can compute the annuities-due values for all iterations, as follows:
\begin{equation}
\ddot{a}^{(j)}_{xt:\angle{n_x}} = \sum_{k=0}^{n_x-1}v^k {}_{k}p^{(j)}_{xt},
\label{eqn:axn_MCMC}
\end{equation}
where $p_{xt}^{(j)}=\exp(-\mu_{xt}^{(j)})$\footnote{The notation adopted in this section differs slightly from standard actuarial notation, particularly with respect to the indexing of iterations, in order to align more closely with conventional statistical practice.}. We can then use them to derive sample properties such as means and variances. For instance, we can then find the MCMC estimated mean as 
\begin{equation}
\hat{\ddot{a}}_{xt:\angle{n_x}} \coloneqq \mathbb{E}[ \ddot{a}_{xt:\angle{n}} \mid \boldsymbol{d}]=\frac{1}{m}\sum_{j=1}^{m}\ddot{a}^{(j)}_{xt:\angle{n_x}} ,
\label{eqn:axn_mean}
\end{equation}
as well as the variance as
\begin{equation}
\text{Var}[ \ddot{a}_{xt:\angle{n_x}} \mid \boldsymbol{d}]=\frac{1}{m}\sum_{j=1}^{m} \left(\ddot{a}^{(j)}_{xt:\angle{n_x}}\right)^2 - (\hat{\ddot{a}}_{xt:\angle{n_x}})^2 ,
\label{eqn:axn_var}
\end{equation}

\begin{figure}[!htbp]
\centering\includegraphics[scale=0.6]{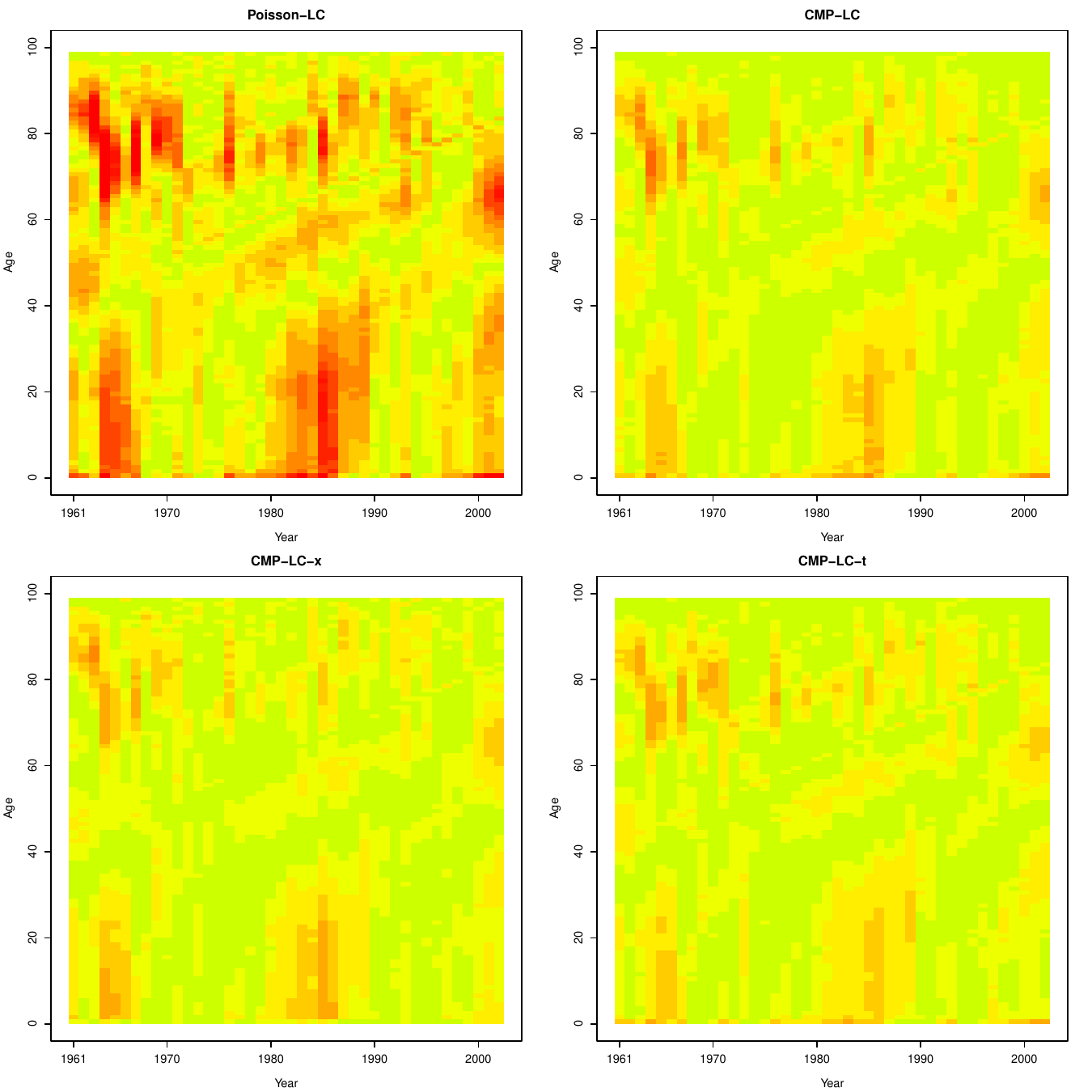}
\includegraphics[width=13.6cm,height=2.0cm]{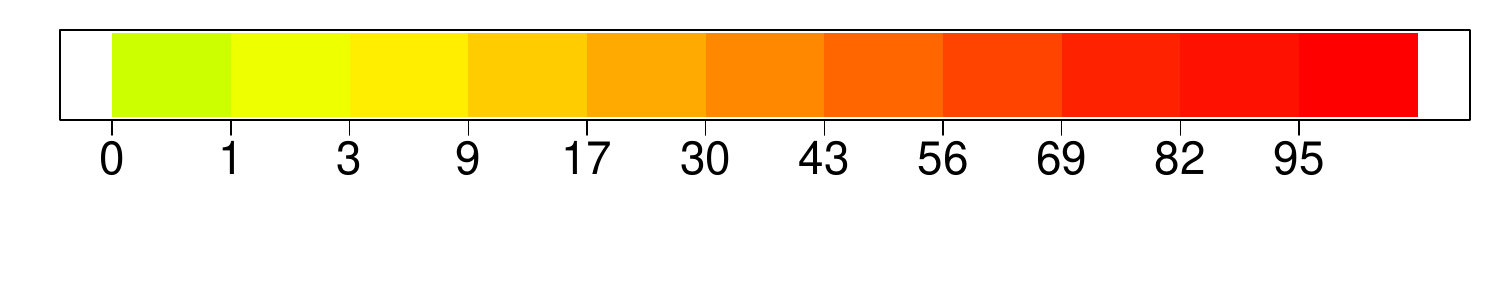}
\caption{\label{fig:heatmap_axt_compare_LC_male_vary_age}Heatmaps of $r_{xt}^2(\ddot{a})$ for the LC models.}
\end{figure}

\begin{figure}[!htbp]
\centering\includegraphics[scale=0.6]{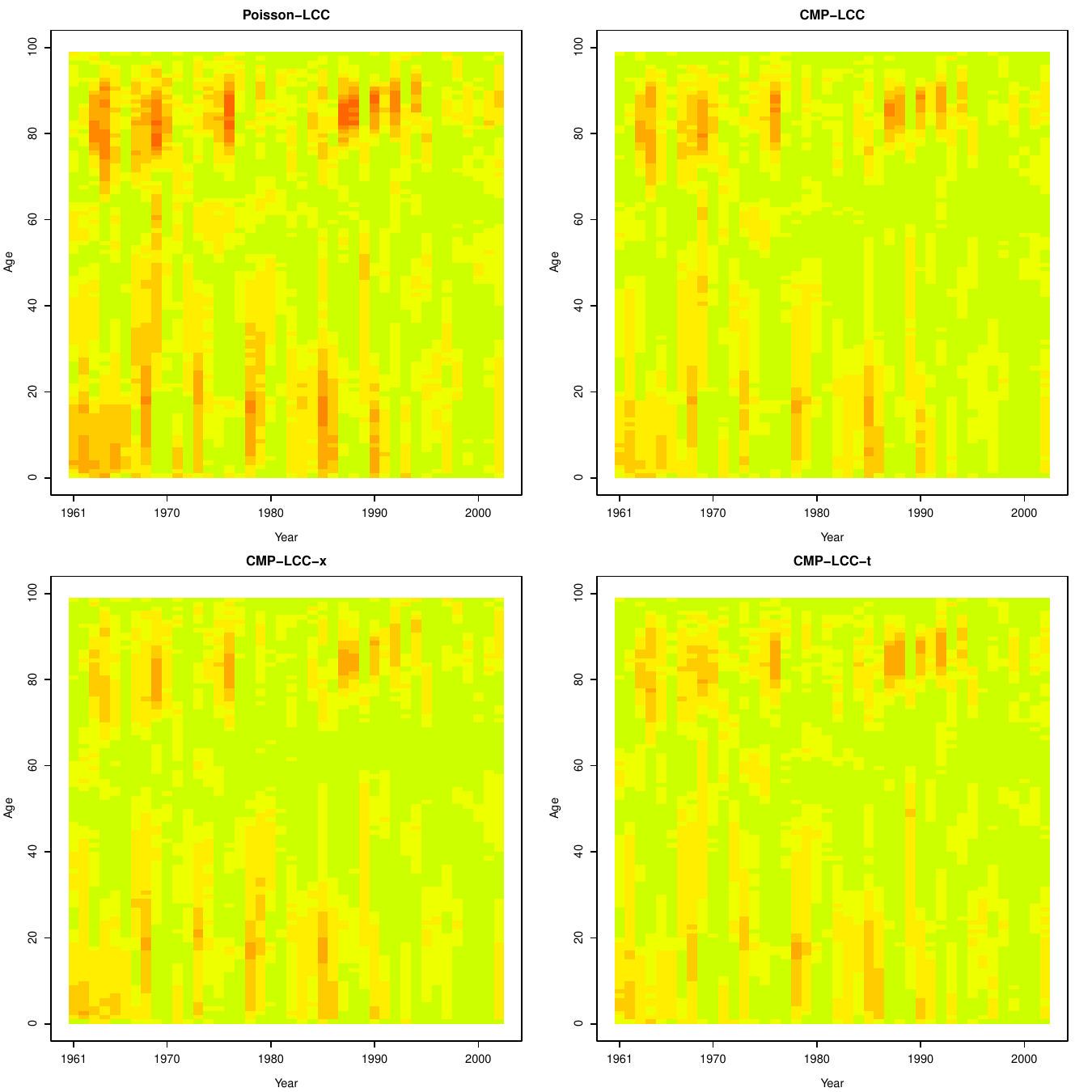}
\includegraphics[width=13.6cm,height=2.0cm]{heatmap_a_ddot_label_male.pdf}
\caption{\label{fig:heatmap_axt_compare_LCC_male_vary_age}Heatmaps of $r_{xt}^2(\ddot{a})$ for the LCC models.}
\end{figure}

To visualize the results, we can construct a heatmap for the annuity-due functions, using calculations similar in spirit to (\ref{eqn:pearson}). Specifically, 
\begin{eqnarray}
r_{xt}^2(\ddot{a})=\frac{(\ddot{a}_{xt:\angle{n_x}}-\mathbb{E}[ \ddot{a}_{xt:\angle{n_x}} \mid \boldsymbol{d}])^2}{\text{Var}[ \ddot{a}_{xt:\angle{n_x}} \mid \boldsymbol{d}]},
\label{eqn:axn_pearson}
\end{eqnarray}
where the means and variances are as given in (\ref{eqn:axn_mean}) and (\ref{eqn:axn_var}). The corresponding heatmaps of $r_{xt}^2(\ddot{a})$ are presented in Figures \ref{fig:heatmap_axt_compare_LC_male_vary_age} and \ref{fig:heatmap_axt_compare_LCC_male_vary_age}. Total residuals in terms of annuities-due, $r^2(\ddot{a})=\sum_{x,t}r_{xt}^2(\ddot{a})$, are illustrated in Table \ref{tab:pearson_residuals_axt}. The best performing model in this context (lowest total residuals in terms of annuities-due) is the CMP--LCC--$t$.  It is also evident that the CMP specifications with varying dispersion provide an improved calibration of the annuity-due values, the improvement of which is substantial over the Poisson specifications. In particular, both the CMP--LCC--$x$ and CMP--LCC--$t$ models produce heat maps that are well-calibrated with mostly small deviations (yellow/green cells), with the latter giving only 4.88\% large deviations. The vertical bands are likely due to accumulated deviations due to the construction of the annuity values (being a summation across ages).

\begin{table}[!htbp]
\caption{The estimated total residuals $r^2(\ddot{a})$ in terms of annuities-due under each of the model specifications. Also included in parentheses are the percentages of cells with poor fit.}
  \label{tab:pearson_residuals_axt}
  \centering
\begin{tabular}{r rrrr}
\hline
Model & Poisson & CMP & CMP (age-specific) & CMP (period-specific) \\
\hline
LC & 51,331.91 & 12,468.61 & 10,764.13 & \textbf{10,726.68} \\
 & $(36.07\%)$ & $(8.31\%)$ & ${(7.05\%)}$ & $\boldsymbol{(6.38\%)}$ \\
 \hline
LCC & 17,591.88 & 10,406.66 & 9,391.48 & \textbf{9,276.92} \\
 & $(13.93\%)$ & $(6.81\%)$ & ${(5.52\%)}$ & $\boldsymbol{(4.88\%)}$ \\
\hline
\end{tabular}
\end{table}


\section{Forecasting time-varying dispersion}
\label{sec:forecast_nu_vary_year}

It is also of interest to explore time-series prior specifications for $\nu_t$, particularly in the context of mortality projection, where they may offer both improved predictive performance and meaningful interpretability. Such a specification allows the evolution of dispersion to be modelled dynamically, thereby capturing potential persistence and structural changes over time. 

Preliminary analyses suggest that an ARIMA(0,1,1) specification provides a good fit to the estimated values of $\nu_t$, indicating a slowly evolving stochastic trend with short-run dependence in the dispersion process. So, for instance, we specify an ARIMA(0,1,1) prior with linear drift on $\nu_t$, that is
\begin{eqnarray}
\left.\begin{array}{l}
    \nu_t-\eta^\nu_t = \nu_{t-1}-\eta^\nu_{t-1}+\epsilon_t^\nu+\phi_1^\nu\epsilon^\nu_{t-1}, \mbox{ for } t=2,3,\ldots,T , \\
    \nu_1=\eta_1+\epsilon_1^\nu ,
    \end{array}\right\}
    \label{eqn:arima_prior_nu}
\end{eqnarray}
where $\eta^\nu_t=\psi_1^\nu+\psi_2^\nu t$ is the linear drift term of $\nu_t$, $\phi_1^\nu$ is the parameter for the MA part, $\epsilon^\nu_t\stackrel{{\rm ind}}{\sim} N(0,\sigma_{\nu}^2)$, $\sigma_{\nu}^2$ is the variance of $\nu_t$. The hyperparameters in (\ref{eqn:arima_prior_nu}) are given the following non-informative priors:
\begin{eqnarray}
\left.\begin{array}{l}
    \psi_1^\nu \sim N(0,100), \\
    \psi_2^\nu \sim N(0,100), \\
    \phi_1^\nu \sim N(0,100), \\
    \sigma_\nu^2 \sim \text{Gamma}(1,0.01). 
    \end{array}
    \right\}
\end{eqnarray}

\begin{figure}[htbp]
\centering
\includegraphics[scale=0.6]{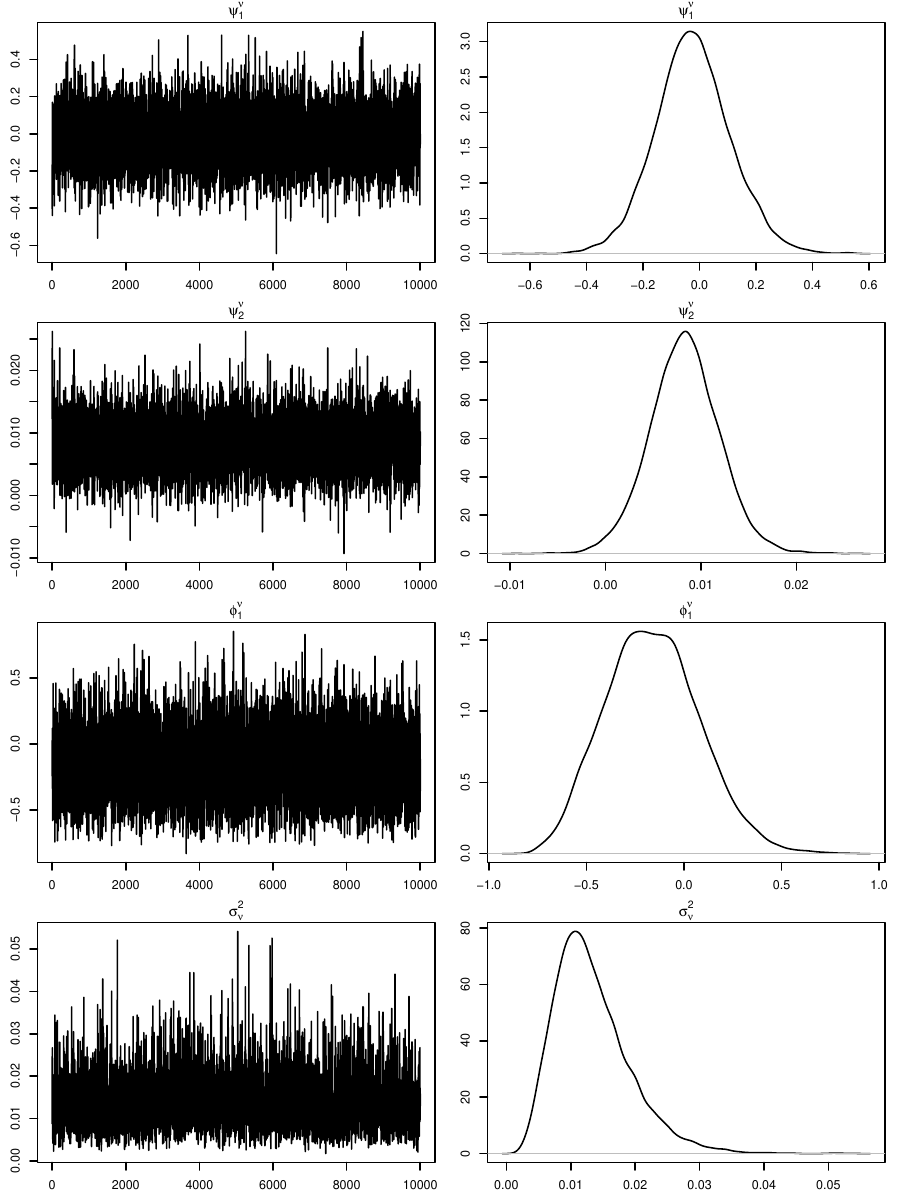}
\caption{\label{fig:arima_param_vary_year_ts}Estimated parameters of the time-series prior of $\nu_t$ under the CMP--LCC--$ts$ model.}
\end{figure}

Figure \ref{fig:arima_param_vary_year_ts} illustrates the estimated parameters of the time-series prior in (\ref{eqn:arima_prior_nu}). The posterior means of the parameters are
$$\psi_1^\nu=-0.02136046 \mbox{\hspace{1cm}} \psi_2^\nu=0.008283345 \mbox{\hspace{1cm}} \phi_1^\nu=-0.1650096 \mbox{\hspace{1cm}} \sigma_\nu^2=0.01318306.$$
Hence, the resulting model is 
\begin{eqnarray}
\left.\begin{array}{l}
    \nu_t = 0.008283345+ \nu_{t-1}+\epsilon_t^\nu-0.1650096\epsilon^\nu_{t-1}, \mbox{ for } t=2,3,\ldots,T , \\
    \nu_1=-0.01307712+\epsilon_1^\nu ,
    \end{array}\right\}
    \label{eqn:arima_prior_nu_est}
\end{eqnarray}
where $\epsilon^\nu_t\stackrel{{\rm ind}}{\sim} N(0,0.01318306)$. The model will be referred to as the CMP--LCC--$ts$ model. 

\begin{figure}[!htbp]
\centering\includegraphics[scale=0.5]{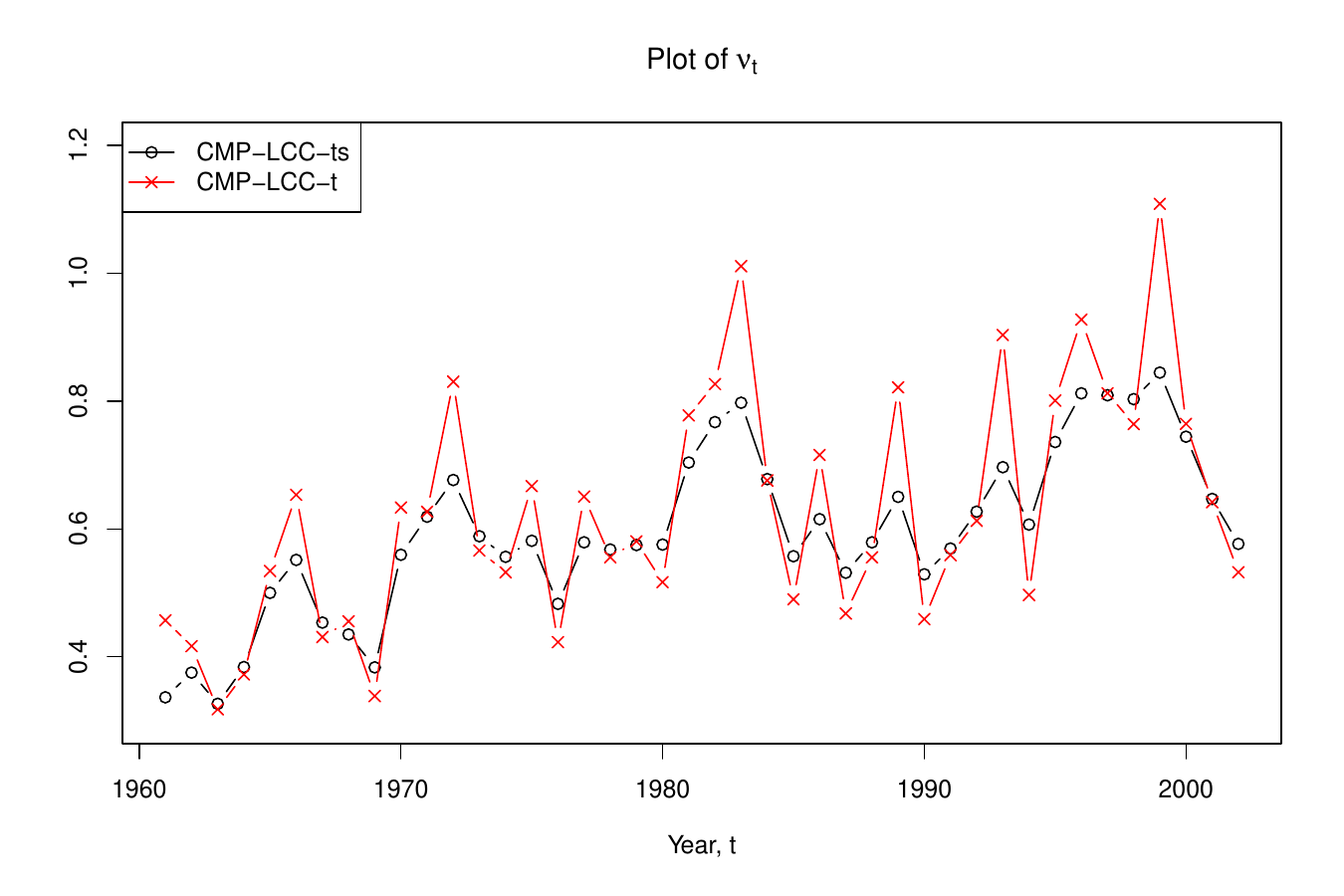}
\caption{\label{fig:posterior_CMP_LCC_nu_vary_year_ts}Posterior means of the period-specific dispersion parameters $\nu_t$ ($t=1,\ldots,T$) under the CMP--LCC--$t$ and CMP--LCC--$ts$ models.}
\end{figure}

Figure \ref{fig:posterior_CMP_LCC_nu_vary_year_ts} indicates that the fitted values of $\nu_t$ under the CMP--LC--$t$ and CMP--LC--$ts$ are rather similar, though there is a slight tendency for smaller variability for the latter. As implied by the findings here, both the gamma and ARIMA(0,1,1) prior specifications for $\nu_t$ produce similar results, but with variability regularized by the temporal correlations imposed by the time series prior.

\begin{figure}[htbp]
\centering
\includegraphics[scale=0.45]{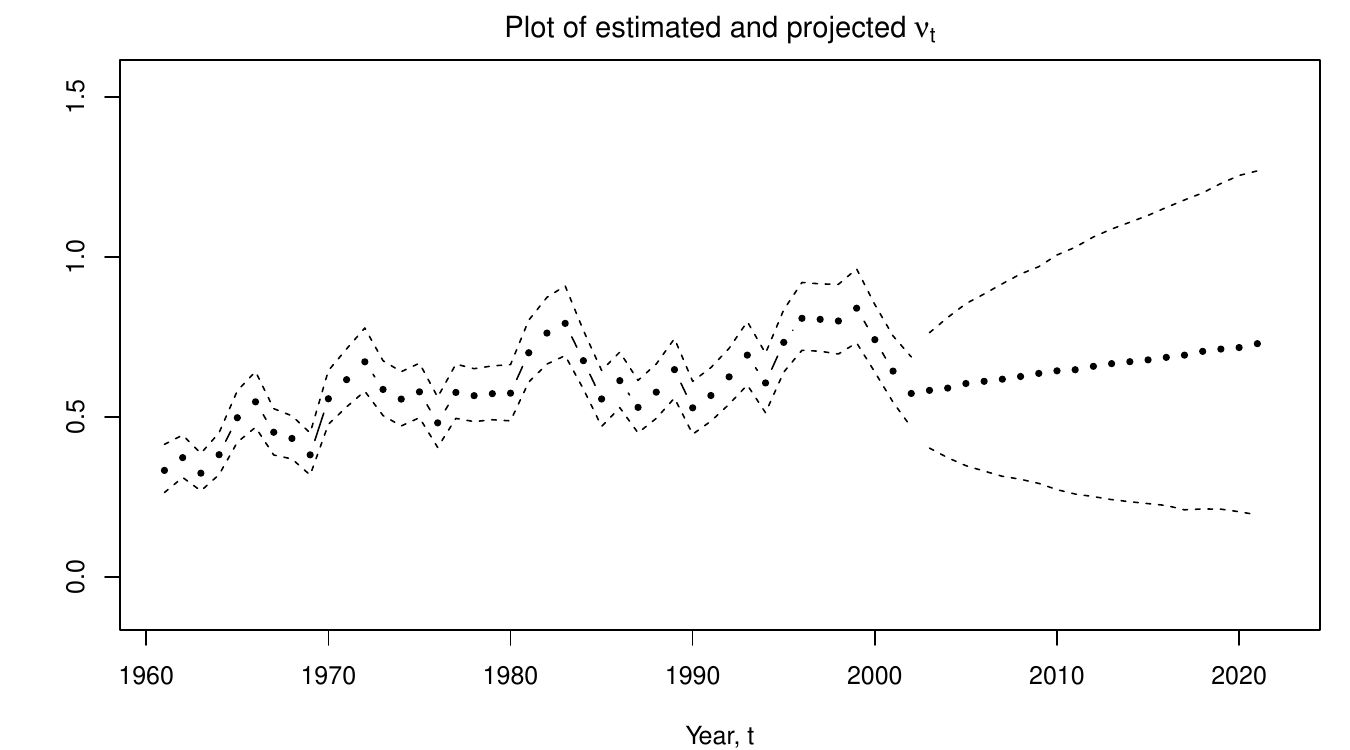}
\caption{\label{fig:forecast_nu_CMP_LCC_vary_year}Plots of the fitted and projected values of $\nu_t$ under the CMP--LCC--$ts$ model.}
\end{figure}

According to Figure \ref{fig:forecast_nu_CMP_LCC_vary_year}, $\nu_t$ demonstrates slowly increasing long-term trend over time, indicating that the level of overdispersion is projected to diminish gradually in the future. The temporal correlation observed in $\nu_t$ is taken into account through both the autoregressive term and the MA parameter $\phi_1^\nu$, which will be propagated into the future via the resulting projection. This is a more sensible approach compared to fixing the value of $\nu_t$ for the projection. There are limitations of such time-series approach for $\nu_t$, first, the linear trend is projected to increase indefinitely into the future, implying that eventually an underdispersion is imposed on the projection. It might be better to assume a long-term value to which $\nu_t$ converges (e.g. $\nu_t \rightarrow 1$ as $t \rightarrow \infty$). Second, the projected values may be negative under the current method, which is not realistic as we require $\nu_t>0$. However, this highlights the potential for other prior specifications that account for temporal correlations in the forecasting of time-varying dispersion.

\section{Conclusion}
\label{sec:conclusion}

In this paper, we introduced the CMP distribution as a flexible framework for modelling structural dispersion in mortality data. Building on the work of \cite{wong2026bayesianCMP}, the proposed framework retains the ability to accommodate underdispersion, equidispersion, and overdispersion through a set of explicit dispersion parameters in the CMP specification, whose values can be inferred directly from the data. Furthermore, the dispersion parameters are designed to vary systematically across the age and time dimensions, forming a flexible framework that allows the underlying structure of heterogeneity to be derived. This offers an advantage relative to models with a single global dispersion parameter, as it facilitates a more localised calibration of dispersion that reflects the potentially dynamic nature of heterogeneity embedded in the mortality data. The framework also accommodates both the LC and LCC mortality rate models and is estimated using a Bayesian approach. Doing so allows all relevant sources of uncertainty to be coherently incorporated, including uncertainty arising from parameter estimation and dispersion. To facilitate Bayesian inference, we adopt a non-informative gamma prior for the dispersion parameters as recommended by \cite{wong2026bayesianCMP} for data-driven inference. 

The proposed methodology was illustrated using England and Wales male mortality data for the whole age range. Empirical results demonstrated that the CMP-based specifications with varying dispersion (age-specific and period-specific) led to superior in-sample and out-of-sample performances, as assessed through residual diagnostics and annuity-due functions, each with their own merits. In particular, the CMP--LCC--$x$ model performs best in terms of residuals based on death counts; whereas the CMP--LCC--$t$ model performs best with respect to residuals derived from annuity-due values. Importantly, the estimated dispersion parameters exhibit substantial variation across both age and time dimensions, encompassing a mixture of under-, equi-, and overdispersion. Such irregular pattern of heterogeneity cannot be adequately captured by a single global dispersion parameter. These findings underscore the importance of combining flexible distributional assumptions with appropriate structural dispersion components in mortality modelling. We further considered a time-series prior specification for the period-specific dispersion parameters $\nu_t$, which was shown to yield sensible results by capturing the temporal trend and correlations observed in the dispersion process. 

Several avenues remain for future research. First, the time-series dynamics of the period and cohort effects, $\kappa_t$ and $\gamma_c$, were specified using standard formulations from the literature. As noted in \cite{wong2026bayesianCMP}, extending this framework to incorporate formal model selection among alternative time-series specifications could further enhance model robustness, particularly when applied to different mortality datasets. Second, the proposed approach could be extended to accommodate dispersion structures that vary systematically along additional dimensions, e.g., the cohort dimension. Finally, alternative time-series prior distributions for $\nu_t$, particularly those with desirable properties such as non-negativity and long-term stability, could be explored. 

Overall, this paper addresses the limitations of the methodology proposed in \cite{wong2026bayesianCMP}. Our study demonstrates that modifying the CMP distribution to incorporate various dispersion structures provides a powerful and flexible approach for analysing death counts. By explicitly modelling the dispersion structure within a coherent Bayesian framework, the proposed methodology delivers improved model fit, richer representation of mortality heterogeneity, and a solid foundation for further methodological and applied work in mortality modelling and projection.

\section*{Acknowledgment}
To be included.

\section*{Funding}
This research received no specific grant from any funding agency in the public, commercial, or not-for-profit sectors.

\section*{Conflict of Interests}
The authors declare that they have no conflicts of interest regarding the publication of this article. No financial or personal relationships that could have influenced the work have been disclosed.

\section*{Data Availability Statement}
The data and code that support the findings of this study are available from the following GitHub repository [to be included if accepted].

\appendix
\setcounter{secnumdepth}{0}
\section{Appendix: Supplementary Material}
Supplementary material related to this research may be found in the online version of the article at the publisher's website.

\bibliographystyle{chicago}
\bibliography{reference}

\end{document}


\maketitle

\noindent Supplementary materials caption:

\begin{itemize}
\item Appendix A: Prior distributions of \texorpdfstring{$\boldsymbol{\kappa}$ and $\boldsymbol{\gamma}$}{}

\item Appendix B: MCMMC Updating Schemes.

\item Appendix C: Some MCMC Convergence Diagnostics.

\end{itemize}

\noindent Figures caption:
\begin{itemize}
\item Figure S1: Proposal variances of $\alpha_x$ and $\beta_x$ for $x=1,\ldots,A$ with the corresponding MCMC acceptance rates for the CMP--LC--$x$ model.

\item Figure S2: Proposal variances of the dispersion parameters, $\sigma_{\nu_x}^2$ and $\sigma_{\nu_t}^2$ with the corresponding MCMC acceptance rates respectively for the CMP--LC/LCC--$x$ and CMP--LC/LCC--$t$ models.

\item Figure S3: Trace plots of estimated parameters for the CMP--LC--$x$ model.

\item Figure S4: Trace plots of estimated parameters for the CMP--LCC--$x$ model.

\item Figure S5: Trace plots of estimated parameters for the CMP--LC--$t$ model.

\item Figure S6: Trace plots of estimated parameters for the CMP--LCC--$t$ model.

\item Figure S7: Trace plots of the estimated age-specific dispersion parameters $\nu_x$ for several randomly selected $x$ under the CMP--LC--$x$ model.

\item Figure S8: Trace plots of the estimated age-specific dispersion parameters $\nu_x$ for several randomly selected $x$ under the CMP--LCC--$x$ model.

\item Figure S9: Trace plots of the estimated year-specific dispersion parameters $\nu_t$ for several randomly selected $x$ under the CMP--LC--$t$ model.

\item Figure S10: Trace plots of the estimated year-specific dispersion parameters $\nu_t$ for several randomly selected $x$ under the CMP--LCC--$t$ model.

\item Figure S11: Density plots of estimated parameters for the CMP--LC--$x$ model.

\item Figure S12: Density plots of estimated parameters for the CMP--LCC--$x$ model.

\item Figure S13: Density plots of estimated parameters for the CMP--LC--$t$ model.

\item Figure S14: Density plots of estimated parameters for the CMP--LCC--$t$ model.

\item Figure S15: Autocorrelation plots of some randomly selected parameters for the CMP--LC--$x$ model.

\item Figure S16: Autocorrelation plots of some randomly selected parameters for the CMP--LCC--$x$ model.

\item Figure S17: Autocorrelation plots of some randomly selected parameters for the CMP--LC--$t$ model.

\item Figure S18: Autocorrelation plots of some randomly selected parameters for the CMP--LCC--$t$ model.

\item Figure S19: Plot of Gelman's PSRF under various model specifications.

\item Figure S20: Plot of Geweke's Z statistics under the CMP--LC--$x$ and CMP--LC--$t$ models.

\item Figure S21: Plot of Geweke's Z statistics under the CMP--LCC--$x$ and CMP--LCC--$t$ models.

\end{itemize}

\noindent Figures caption:
\begin{itemize}
\item Table S1: The numerically calibrated proposal variances of the dispersion parameters, $\sigma_{\nu_x}^2$ for the CMP--LC--$x$ and CMP--LCC--$x$ models.

\item Table S2: The numerically calibrated proposal variances of the dispersion parameters, $\sigma_{\nu_t}^2$ for the CMP--LC--$t$ and CMP--LCC--$t$ models.

\end{itemize}


\section{Prior distributions of \texorpdfstring{$\boldsymbol{\kappa}$ and $\boldsymbol{\gamma}$}{}}

Following the formulation of \cite{bayespoisson}, we impose for $\kappa_t$
\begin{eqnarray}
\left.\begin{array}{l c}
\kappa_{t}-\eta_{t}=\rho(\kappa_{t-1}-\eta_{t-1})+\epsilon_{t} , & \mbox{ for } t=2,3,\ldots,T, \\
\kappa_1=\eta_1+\epsilon_1, & \end{array}\right\}
\label{eqn:ar1}
\end{eqnarray}
where $\eta_{t}=\psi_1+\psi_2 t$ denotes the linear drift and $\epsilon_t\stackrel{{\rm ind}}{\sim} N(0,\sigma_{\kappa}^2)$. Then, depending on the value of $\rho$, this forms either a first-order autoregressive (AR(1)) model ($|\rho|<1$) or a random walk model ($\rho=1$) for $\kappa_t$. The model in (\ref{eqn:ar1}) can be expressed multivariately (along with the constraint) as
\begin{eqnarray}
\left.\begin{array}{l}\boldsymbol{\kappa}|\rho,\vect{\psi},\sigma_\kappa^2 \sim N\left(\vect{\mu}_\kappa,\sigma_\kappa^2 {V}\right), \\
\kappa_1=-\sum_{t=2}^{T} \kappa_t, \end{array}\right\}
\label{eqn:mult_ar1}
\end{eqnarray}
where $\vect{\kappa}=(\kappa_2,\ldots,\kappa_T)^\top$, $\vect{\mu}_\kappa=({I}_{T-1}-{B}_{21}{B}_{11}^{-1}\vect{1}_{T-1}^\top)\cdot{X}\boldsymbol{\psi}$,
${X}=\left(\begin{array}{c c c c}
1 & 1 & \cdots & 1 \\
2 & 3 & \cdots & T  \end{array}\right)^\top\, ,$ $\vect{\psi}=(\psi_1,\psi_2)^\top$, ${V}={B}^{22}-{B}^{21}({B}^{11})^{-1}{B}^{12}$, ${B}={A}{Q}^{-1}{A}^\top$ and is partitioned such that \\
${B}=\left(\begin{array}{c c}
{B}^{11}_{{1\times 1}} & {B}^{12}_{{1\times (T-1)}} \\
{B}^{21}_{{(T-1)\times 1}} & {B}^{22}_{{(T-1)\times (T-1)}}
\end{array}\right)$,  ${Q}=({I}_T-{P})^\top ({I}_T-{P})$,  
${P}=\left(\begin{array}{c c c c c}
0 & 0 & \cdots & \cdots & 0 \\
\rho & 0 & & & \vdots \\
0 & \rho & \ddots & & \vdots \\
\vdots & \ddots & \ddots & \ddots & \vdots \\
0 & \cdots & 0 & \rho & 0 \end{array}\right)_{T\times T}, \\ 
{A}=\left(\begin{array}{c c c c c}
1 & 1 & 1 & \cdots & 1 \\
0 & 1 & 0 & \cdots & 0 \\
0 & 0 & 1 & \cdots & 0 \\
\vdots & \vdots & \vdots & \ddots & \vdots \\
0 & 0 & 0 & \cdots & 1
\end{array}\right)_{T\times T}\,.
$

For the cohort parameters $\gamma_c$, we use an ARIMA(1,1,0) as adopted by \cite{stmomo}, i.e.
\begin{eqnarray}
\left.\begin{array}{l l}
(\gamma_c-\gamma_{c-1})=\rho_\gamma (\gamma_{c-1}-\gamma_{c-2})+\epsilon_c^\gamma , & \mbox{ for } c=3,\ldots,C, \\
\gamma_2-\gamma_1=\frac{1}{\sqrt{1-\rho_\gamma^2}}\epsilon_2^\gamma, & \\
\gamma_1= 100\epsilon_1^\gamma,
\end{array}\right\}
\label{eqn:gamma_projection_model}
\end{eqnarray}
where $\epsilon_c^{\gamma}\stackrel{{\rm ind}}{\sim} N(0,\sigma_\gamma^2)$ for $c=1,\ldots,C$, and $\rho_\gamma$ and $\sigma_\gamma^2$ are hyperparameters. Applying the constraints $\sum_c \gamma_c=\sum_c c\gamma_c=\sum_c c^2\gamma_c=0$ on (\ref{eqn:gamma_projection_model}), we write
\begin{eqnarray}
\vect{\gamma}\sim N(\vect{0},\sigma_\gamma^2{W}_\gamma),
\label{eqn:gamma_prior_withcons}
\end{eqnarray}
where $\vect{\gamma}=(\gamma_2,\ldots,\gamma_{71},\gamma_{73},\ldots,\gamma_{C-1})^\top$, and ${W}_\gamma=[{E}^{22}_\gamma-{E}^{21}_\gamma({E}^{11}_\gamma)^{-1}{E}^{12}_\gamma]$, ${E}_\gamma={D}_\gamma{Q}_\gamma^{-1}{D}_\gamma^\top$ which is partitioned such that
$${E}_\gamma=\left(\begin{array}{c c}
{E}_{\gamma_{\ \ 3\times 3}}^{11} & {E}^{12}_{\gamma_{\ \ 3\times (C-3)}} \\
{E}^{21}_{\gamma_{\ \ (C-3)\times 3}} & {E}^{22}_{\gamma_{\ \ (C-3)\times (C-3)}}
\end{array}\right),$$
\begin{eqnarray*}
{D}_\gamma=\bordermatrix{
 & & & & & & & & & \cr
\mbox{ row } 1 & 1 & 1 & 1 & 1 & 1 & \cdots & \cdots & \cdots & \cdots & 1 \cr
\mbox{ row } 2 & 1 & 2 & 3 & 4 & 5 & \cdots & \cdots & \cdots & \cdots & C \cr
\mbox{ row } 3 & 1^2 & 2^2 & 3^2 & 4^2 & 5^2 & \cdots & \cdots & \cdots & \cdots & C^2 \cr
\mbox{ row } 4 & 0 & 1 & 0 & 0 & 0 & \cdots & \cdots & \cdots & \cdots & 0 \cr
\mbox{ row } 5 & 0 & 0 & 1 & 0 & 0 & \cdots & \cdots & \cdots & \cdots & 0 \cr
\mbox{\hspace{0.5cm} $\vdots$} & \vdots & \vdots & \ddots & \ddots & \ddots & \ddots & & & & \vdots \cr
\mbox{ row } 73 & 0 & 0 & \cdots & 0 & 1 & 0 & 0 & \cdots & \cdots & 0 \cr
\mbox{ row } 74 & 0 & \cdots & \cdots & \cdots & 0 & 0 & 1 & 0 & \cdots & 0\cr
\mbox{\hspace{0.5cm} $\vdots$} & \vdots & & & & & \ddots & \ddots & \ddots & & \vdots \cr
\mbox{ row } C & 0 & 0 & 0 & 0 & 0 & \cdots & \cdots & 0 & 1 &0
}_{C\times C} \ ,
\end{eqnarray*}
${Q}_\gamma={R}_\gamma^\top{R}_\gamma$, with
\begin{eqnarray*}
{R}_\gamma=\left(\begin{array}{c c c c c c c c c}
1/100 & 0 & 0 & 0 & 0 & \cdots & \cdots & \cdots & 0 \\
-\sqrt{1-\rho_\gamma^2} & \sqrt{1-\rho_\gamma^2} & 0 & 0 & 0 & \cdots & \cdots & \cdots & 0 \\
\rho_\gamma & -(1+\rho_\gamma) & 1 & 0 & 0 & \cdots & \cdots & \cdots & 0 \\
0 & \ddots & \ddots & \ddots & 0 & \cdots & \cdots & \cdots & 0 \\
\vdots & \ddots & \ddots & \ddots & \ddots & \ddots & & & \vdots \\
\\
0 & \cdots & 0 & \rho_\gamma & -(1+\rho_\gamma) & 1 & 0 & \cdots & 0
\end{array}\right)_{C\times C}\ .
\end{eqnarray*}

The remaining cohort components $\{\gamma_1,\gamma_{72},\gamma_C\}$ can then be derived as follows,
\begin{eqnarray*}
\gamma_1 &=& \frac{1}{71\times (C-1)}\sum_{c\neq 1,72,C} (c-72)(C-c)\gamma_c, \\
\gamma_{72} &=& -\frac{1}{69\times 71}\sum_{c\neq 1,72,C} (C-c)(c-1)\gamma_c, \\
\gamma_C &=& \frac{1}{69\times (C-1)}\sum_{c\neq 1,72,C} (c-1)(72-c)\gamma_c.
\end{eqnarray*}

\section{MCMC Updating Schemes}
Here, we provide a brief description of the MCMC updating schemes developed for all other parameters. For more details, see \cite{wong2026bayesianCMP}. Hereon, we denote $f(\theta|\bold{d},rest)$ as the density of $\theta$ conditional on the observed death data $\bold{d}$ and the rest of the parameters (also known as the conditional posterior density). For illustrative purposes, we focus the following exposition on the CMP specification with age-specific dispersion, namely the CMP--LC--$x$ and CMP--LCC--$x$ models.

In general, the MH updating step proceeds as follows: 
\begin{enumerate}[i.]
\item Set $i=1$ with the initial value $\theta^{(0)}$.
\item Propose at the $i^{th}$ iteration
$$\theta^* \sim N(\theta^{(i-1)},\sigma_\theta^2),$$
where $\theta^{(i-1)}$ is the current iterate and $\sigma_\theta^2$ is the proposal variance to be specified.
\item Compute the acceptance ratio as 
\begin{eqnarray}
a(\theta^*|\theta^{(i-1)}) = \min\left\{1,\frac{f(\theta^*|\bold{d},rest)\times \pi(\theta^{(t-1)})}{f(\theta^{(t-1)}|\bold{d},rest)\times \pi(\theta^{*})}\right\}
\label{eqn:MH_accept_ratio}
\end{eqnarray}
\item Simulate a uniform random variate $u\sim U[0,1]$.
\item Accept the proposal using the probability given in (\ref{eqn:MH_accept_ratio}), i.e.
\begin{eqnarray*}
    \theta^{(i)}=\left\{\begin{array}{cc}
       \theta^*  & \text{ if } u< a(\theta^*|\theta^{(i-1)}) \\
       \theta^{(i-1)}  & \text{ otherwise}
    \end{array}\right..
\end{eqnarray*}
\item Set $i=i+1$ and repeat from Step ii.
\end{enumerate}
Note that the above works if we are to perform updating in block, just need to swap to multivariate normal proposal in Step 2, i.e. $\theta^* \sim N(\theta^{(i-1)},\Sigma_\theta)$
where $\Sigma_\theta$ is the proposal variance matrix which should be specified appropriately to maximize the efficiency of posterior explorations (see below).

\subsection{MH updating for \texorpdfstring{$\alpha_x$ and $\beta_x$}{}}

The conditional posterior densities of $\alpha_x$, and $\beta_x$ are
\begin{enumerate}[i.]
\item For $x=1,\ldots,A$,
$$f(\alpha_x|\bold{d},rest)\propto \prod_{t} \left\{\frac{1}{Z(\lambda_{xt},\nu_x)}\frac{(\lambda_{xt})^{d_{xt}}}{(d_{xt}!)^{\nu_x}}\right\} \times \exp\left(-\frac{(\alpha_x+5)^2}{8}\right).$$
\item For $x=2,\ldots,A$,
\begin{eqnarray*}
f(\beta_x|rest)&\propto& \prod_{t} \left\{\frac{1}{Z(\lambda_{xt},\nu_x)}\frac{(\lambda_{xt})^{d_{xt}}}{(d_{xt}!)^{\nu_x}} \times \frac{1}{Z(\lambda_{1t},\nu)}\frac{(\lambda_{1t})^{d_{1t}}}{(d_{1t}!)^{\nu_x}}\right\} \\
&& \times \exp\left(-(\bold{\beta}-\frac{1}{A}\bold{1}_{A-1})^\top (0.005({I}_{A-1}-\frac{1}{A}{J}_{A-1}))^{-1} (\bold{\beta}-\frac{1}{A}\bold{1}_{A-1}) \right),
\end{eqnarray*}
where $\bold{\beta}=(\beta_2,\ldots,\beta_A)^\top$ and $\beta_1=1-\sum_{x=2}^{A}\beta_x$.
\end{enumerate}

\begin{figure}[htbp]
\centering
\includegraphics[scale=0.7]{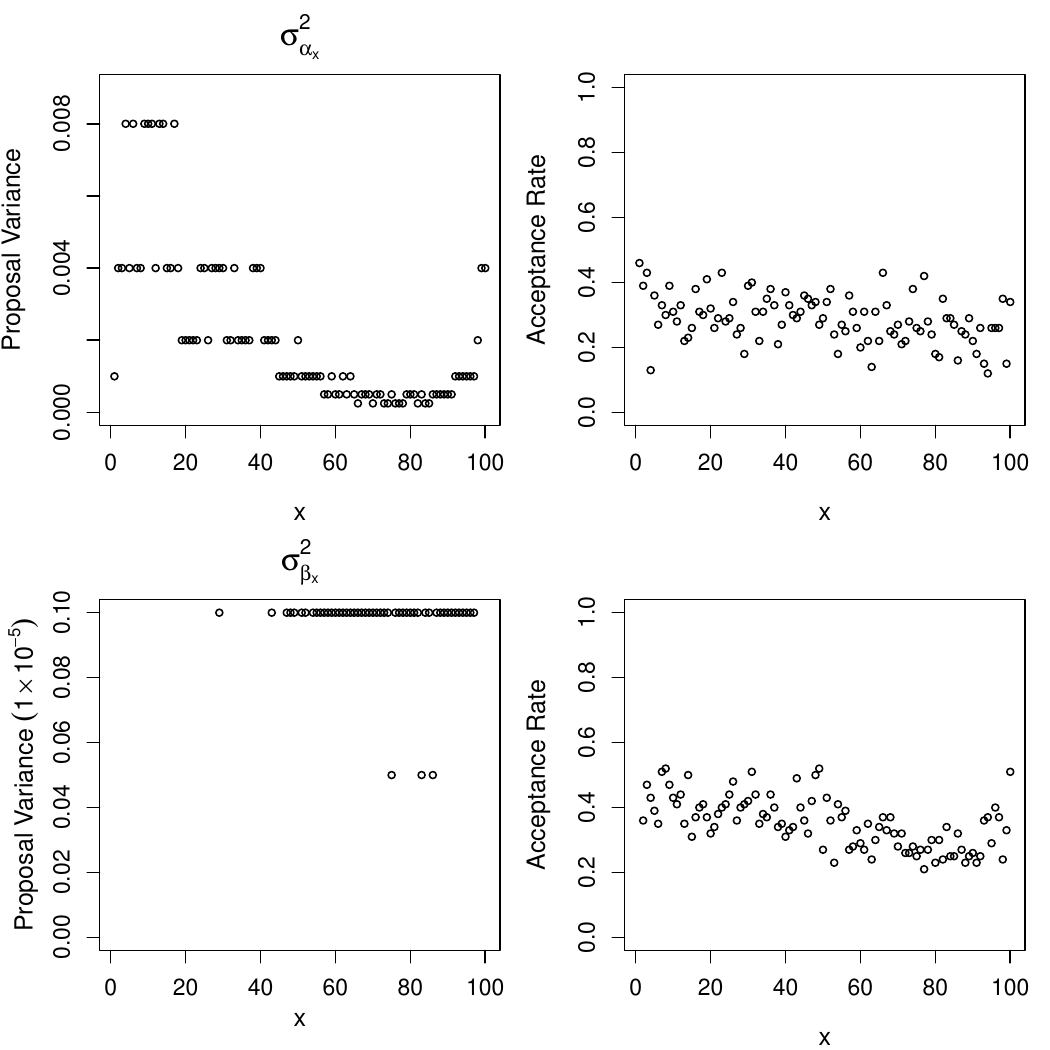}
\caption{Proposal variances of $\alpha_x$ and $\beta_x$ for $x=1,\ldots,A$ with the corresponding MCMC acceptance rates for the CMP--LC--$x$ model.}
\label{fig:proposal_var_ab_vary_age}
\end{figure}

We adopted the search algorithm described by \cite{wong_2017} for determining the appropriate proposal variances (so that the acceptance rates are between 0.15-0.45) based on pilot runs. As an illustration, the calibrated proposal variances for the CMP--LC--$x$ are illustrated in Figure \ref{fig:proposal_var_ab_vary_age}.

\subsection{MH updating for \texorpdfstring{$\kappa_t$}{}}

Due to the correlations induced by the constraints, it is more efficient to perform MCMC updating on $\bold{\kappa}$ in a single block as suggested by \cite{blocking}. The conditional posterior density of $\bold{\kappa}$ is
$$f(\bold{\kappa}|\bold{d},rest)
\propto \prod_{x,t} \left\{\frac{1}{Z(\lambda_{xt},\nu_x)}\frac{(\lambda_{xt})^{d_{xt}}}{(d_{xt}!)^{\nu_x}}\right\} \times \exp\left(-\frac{1}{2\sigma_\kappa^2}(\bold{\kappa}-\bold{\mu}_{\kappa})^\top {V}^{-1}(\bold{\kappa}-\bold{\mu}_{\kappa})\right).$$

After some numerical tuning, the proposal variance matrix used for updating $\bold{\kappa}$ is
$\Sigma_\kappa=\frac{5.38^2}{41}\times \bold{H}_{\kappa},$ where $\bold{H}_\kappa$ is the sub-matrix of $[-\bold{H}]^{-1}$ corresponding to $\bold{\kappa}$ evaluated at the MLE of the likelihood function of the Poisson-LC model, and $\bold{H}$ is the Hessian matrix of the likelihood function\footnote[2]{This choice was motivated by \cite{acceptrate}.}. In practice, $\bold{H}$ can be estimated by fitting the Poisson-LC model in the frequentist framework (for example, using the function \textit{glm} iteratively in R), then extracting the corresponding covariance matrix. Pilot runs indicate an acceptance rate of around $0.37$ for the MH updating of $\bold{\kappa}$.

\subsection{MH updating for \texorpdfstring{$\sigma_\kappa^2$ and $\rho$}{}}

The conditional posterior densities of $\sigma_{\kappa}^{2}$ and $\rho$ are
\begin{enumerate}[i.]
\item $\sigma_{\kappa}^{-2}|\bold{d},rest \sim \mbox{Gamma}\left(1+\frac{T-1}{2},0.0001+\frac{1}{2}(\bold{\kappa}-\bold{\mu}_\kappa)^\top {V}^{-1}(\bold{\kappa}-\bold{\mu}_\kappa)\right).$
\item $f(\rho|\bold{d},rest)
\propto |{V}|^{-\frac{1}{2}}\exp\left\{-\frac{1}{2\sigma_\kappa^2}(\bold{\kappa}-\bold{\mu}_{\kappa})^\top {V}^{-1}(\bold{\kappa}-\bold{\mu}_{\kappa})\right\} \times (1+\rho)^2(1-\rho) \times I_{(-1,1)}(\rho),$

where $I_{(-1,1)}(\rho)$ is the indicator function. This corresponds to fitting a stationary AR(1) model on $\kappa_t$. Manually setting $\rho=1$ within the MCMC gives the random walk model. 
\end{enumerate}

\subsection{MH updating for \texorpdfstring{$\gamma_c$}{}}

The conditional posterior density of $\bold{\gamma}$ is 
$$f(\bold{\gamma}|\bold{d},rest)\propto \prod_{x,t} \left\{\frac{1}{Z(\lambda_{xt},\nu_x)}\frac{(\lambda_{xt})^{d_{xt}}}{(d_{xt}!)^{\nu_x}}\right\}  \times \exp\left(-\frac{1}{2\sigma_\gamma^2}\bold{\gamma}^\top{W}_\gamma^{-1}\bold{\gamma}\right).$$
$\bold{\gamma}$ is updated using random walk MH in a single block. Using the proposal variance matrix
${\Sigma}_\gamma=\frac{2.38^2}{138}\times\bold{H}_{\gamma},$ where $\bold{H}_{\gamma}$ is the sub-matrix of $[-\bold{H}]^{-1}$ corresponding to $\bold{\gamma}$, evaluated at the MLE of the likelihood function of the Poisson-LCC model, and $\bold{H}$ is the Hessian matrix of the likelihood function (again, fitted using the \textit{glm} function in R). Pilot runs indicate an acceptance rate of around $0.26$ for the MH updating of $\bold{\gamma}$.

\subsection{MH updating for \texorpdfstring{$\sigma_\gamma$ and $\rho_\gamma$}{}}

The conditional posterior densities of $\sigma_\gamma$ and $\rho_\gamma$ are 
\begin{enumerate}[i.]
\item $f(\sigma_\gamma|\bold{d},rest) \propto \sigma_\gamma^{-(C-3)}\times \exp\left(-\frac{1}{2\sigma_\gamma^2}\bold{\gamma}^\top {W}_\gamma^{-1}\bold{\gamma}\right)\times I_{(0,0.1)}(\sigma_\gamma).$
\item $f(\rho_\gamma|\bold{d},rest)\propto |{W}_\gamma|^{-\frac{1}{2}} \times \exp\left(-\frac{1}{2\sigma_\gamma^2}\bold{\gamma}^\top {W}_\gamma^{-1}\bold{\gamma}-\frac{1}{2}\rho_\gamma^2\right).$
\end{enumerate}

\subsection{Proposal variances of dispersion parameters}

\begin{figure}[htbp]
\centering
\includegraphics[scale=0.7]{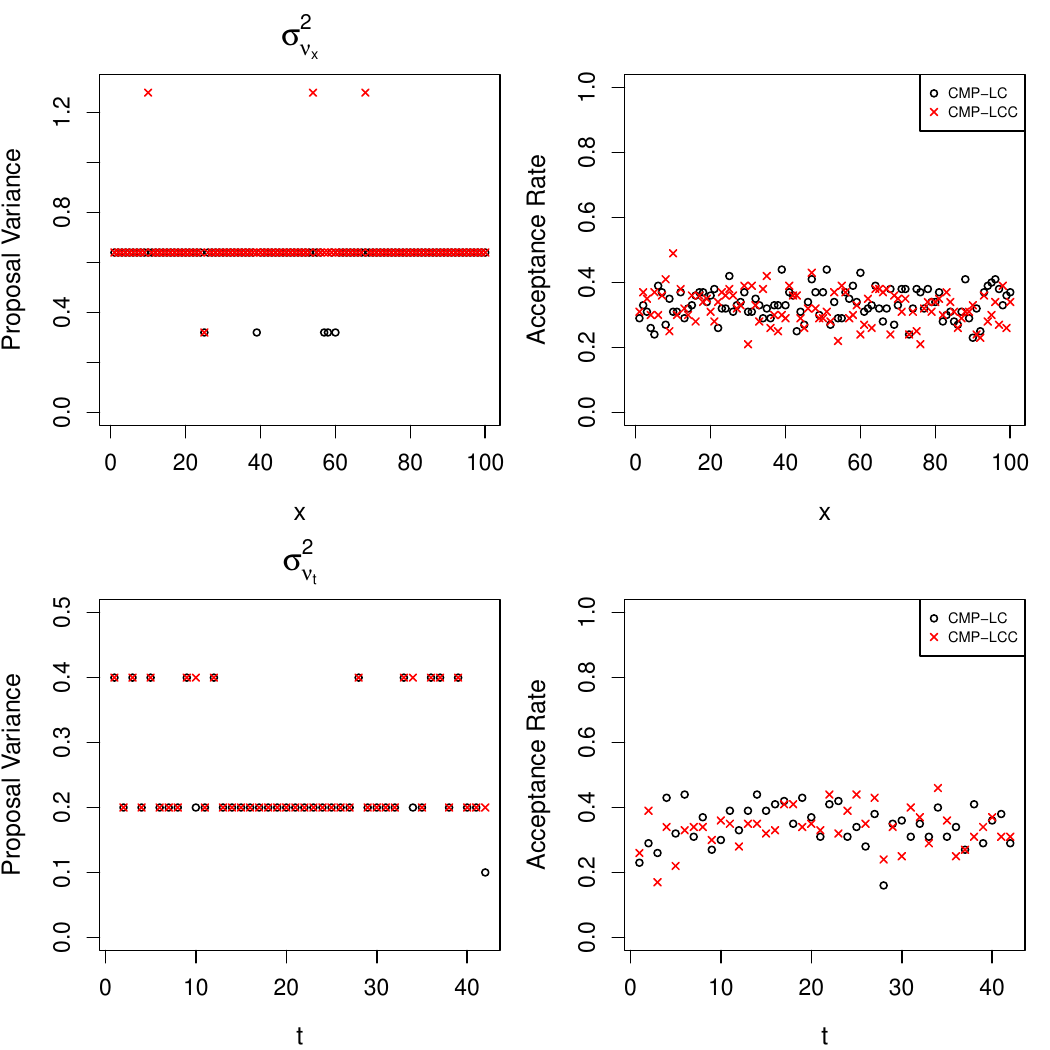}
\caption{Proposal variances of the dispersion parameters, $\sigma_{\nu_x}^2$ and $\sigma_{\nu_t}^2$ with the corresponding MCMC acceptance rates respectively for the CMP--LC/LCC--$x$ and CMP--LC/LCC--$t$ models.}
\label{fig:proposal_var_nu}
\end{figure}

The numerically determined $\sigma_{\nu_x}^2$ ($x=1,\ldots,A$) and $\sigma_{\nu_t}^2$ ($t=1,\ldots,T$) are tabulated in Tables \ref{tab:proposal_var_nu_vary_age} and \ref{tab:proposal_var_nu_vary_year}. They are also illustrated alongside the corresponding MCMC acceptance rates in Figure \ref{fig:proposal_var_nu}.

\begin{table}[!htbp]
\caption{The numerically calibrated proposal variances of the dispersion parameters, $\sigma_{\nu_x}^2$ for the CMP--LC--$x$ and CMP--LCC--$x$ models.}
\label{tab:proposal_var_nu_vary_age}
\centering
\begin{tabular}{rrr|rrr}
\hline
$x$ & CMP--LC--$x$ & CMP--LCC--$x$ 
 & $x$ & CMP--LC--$x$ & CMP--LCC--$x$ \\
\hline
1  & 0.64 & 0.64 & 51  & 0.64 & 0.64 \\
2  & 0.64 & 0.64 & 52  & 0.64 & 0.64 \\
3  & 0.64 & 0.64 & 53  & 0.64 & 0.64 \\
4  & 0.64 & 0.64 & 54  & 0.64 & 1.28 \\
5  & 0.64 & 0.64 & 55  & 0.64 & 0.64 \\
6  & 0.64 & 0.64 & 56  & 0.64 & 0.64 \\
7  & 0.64 & 0.64 & 57  & 0.32 & 0.64 \\
8  & 0.64 & 0.64 & 58  & 0.32 & 0.64 \\
9  & 0.64 & 0.64 & 59  & 0.64 & 0.64 \\
10 & 0.64 & 1.28 & 60  & 0.32 & 0.64 \\
11 & 0.64 & 0.64 & 61  & 0.64 & 0.64 \\
12 & 0.64 & 0.64 & 62  & 0.64 & 0.64 \\
13 & 0.64 & 0.64 & 63  & 0.64 & 0.64 \\
14 & 0.64 & 0.64 & 64  & 0.64 & 0.64 \\
15 & 0.64 & 0.64 & 65  & 0.64 & 0.64 \\
16 & 0.64 & 0.64 & 66  & 0.64 & 0.64 \\
17 & 0.64 & 0.64 & 67  & 0.64 & 0.64 \\
18 & 0.64 & 0.64 & 68  & 0.64 & 1.28 \\
19 & 0.64 & 0.64 & 69  & 0.64 & 0.64 \\
20 & 0.64 & 0.64 & 70  & 0.64 & 0.64 \\
21 & 0.64 & 0.64 & 71  & 0.64 & 0.64 \\
22 & 0.64 & 0.64 & 72  & 0.64 & 0.64 \\
23 & 0.64 & 0.64 & 73  & 0.64 & 0.64 \\
24 & 0.64 & 0.64 & 74  & 0.64 & 0.64 \\
25 & 0.32 & 0.32 & 75  & 0.64 & 0.64 \\
26 & 0.64 & 0.64 & 76  & 0.64 & 0.64 \\
27 & 0.64 & 0.64 & 77  & 0.64 & 0.64 \\
28 & 0.64 & 0.64 & 78  & 0.64 & 0.64 \\
29 & 0.64 & 0.64 & 79  & 0.64 & 0.64 \\
30 & 0.64 & 0.64 & 80  & 0.64 & 0.64 \\
31 & 0.64 & 0.64 & 81  & 0.64 & 0.64 \\
32 & 0.64 & 0.64 & 82  & 0.64 & 0.64 \\
33 & 0.64 & 0.64 & 83  & 0.64 & 0.64 \\
34 & 0.64 & 0.64 & 84  & 0.64 & 0.64 \\
35 & 0.64 & 0.64 & 85  & 0.64 & 0.64 \\
36 & 0.64 & 0.64 & 86  & 0.64 & 0.64 \\
37 & 0.64 & 0.64 & 87  & 0.64 & 0.64 \\
38 & 0.64 & 0.64 & 88  & 0.64 & 0.64 \\
39 & 0.32 & 0.64 & 89  & 0.64 & 0.64 \\
40 & 0.64 & 0.64 & 90  & 0.64 & 0.64 \\
41 & 0.64 & 0.64 & 91  & 0.64 & 0.64 \\
42 & 0.64 & 0.64 & 92  & 0.64 & 0.64 \\
43 & 0.64 & 0.64 & 93  & 0.64 & 0.64 \\
44 & 0.64 & 0.64 & 94  & 0.64 & 0.64 \\
45 & 0.64 & 0.64 & 95  & 0.64 & 0.64 \\
46 & 0.64 & 0.64 & 96  & 0.64 & 0.64 \\
47 & 0.64 & 0.64 & 97  & 0.64 & 0.64 \\
48 & 0.64 & 0.64 & 98  & 0.64 & 0.64 \\
49 & 0.64 & 0.64 & 99  & 0.64 & 0.64 \\
50 & 0.64 & 0.64 & 100 & 0.64 & 0.64 \\
\hline
\end{tabular}
\end{table}

\begin{table}[!htbp]
\caption{The numerically calibrated proposal variances of the dispersion parameters, $\sigma_{\nu_t}^2$ for the CMP--LC--$t$ and CMP--LCC--$t$ models.}
\label{tab:proposal_var_nu_vary_year}
\centering
\begin{tabular}{rrr|rrr}
\hline
 & CMP--LC--$t$ & CMP--LCC--$t$ 
 &  & CMP--LC--$t$ & CMP--LCC--$t$ \\
\hline
1  & 0.40 & 0.40 & 22 & 0.20 & 0.20 \\
2  & 0.20 & 0.20 & 23 & 0.20 & 0.20 \\
3  & 0.40 & 0.40 & 24 & 0.20 & 0.20 \\
4  & 0.20 & 0.20 & 25 & 0.20 & 0.20 \\
5  & 0.40 & 0.40 & 26 & 0.20 & 0.20 \\
6  & 0.20 & 0.20 & 27 & 0.20 & 0.20 \\
7  & 0.20 & 0.20 & 28 & 0.40 & 0.40 \\
8  & 0.20 & 0.20 & 29 & 0.20 & 0.20 \\
9  & 0.40 & 0.40 & 30 & 0.20 & 0.20 \\
10 & 0.20 & 0.40 & 31 & 0.20 & 0.20 \\
11 & 0.20 & 0.20 & 32 & 0.20 & 0.20 \\
12 & 0.40 & 0.40 & 33 & 0.40 & 0.40 \\
13 & 0.20 & 0.20 & 34 & 0.20 & 0.40 \\
14 & 0.20 & 0.20 & 35 & 0.20 & 0.20 \\
15 & 0.20 & 0.20 & 36 & 0.40 & 0.40 \\
16 & 0.20 & 0.20 & 37 & 0.40 & 0.40 \\
17 & 0.20 & 0.20 & 38 & 0.20 & 0.20 \\
18 & 0.20 & 0.20 & 39 & 0.40 & 0.40 \\
19 & 0.20 & 0.20 & 40 & 0.20 & 0.20 \\
20 & 0.20 & 0.20 & 41 & 0.20 & 0.20 \\
21 & 0.20 & 0.20 & 42 & 0.10 & 0.20 \\
\hline
\end{tabular}
\end{table}

\clearpage

\newpage

\section{Some MCMC Convergence Diagnostics}
\label{sec:appendix_convergence}
\subsection{Trace plots}
\begin{figure}[!htbp]
\centering\includegraphics[scale=0.75]{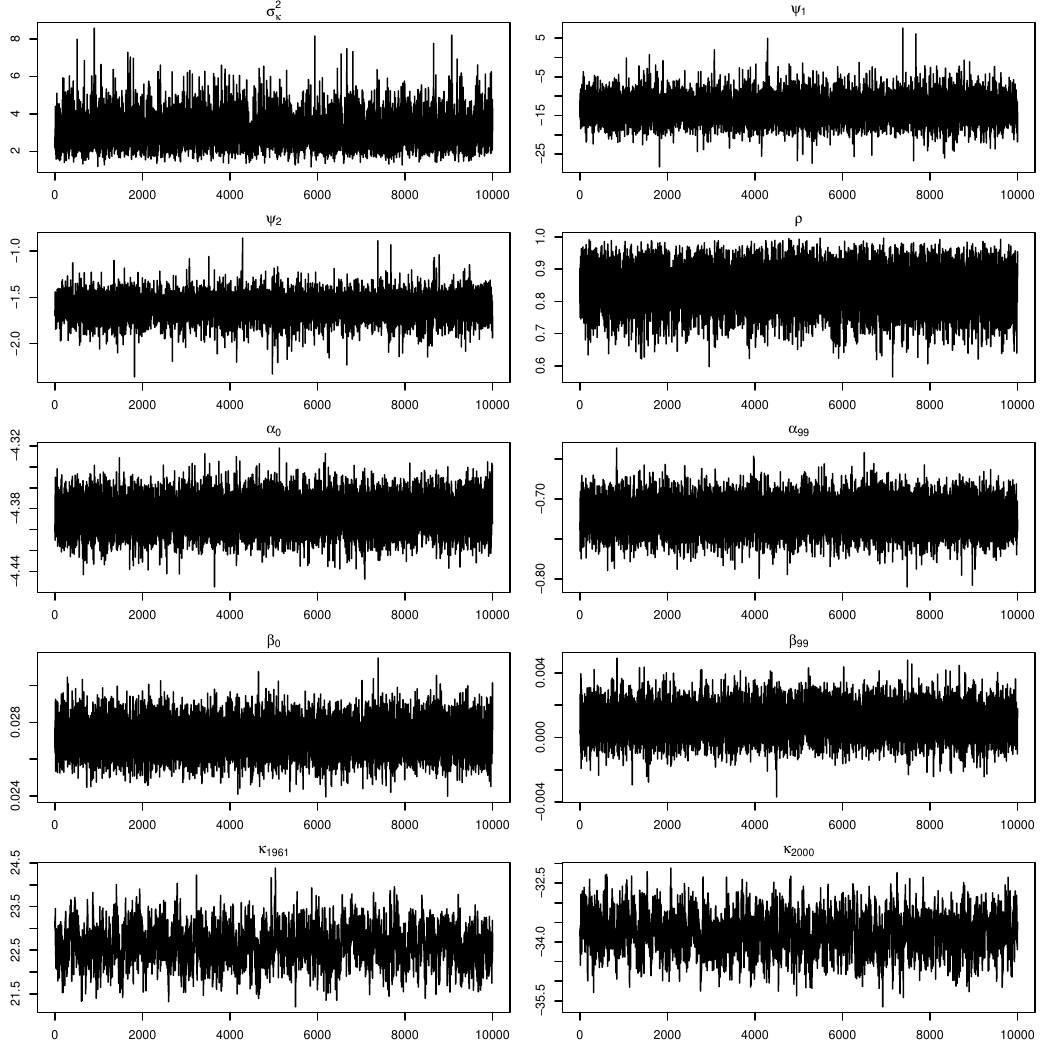}
\caption{\label{fig:trace_CMP_LC_vary_age}Trace plots of estimated parameters for the CMP--LC--$x$ model.}
\end{figure}

\begin{figure}[!htbp]
\centering\includegraphics[scale=0.8]{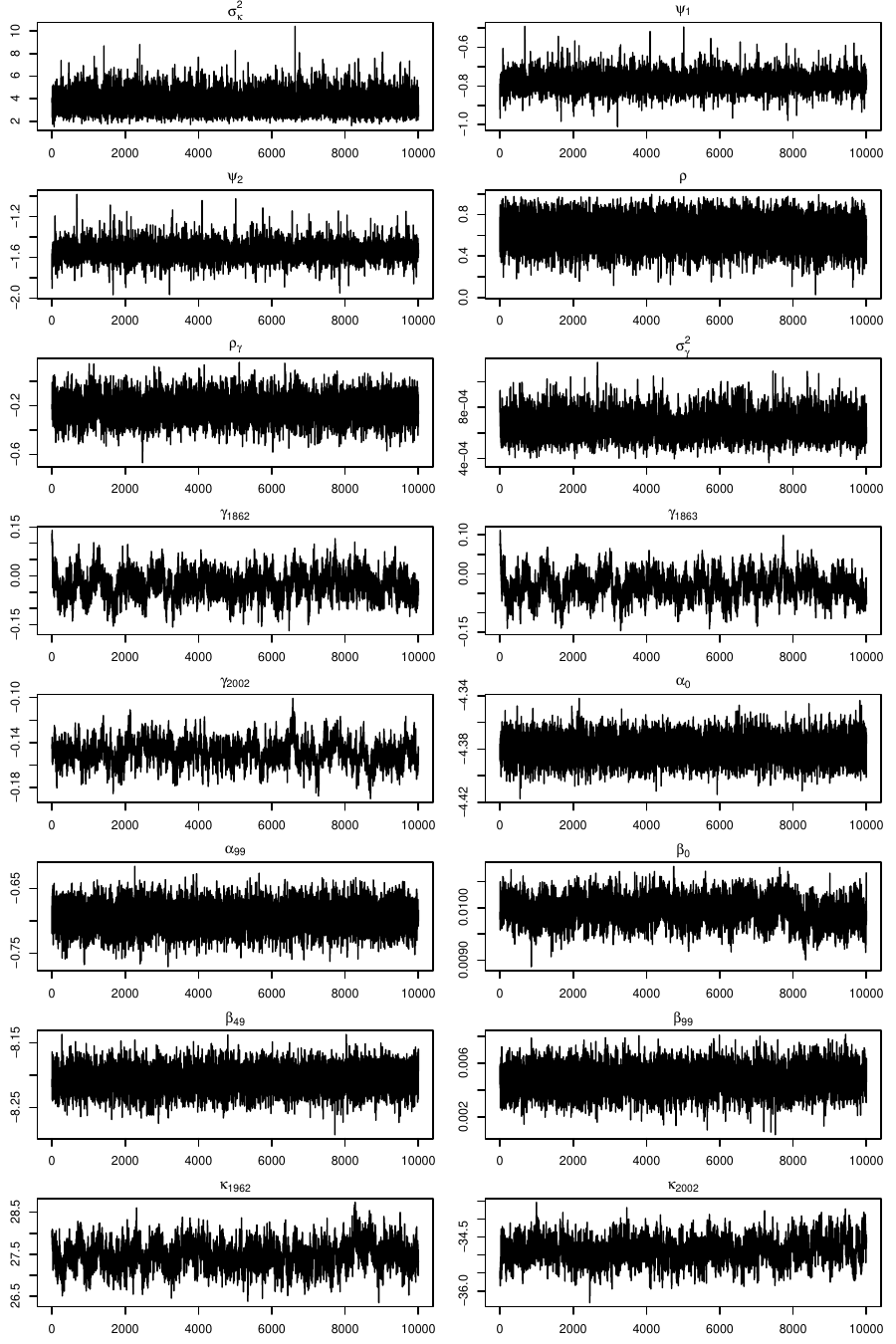}
\caption{\label{fig:trace_CMP_LCC_vary_age}Trace plots of estimated parameters for the CMP--LCC--$x$ model.}
\end{figure}

\begin{figure}[!htbp]
\centering\includegraphics[scale=0.8]{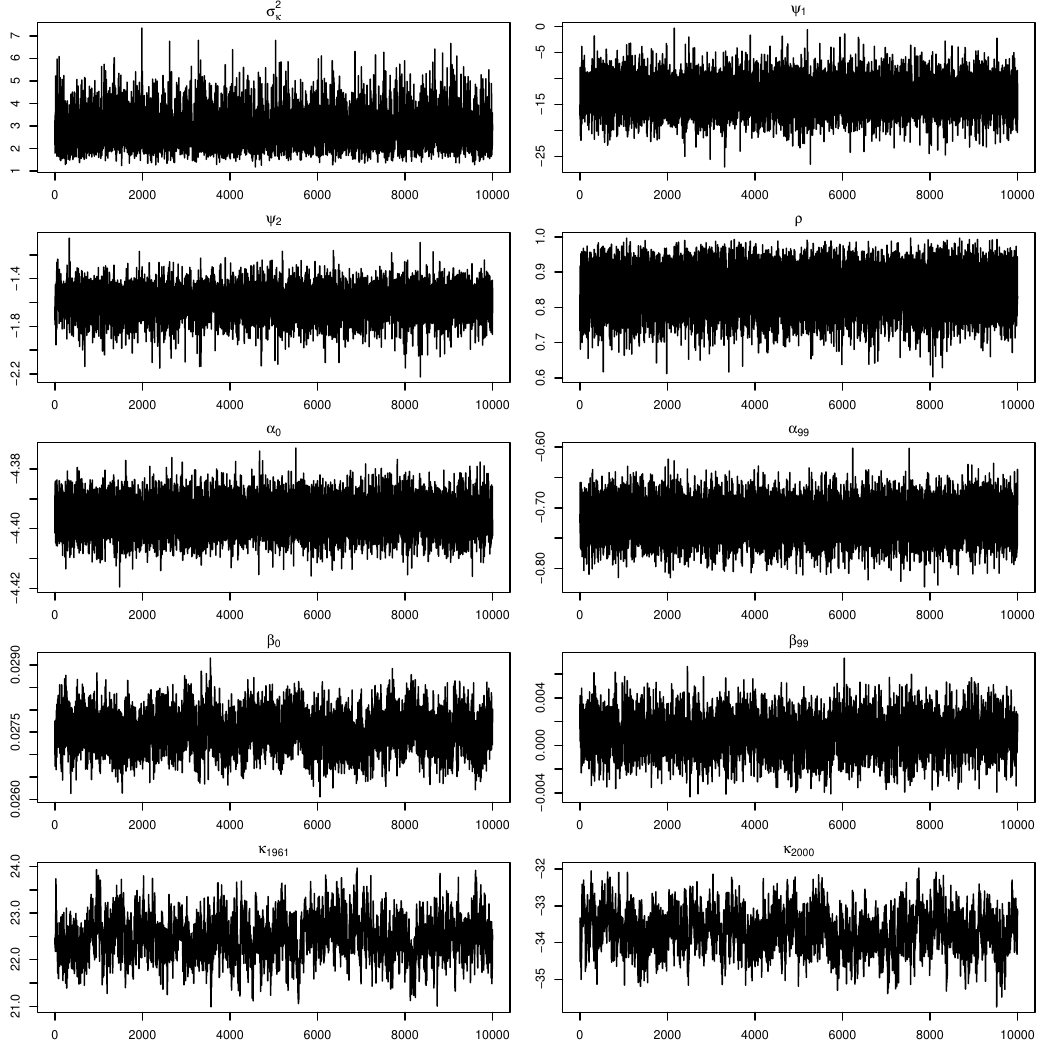}
\caption{\label{fig:trace_CMP_LC_vary_year}Trace plots of estimated parameters for the CMP--LC--$t$ model.}
\end{figure}

\begin{figure}[!htbp]
\centering\includegraphics[scale=0.8]{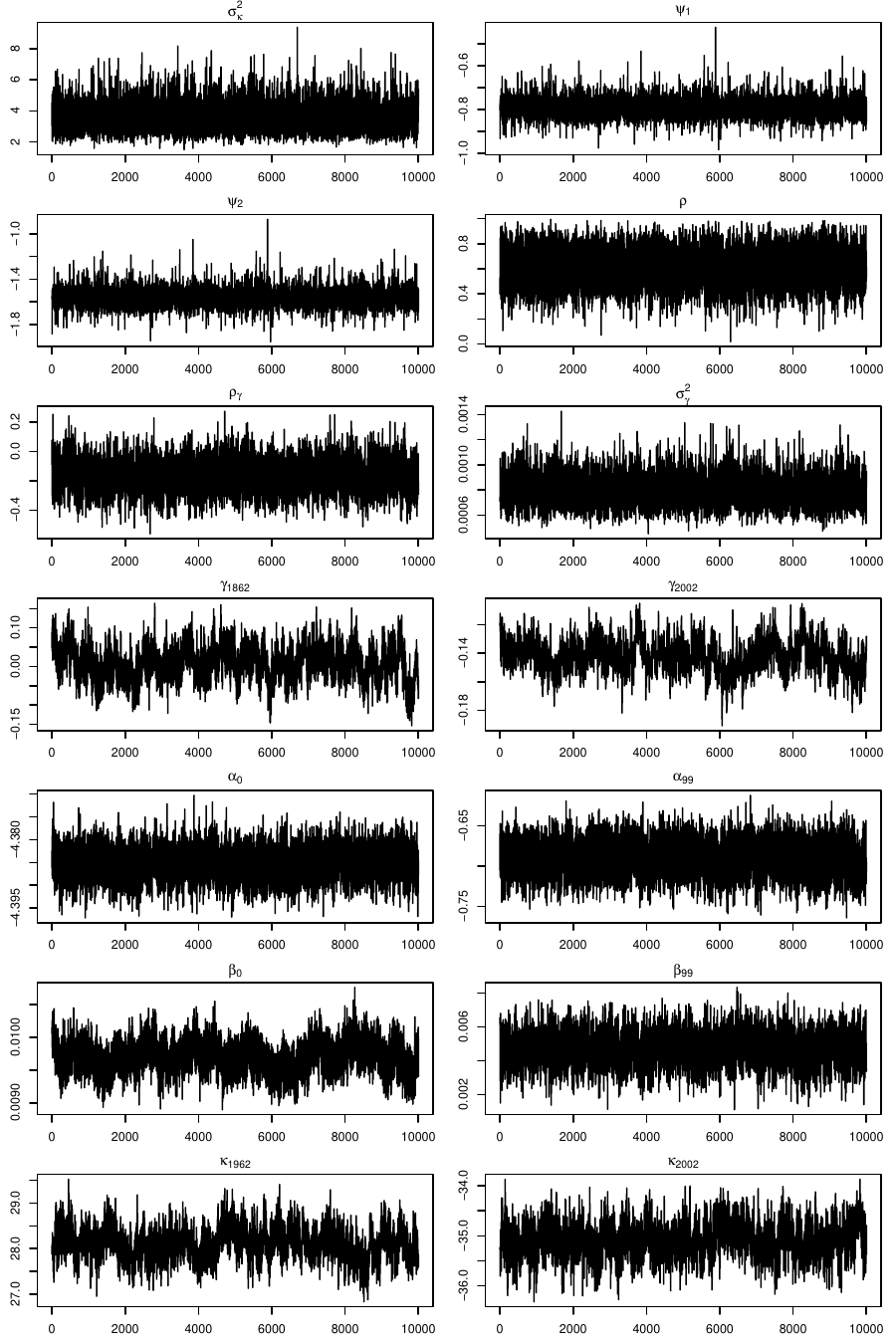}
\caption{\label{fig:trace_CMP_LCC_vary_year}Trace plots of estimated parameters for the CMP--LCC--$t$ model.}
\end{figure}

\begin{figure}[!htbp]
\centering
\includegraphics[scale=0.7]{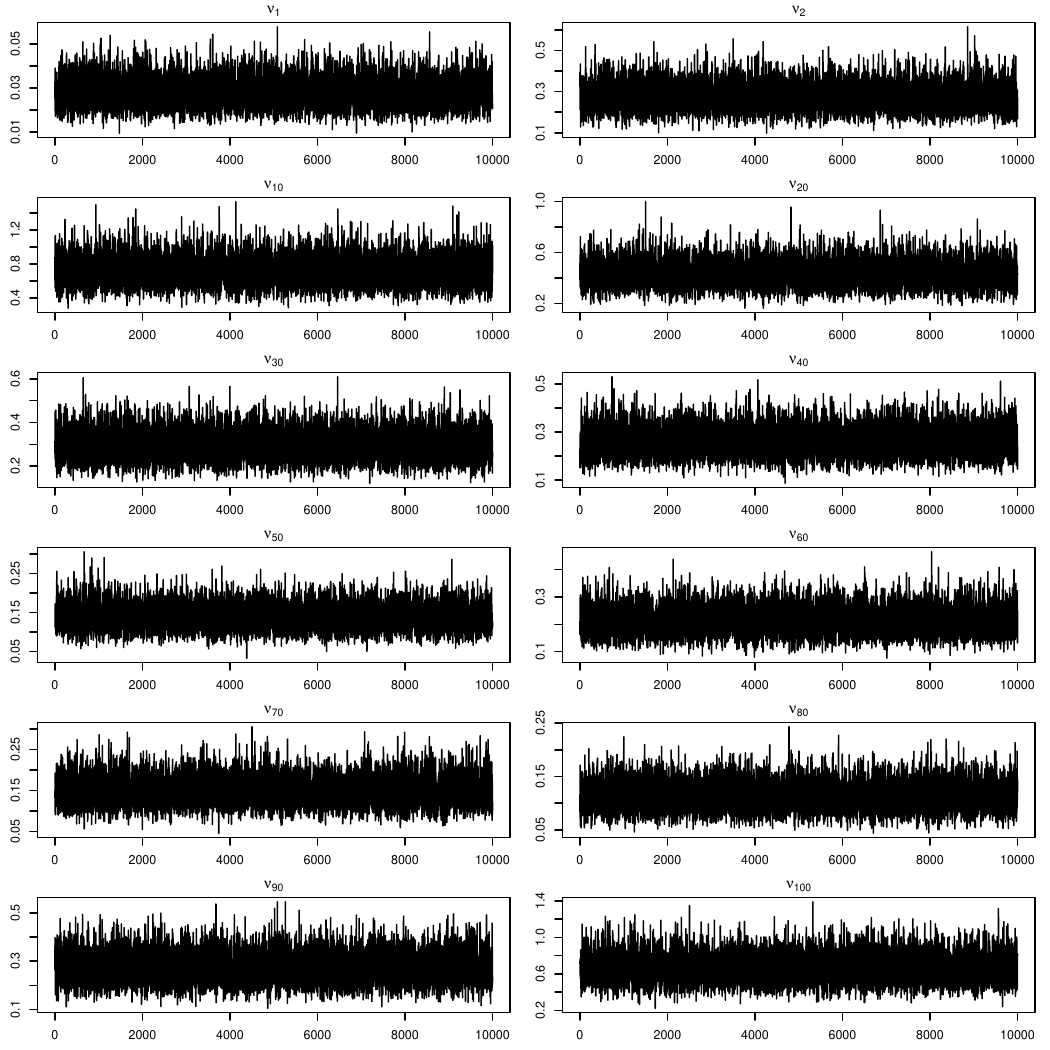}
\caption{\label{fig:CMP_LC_nu_vary_age_trace}Trace plots of the estimated age-specific dispersion parameters $\nu_x$ for several randomly selected $x$ under the CMP--LC--$x$ model.}
\end{figure}

\begin{figure}[!htbp]
\centering
\includegraphics[scale=0.7]{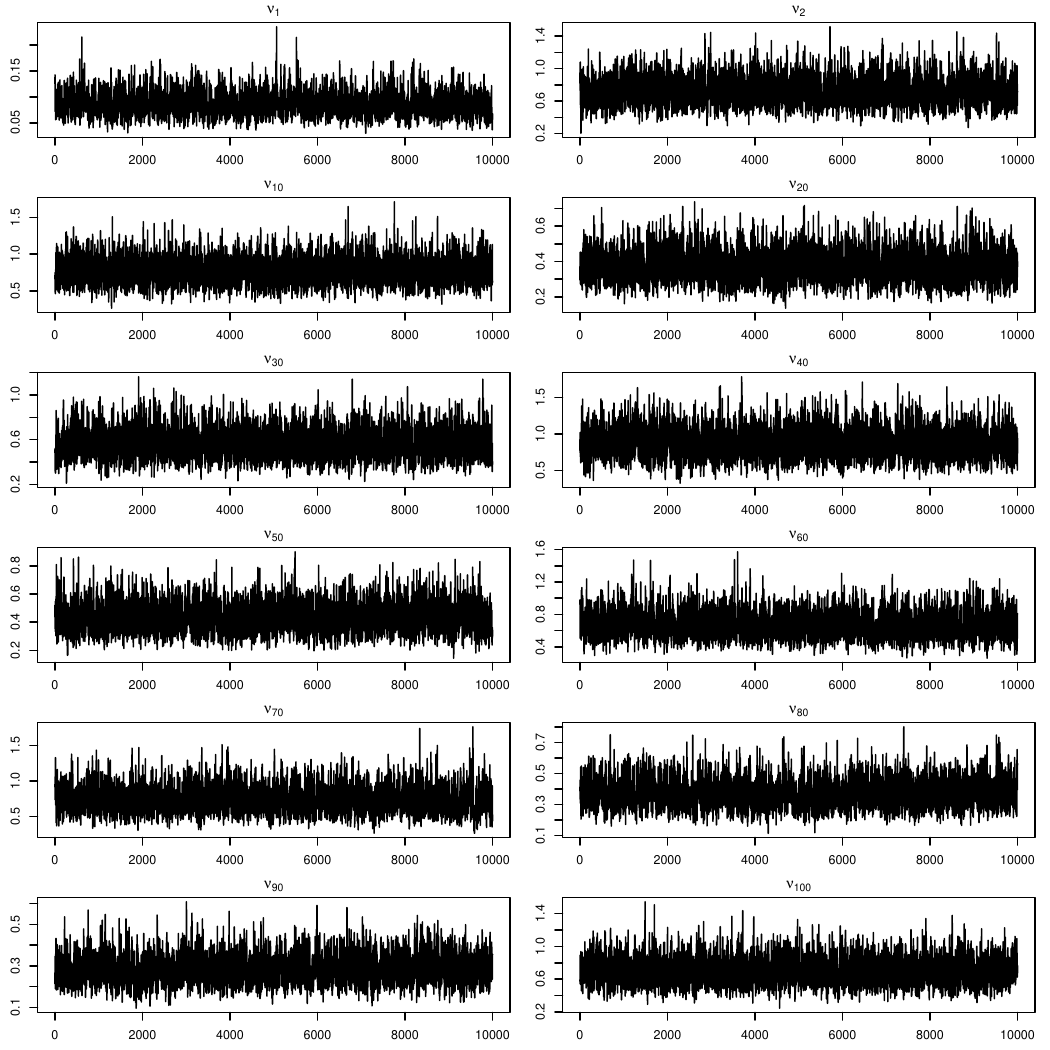}
\caption{\label{fig:CMP_LCC_nu_vary_age_trace}Trace plots of the estimated age-specific dispersion parameters $\nu_x$ for several randomly selected $x$ under the CMP--LCC--$x$ model.}
\end{figure}

\begin{figure}[!htbp]
\centering
\includegraphics[scale=0.7]{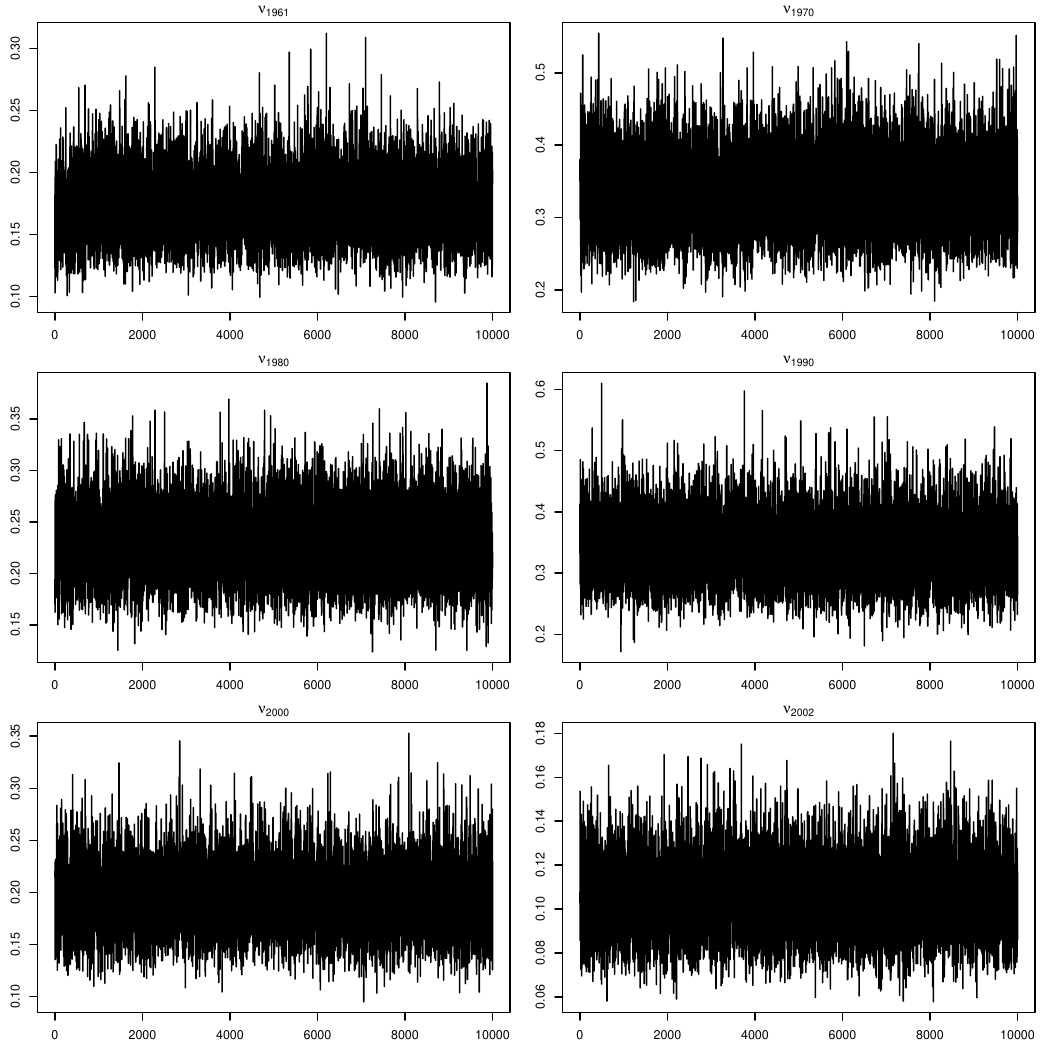}
\caption{\label{fig:CMP_LC_nu_vary_year_trace}Trace plots of the estimated year-specific dispersion parameters $\nu_t$ for several randomly selected $t$ under the CMP--LC--$t$ model.}
\end{figure}

\begin{figure}[!htbp]
\centering
\includegraphics[scale=0.7]{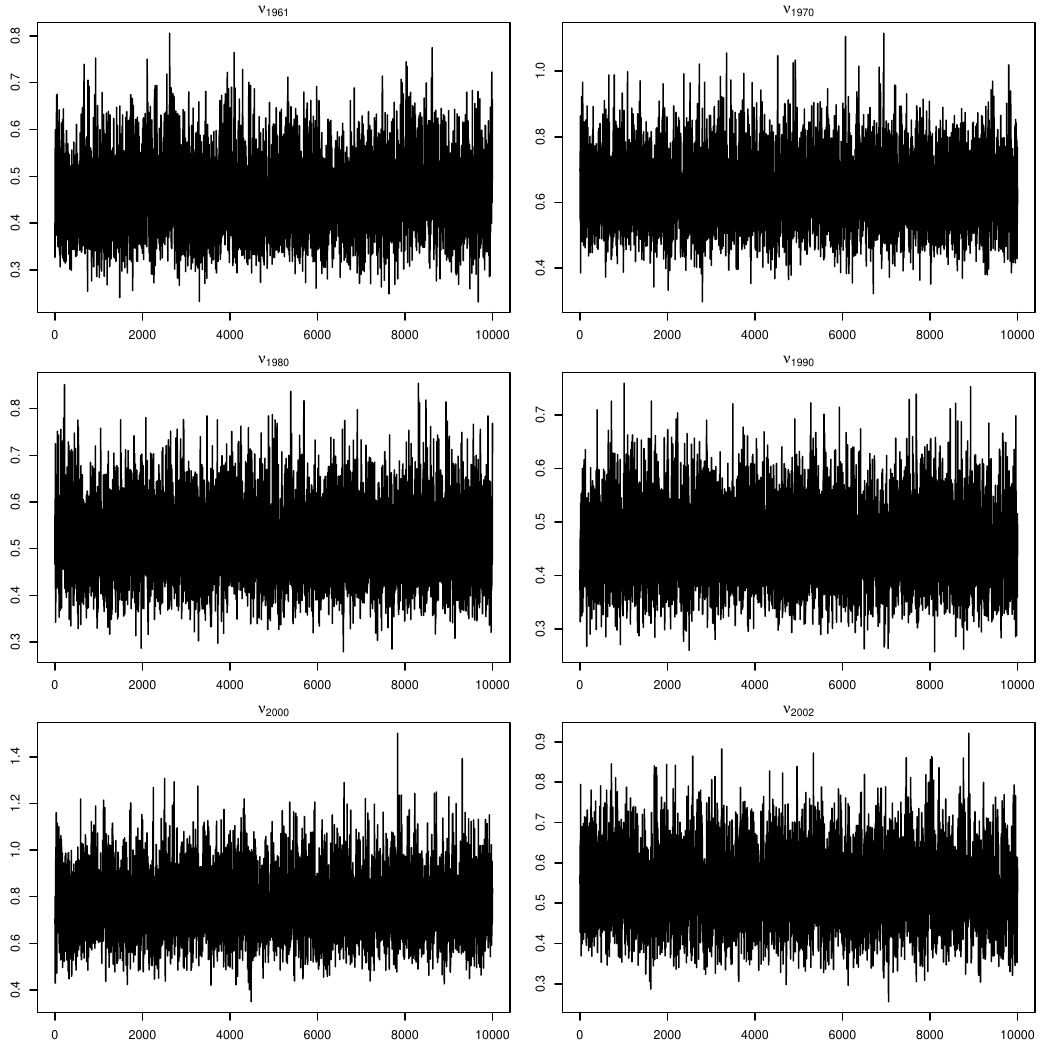}
\caption{\label{fig:CMP_LCC_nu_vary_year_trace}Trace plots of the estimated age-specific dispersion parameters $\nu_t$ for several randomly selected $t$ under the CMP--LCC--$t$ model.}
\end{figure}

\newpage

\subsection{Density plots}

\begin{figure}[!htbp]
\centering\includegraphics[scale=0.8]{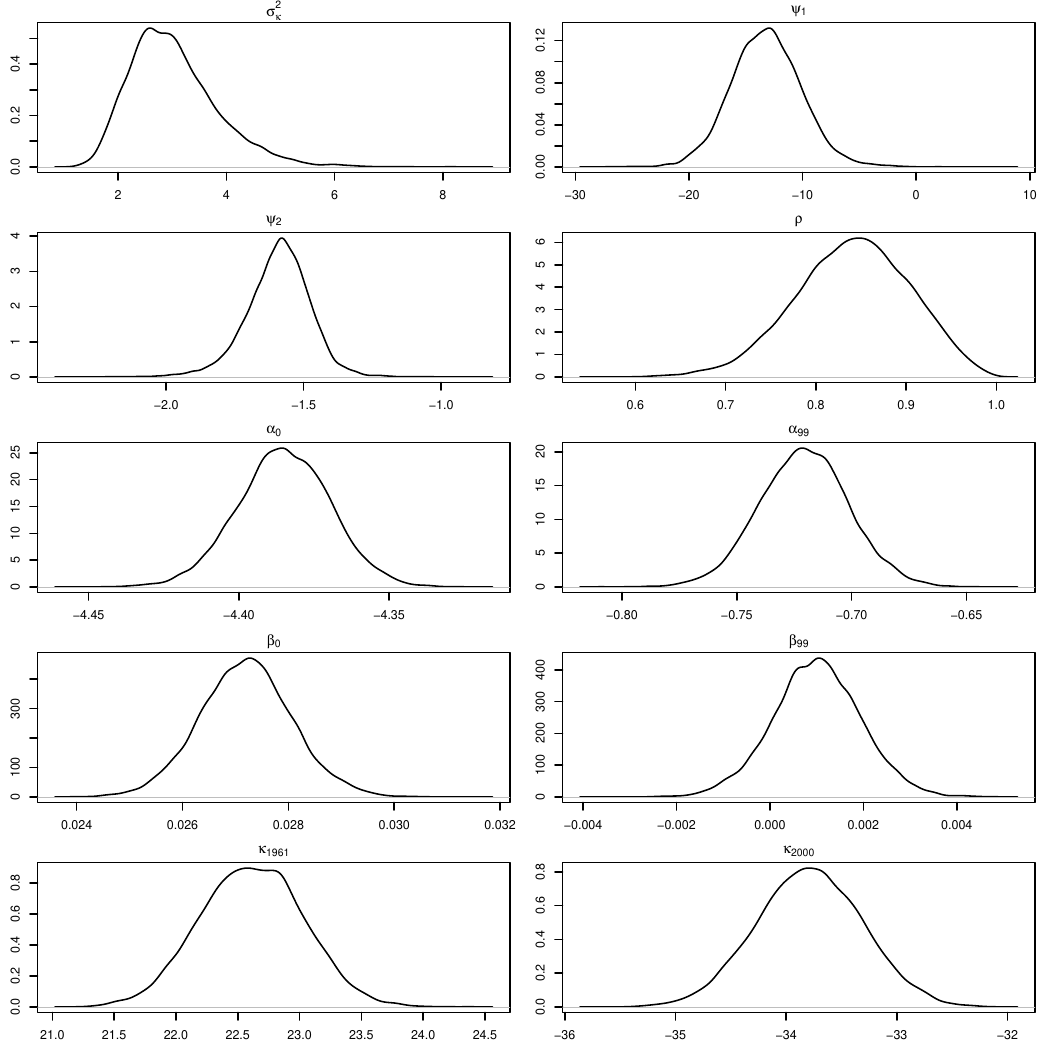}
\caption{\label{fig:density_CMP_LC_vary_age}Density plots of estimated parameters for the CMP--LC--$x$ model.}
\end{figure}

\begin{figure}[!htbp]
\centering\includegraphics[scale=0.8]{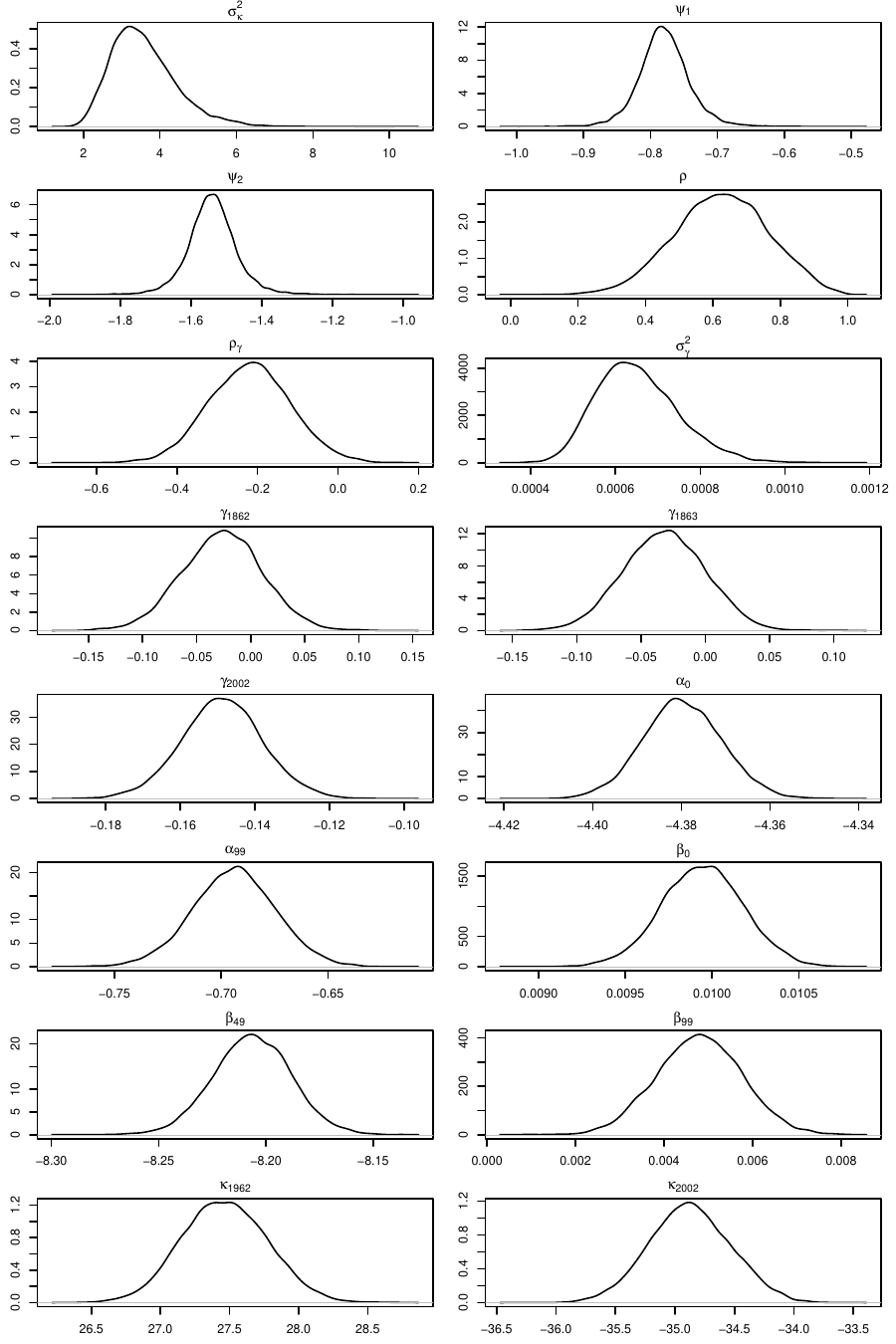}
\caption{\label{fig:density_CMP_LCC_vary_age}Density plots of estimated parameters for the CMP--LCC--$x$ model.}
\end{figure}

\begin{figure}[!htbp]
\centering\includegraphics[scale=0.8]{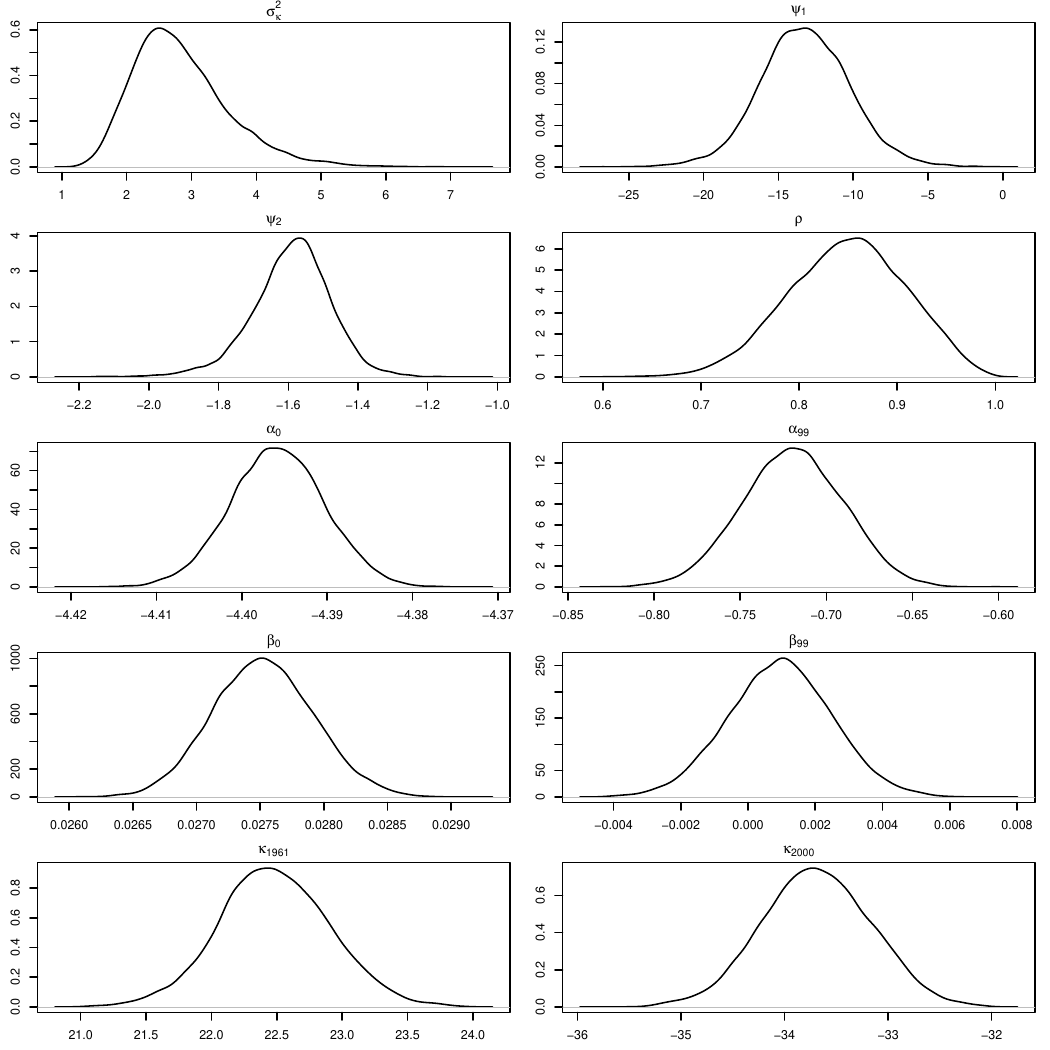}
\caption{\label{fig:density_CMP_LC_vary_year}Density plots of estimated parameters for the CMP--LC--$t$ model.}
\end{figure}

\begin{figure}[!htbp]
\centering\includegraphics[scale=0.8]{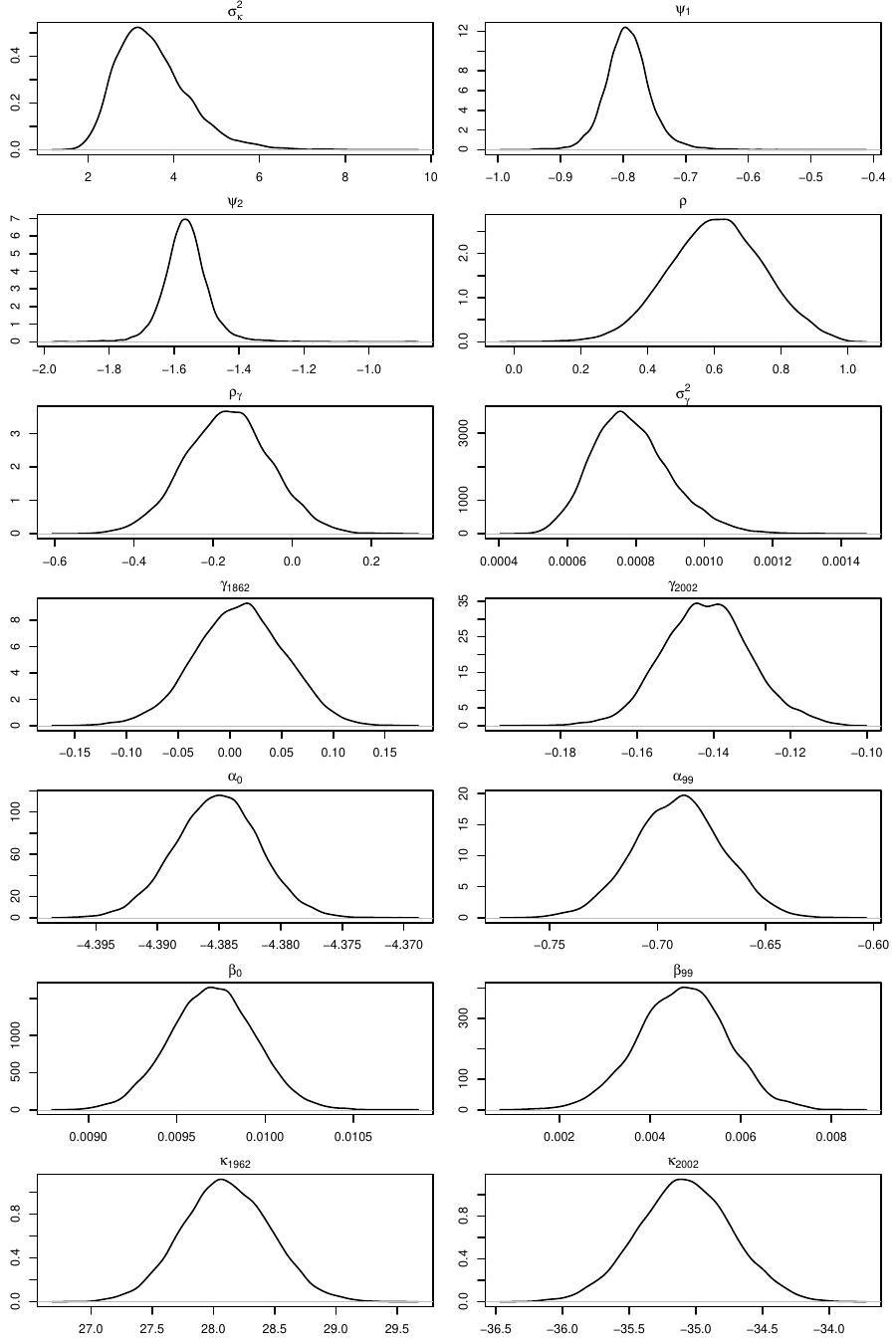}
\caption{\label{fig:density_CMP_LCC_vary_year}Density plots of estimated parameters for the CMP--LCC--$t$ model.}
\end{figure}




\newpage

\subsection{Autocorrelation plots}

\begin{figure}[htbp]
\centering\includegraphics[scale=0.75]{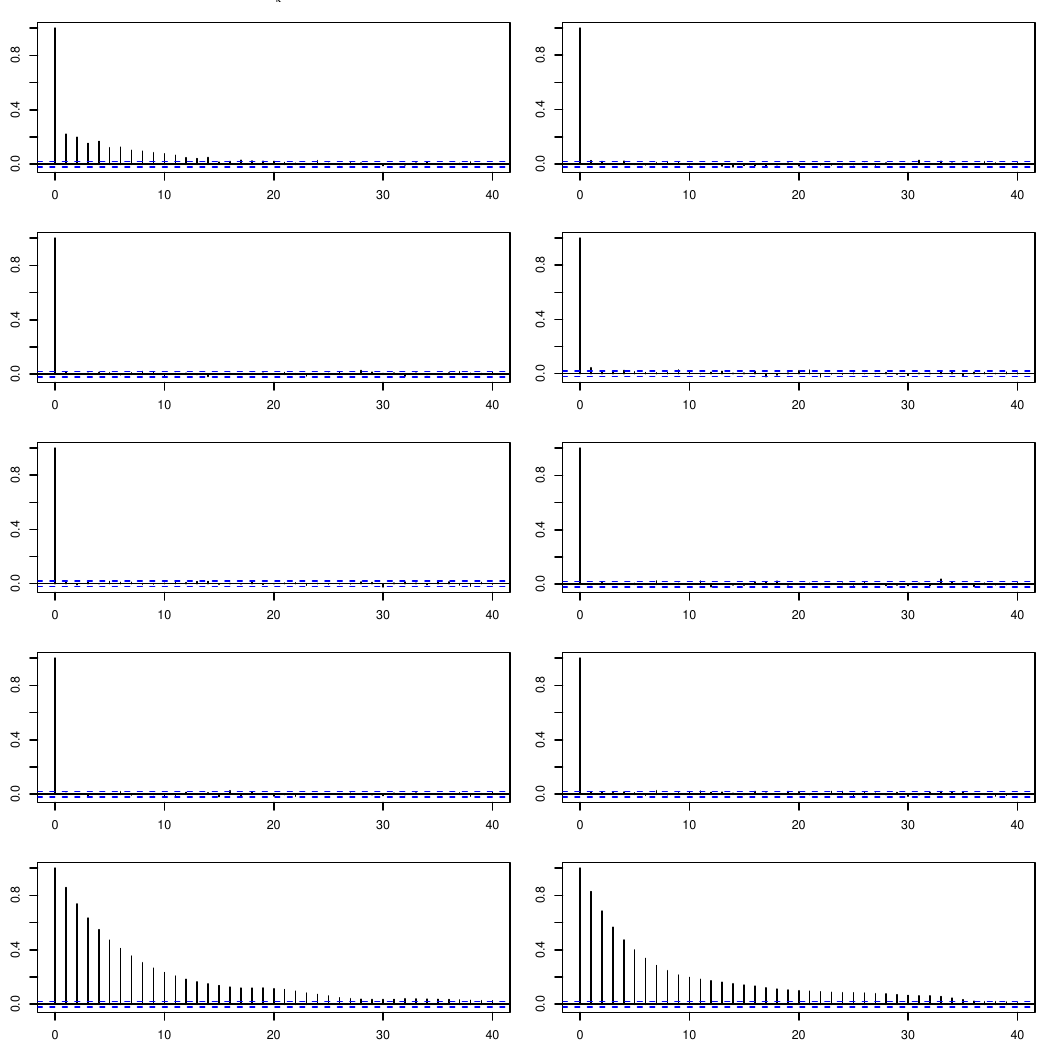}
\caption{\label{fig:acf_LC_vary_age}Autocorrelation plots of some randomly selected parameters for the CMP--LC--$x$ model.}
\end{figure}

\begin{figure}[htbp]
\centering\includegraphics[scale=0.75]{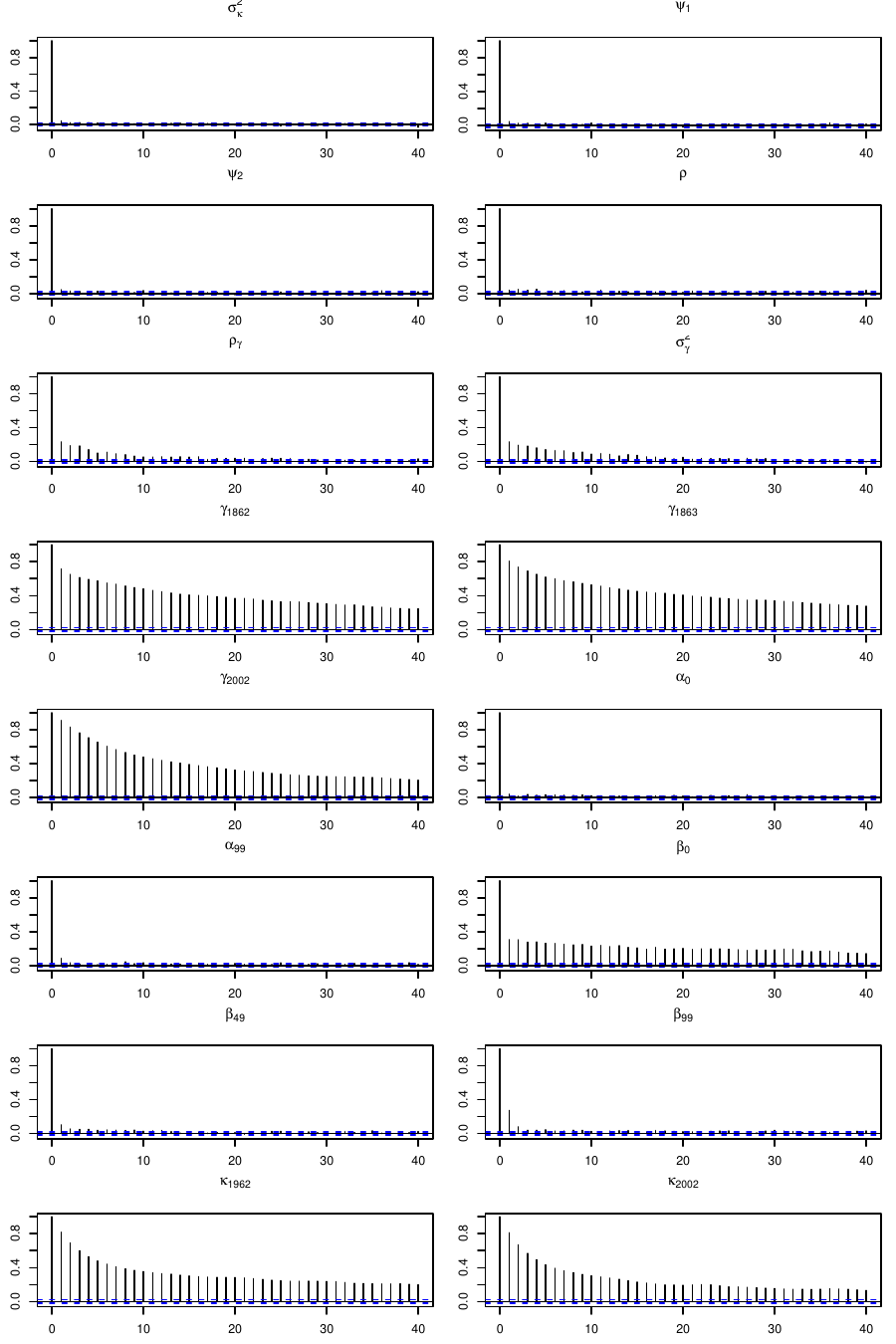}
\caption{\label{fig:acf_LCC_vary_age}Autocorrelation plots of some randomly selected parameters for the CMP--LCC--$x$ model.}
\end{figure}

\begin{figure}[htbp]
\centering\includegraphics[scale=0.75]{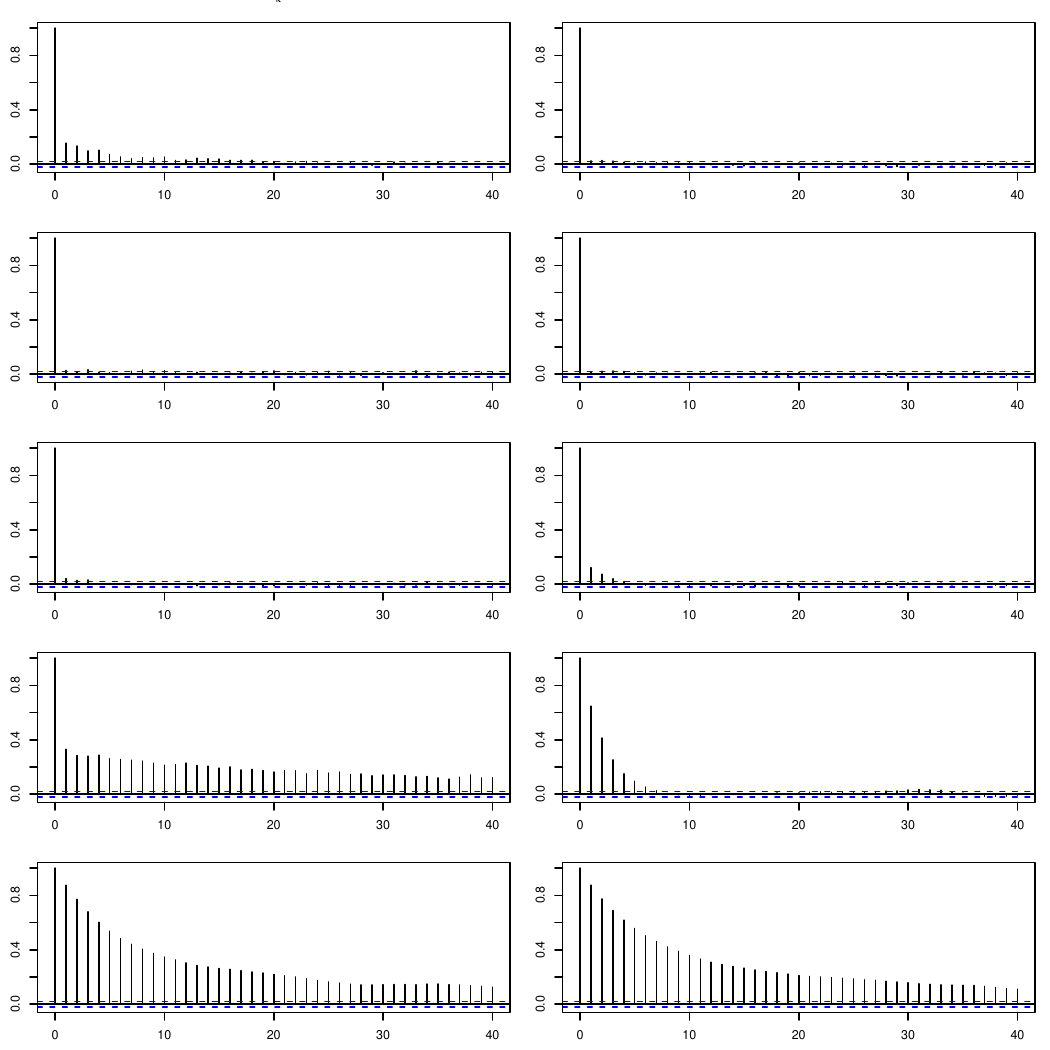}
\caption{\label{fig:acf_LC_vary_year}Autocorrelation plots of some randomly selected parameters for the CMP--LC--$t$ model.}
\end{figure}

\begin{figure}[htbp]
\centering\includegraphics[scale=0.75]{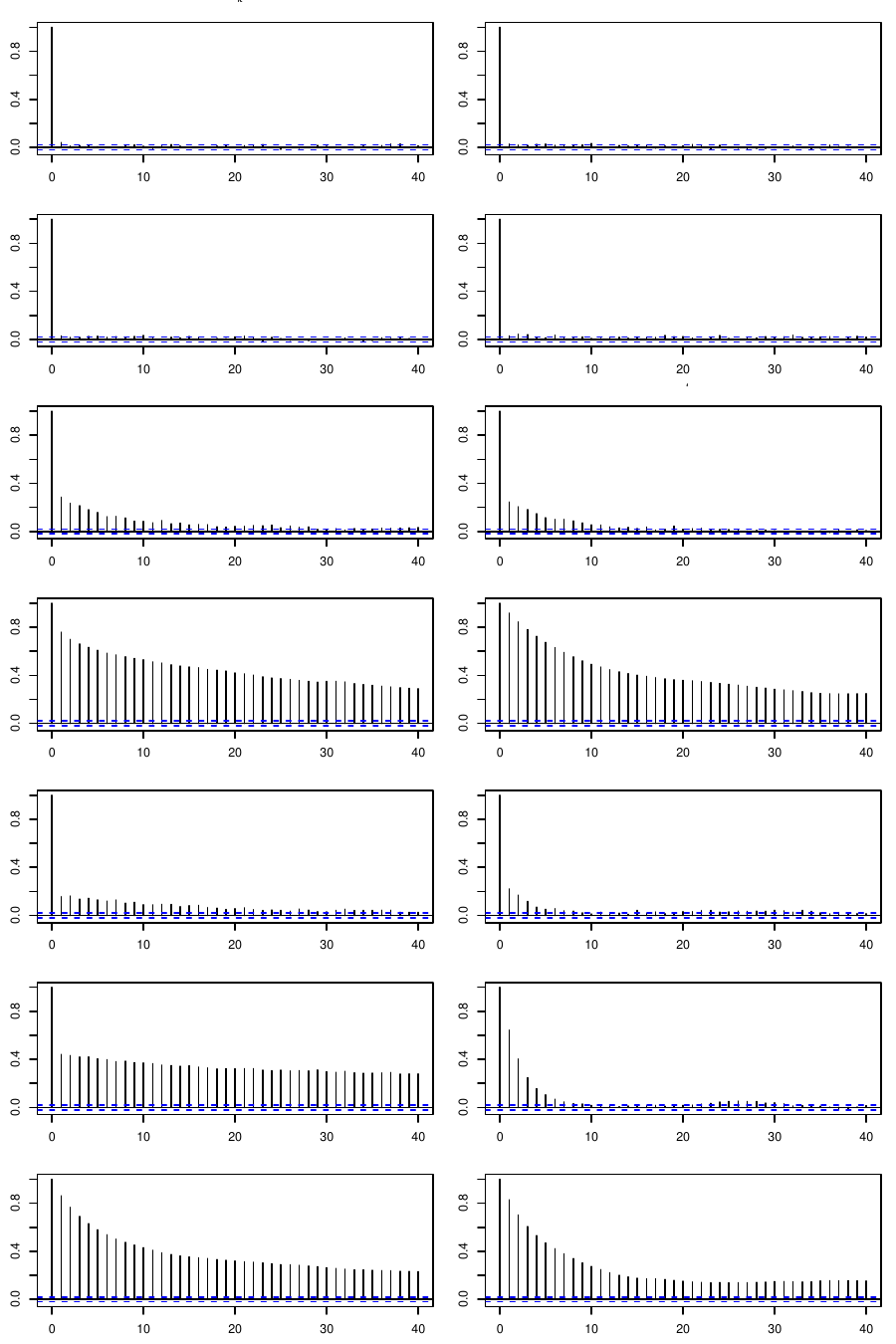}
\caption{\label{fig:acf_LCC_vary_year}Autocorrelation plots of some randomly selected parameters for the CMP--LCC--$t$ model.}
\end{figure}

\newpage

\subsection{Gelman's PSRF}

\begin{figure}[htbp]
\centering\includegraphics[scale=0.5]{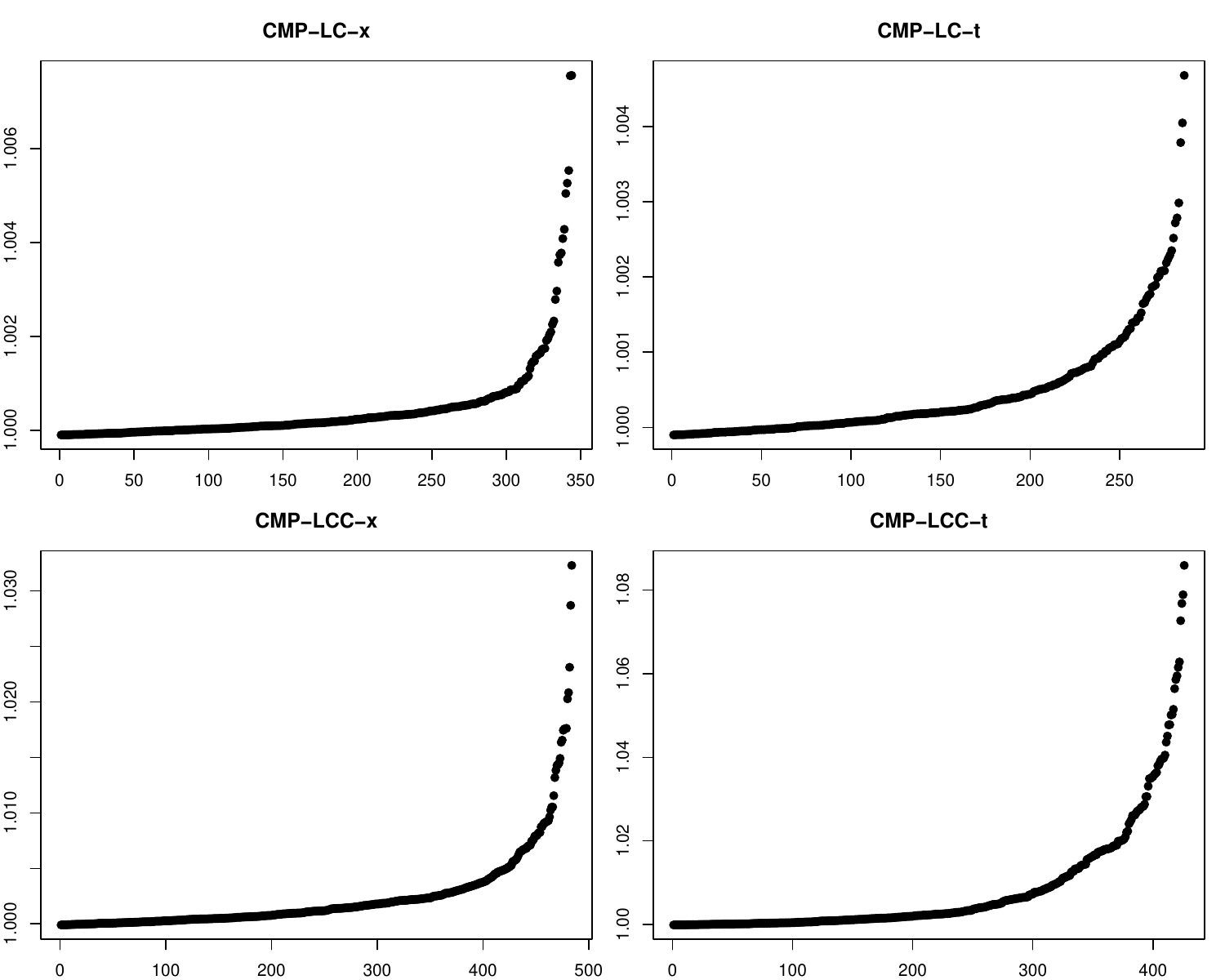}
\caption{\label{fig:gelman}Plot of Gelman's PSRF under various model specifications.}
\end{figure}

\newpage

\subsection{Geweke's Z statistic}

\begin{figure}[htbp]
\centering\includegraphics[scale=0.375]{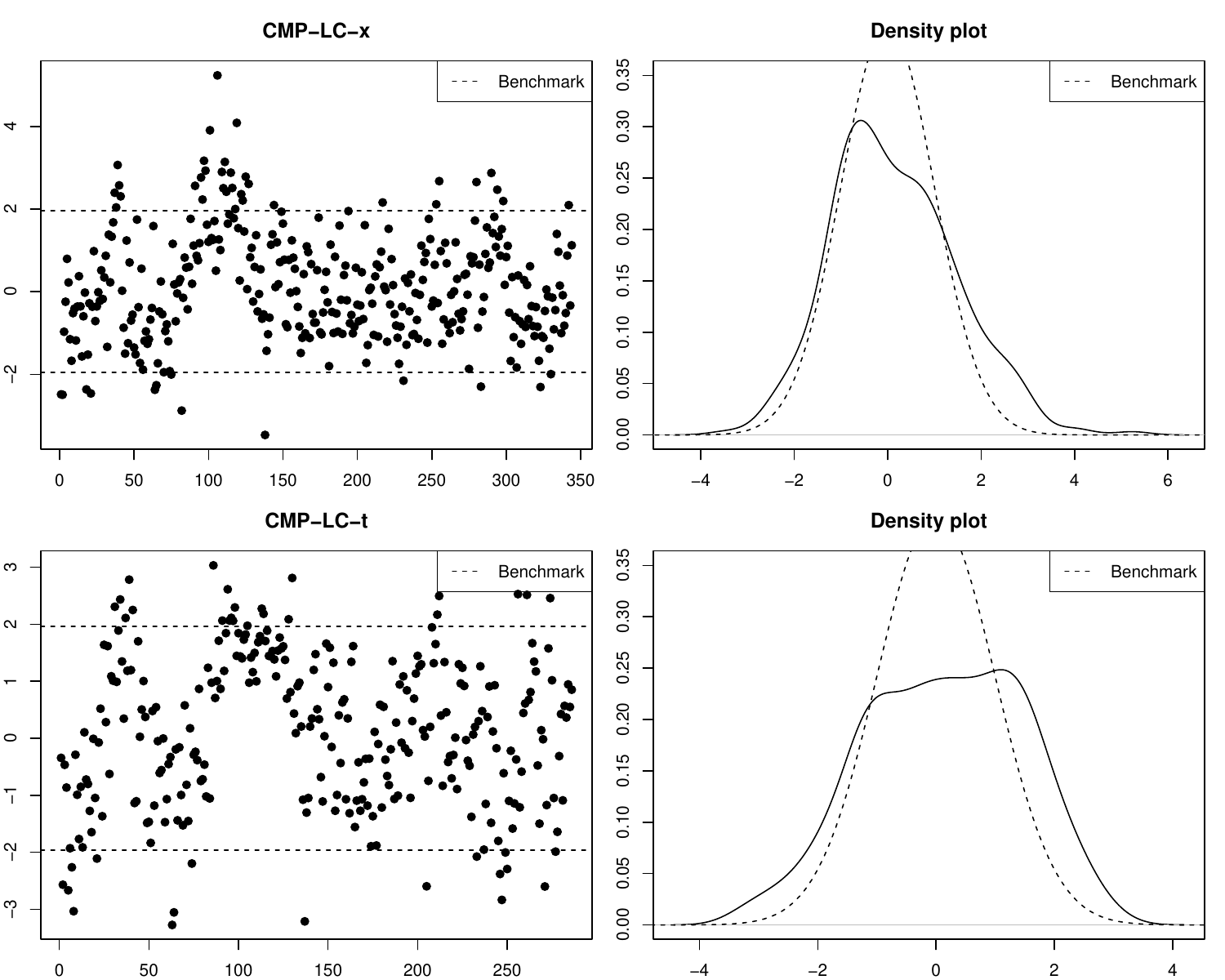}
\caption{\label{fig:geweke_LC}Plot of Geweke's Z statistics under the CMP--LC--$x$ and CMP--LC--$t$ models.}
\end{figure}

\begin{figure}[htbp]
\centering\includegraphics[scale=0.375]{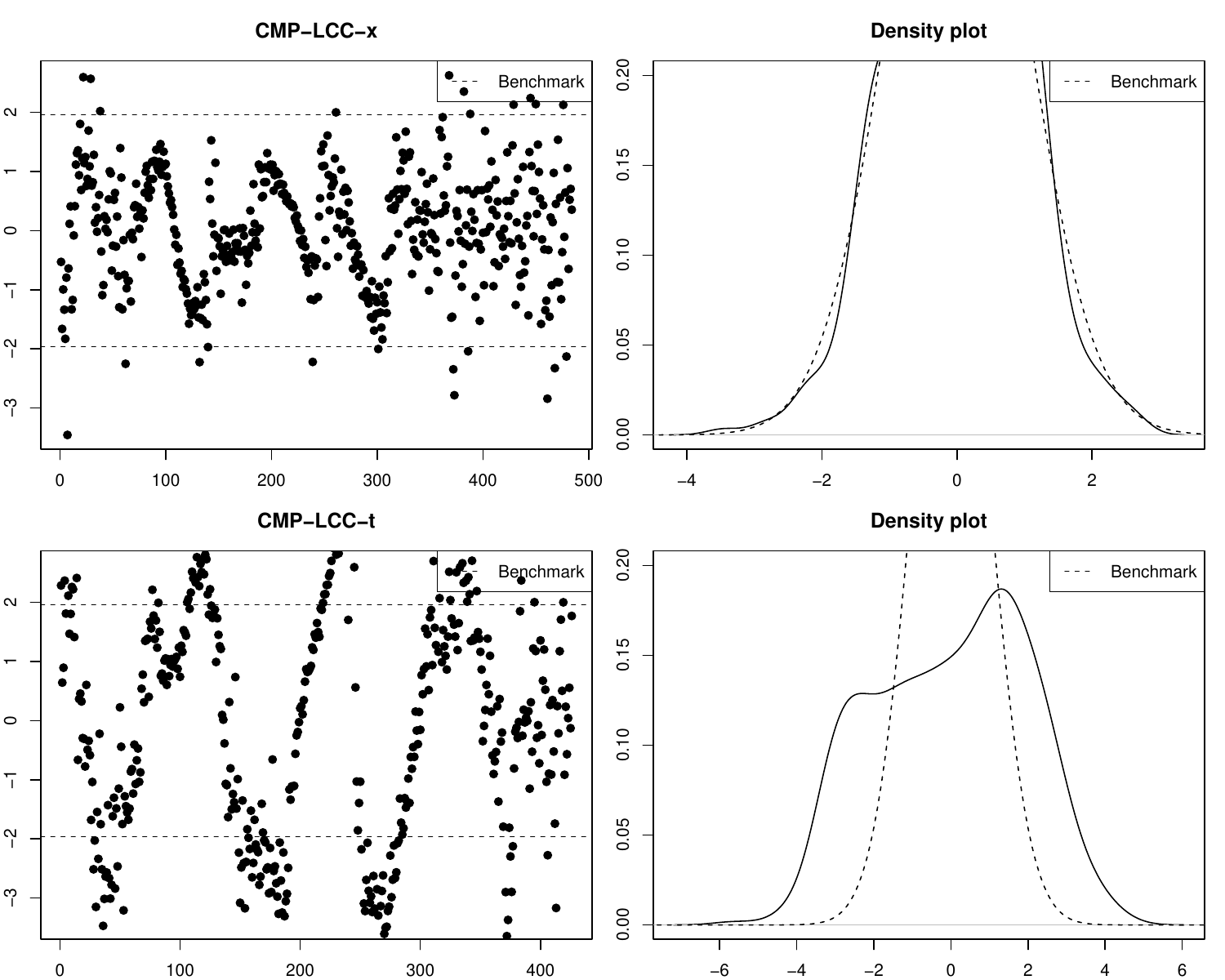}
\caption{\label{fig:geweke_LCC}Plot of Geweke's Z statistics under the CMP--LCC--$x$ and CMP--LCC--$t$ models.}
\end{figure}

\newpage

\bibliographystyle{chicago}
\bibliography{reference}